\documentclass[12pt,usenames,dvipsnames]{article}

\usepackage{latexsym}
\usepackage{amssymb,amsfonts,amsmath}
\usepackage{graphicx} 
\usepackage{indentfirst}
\usepackage{bbm}
\usepackage{amssymb}
\usepackage{verbatim}
\usepackage{amsmath, amsthm,amssymb}
\usepackage{mathrsfs}
\usepackage{hyperref}
\usepackage{amsfonts}
\usepackage{dsfont}
\usepackage{cite}
\usepackage{xcolor}
\usepackage{enumerate}
\usepackage{cleveref}

\parskip=\medskipamount

\newcommand {\cD}{{\cal D}}
\newcommand {\cE}{{\cal E}}

\newcommand {\cK}{{\cal K}}
\newcommand {\cL}{{\cal L}}
\newcommand {\cM}{{\cal M}}
\newcommand {\cN}{{\cal N}}
\newcommand {\cO}{{\cal O}}

\newcommand {\cR}{{\cal R}}
\newcommand {\cS}{{\cal S}}
\newcommand {\cT}{{\cal T}}
\newcommand {\cU}{{\cal U}}

\newcommand {\cW}{{\cal W}}

\def\a{\alpha}

\def\b{\beta}

\def\d{\delta}
\def\e{\epsilon}
\def\f{\phi}
\def\g{\gamma}
\def\G{\Gamma}

\def\j{\psi}
\def\k{\kappa}
\def\l{\lambda}
\def\m{\mu}
\def\n{\nu}
\def\o{\omega}

\def\q{\theta}
\def\r{\rho}
\def\s{\sigma}
\def\t{\tau}

\def\x{\xi}
\def\z{\zeta}
\def\D{\Delta}
\def\F{\Phi}
\def\J{\Psi}
\def\L{\Lambda}
\def\O{\Omega}

\def\S{\Sigma}
\def\U{\Upsilon}
\def\X{\Xi}
\def\tr{{\rm tr}}
\def\rd{{\rm d}}
\def\ri{{\rm i}}
\def\re{{\rm e}}

\newcommand{\ad}{{\dot{\alpha}}}                           %
\newcommand{\bd}{{\dot{\beta}}}                            %
\newcommand{\ve}{\varepsilon}                            %
\newcommand{\cDB}{{\bar\cD}}                            %

\renewcommand{\aa}{{\a\ad}}
\newcommand{\bb}{{\b\bd}}
\newcommand{\pa}{\partial}                           %
\newcommand{\hf}{\frac12}

\newcommand{\vf}{\varphi}

\newcommand{\be}{\begin{equation}}
\newcommand{\ee}{\end{equation}}
\newcommand{\bea}{\begin{eqnarray}}
\newcommand{\eea}{\end{eqnarray}}
\newcommand{\non}{\nonumber}

\newcommand{\bm}[1]{\mbox{\boldmath$#1$}}

\def\double #1{#1{\hbox{\kern-2pt $#1$}}}

\newcommand{\gd}{{\dot\g}}
\newcommand{\dd}{{\dot\d}}

\newcommand{\ts}{{\tilde{\s}}}

\newif\ifdtup

\newcommand{\bsubeq}{\begin{subequations}}
\newcommand{\esubeq}{\end{subequations}}

\newcommand{\eol}{\notag \\}

\numberwithin{equation}{section}

\newcommand{\sSp}{\mathsf{Sp}}
\newcommand{\sSU}{\mathsf{SU}}
\newcommand{\sSL}{\mathsf{SL}}
\newcommand{\sGL}{\mathsf{GL}}
\newcommand{\sSO}{\mathsf{SO}}
\newcommand{\sO}{\mathsf{O}}
\newcommand{\sU}{\mathsf{U}}

\begin{document}

\begin{titlepage}
\begin{flushright}
February, 2026 \\
Revised version: May, 2026
\end{flushright}
\vspace{5mm}

\begin{center}
{\Large \bf 
Nonlinear self-duality for arbitrary spin, superspin, \\
and supersymmetry 
type}\footnote{Based in part on the talks given at the Workshop 
{\it Conformal higher spins, twistors and boundary calculus} (University of Mons, Belgium, 30 June -- 04 July, 2025) and at the Arnold Sommerfeld Center for Theoretical Physics, the University of Munich.}
\end{center}

\begin{center}

{\bf Sergei M. Kuzenko 
} \\
\vspace{5mm}

\footnotesize{
{\it Department of Physics M013, The University of Western Australia,\\
35 Stirling Highway, Perth W.A. 6009, Australia}}  
\\
\vspace{2mm}

\end{center}

\begin{abstract}
\baselineskip=14pt
We review the general formalism of duality rotations for $\cal N$-extended (super)conformal gauge multiplets of arbitrary (super)spin in four dimensions, with ${\cal N} \geq 0$. Self-dual models for a vector field (${\cal N}=0$) and for ${\cal N}=1$ and ${\cal N}=2$ vector supermultiplets are naturally formulated on general (super)gravity backgrounds. 
For all other (super)spin values, the corresponding self-dual systems are realised on arbitrary conformally flat  backgrounds. Every $\mathsf{U}(1)$ duality-invariant model  is demonstrated to be self-dual with respect to a Legendre transformation. Methods are described to generate such self-dual  models including superconformal ones. We show that every model for self-dual nonlinear electrodynamics admits a higher-spin extension. Throughout the review, we make use of the formalism of conformal (super)space, that is the geometric setting to describe the gauge theory of the (super)conformal group.
\end{abstract}
\vspace{5mm}

\begin{center}
{\it Dedicated to the memory of Igor Viktorovich Tyutin $\qquad{}$} \\
\end{center}

\vfill

\vfill
\end{titlepage}

\newpage
\renewcommand{\thefootnote}{\arabic{footnote}}
\setcounter{footnote}{0}

\tableofcontents{}
\vspace{1cm}
\bigskip\hrule

\allowdisplaybreaks

\section{Introduction}

Since the construction of simple supergravity in four dimensions \cite{FvNF,DZ}, 
its extended versions and generalisations beyond four dimensions (see, e.g., \cite{SalamSezgin} for a review), the field-theoretic landscape has changed dramatically. Progress in supergravity has led to conceptually new developments in theoretical high-energy physics including the following: 
\begin{itemize}

\item Pure $\cN=2$ supergravity in four dimensions \cite{FvN} fulfilled Einstein's dream of unifying gravity and electromagnetism, albeit using a symmetry principle that was not known to Einstein --  local supersymmetry.

\item New types of gauge theories 
(as compared with the standard Yang-Mills theories) were introduced. Their specific features in the Lagrangian formalism are: (i) open gauge algebra; and/or (ii) linearly dependent gauge generators. These imply that covariant quantisation of such theories cannot be carried out using the Faddeev-Popov approach \cite{Faddeev:1967fc}. 
A powerful formalism to quantise general reducible gauge theories with open algebra has been developed 
\cite{BRS, Tyutin:1975qk, Batalin:1981jr, BV} known as the BRST-BV or antifield formalism
(see \cite{HT} for a review).

\item
  Supergravity was argued to be the only viable Kaluza-Klein theory  \cite{Duff:1986hr, Appelquist:1987nr}, see also 
 \cite{Duff:2025tot} for a recent reassessment. 

\item Following the patterns of electric-magnetic duality invariance observed in extended supergravity 
\cite{FSZ, Cremmer:1977tt, CJ},
the general formalism of nonlinear self-duality was developed in four \cite{GZ1, Zumino, GR1, GR2, GZ2, GZ3}
and higher \cite{GR1, Tanii, ArakiT, BMZ,  ABMZ} dimensions for non-supersymmetric 
theories.
Supersymmetric extensions of the formalism were given in \cite{KT1,KT2}.

\item Supergravity stimulated the construction of gauge-invariant models for free massless higher-spin fields
\cite{Fronsdal, FF, Fronsdal2, FF2}, interacting theories for massless higher-spin fields in AdS$_4$
\cite{Fradkin:1986qy, Fradkin:1987ks, Vasiliev:1990en}, and the development of conformal higher-spin theory
\cite{FT, Segal, Tseytlin}. 

\end{itemize}
Renaissance of electric-magnetic duality (in the form of nonlinear self-duality) is one of the many remarkable developments inspired by the progress of supergravity. This review is devoted to the generalisations of the concept of nonlinear self-duality to 
higher-spin fields and supermultiplets proposed in  \cite{KR21-2, Kuzenko:2023ebe}.
To start with, it is worth giving a brief history of duality invariance in (nonlinear) electrodynamics. 

Maxwell's electrodynamics in Minkowski space ${\mathbb M}^4$ is the simplest and oldest example of a duality-invariant theory. 
Its Lagrangian is constructed in terms of the electromagnetic field strength\footnote{In the literature, the two-form field strength 
$F_{mn}$ is often called the {\it Maxwell tensor} or the {\it Faraday tensor}. In fact, to the best of our knowledge, it was introduced for the first time by Minkowski in 1908 \cite{Minkowski} who also rewrote the Maxwell equations in the modern relativistic form, including the equations for free electromagnetic field \eqref{freeMaxwell}.}  
$F_{mn}=-F_{nm}$ and has the form
\bea
L_{\rm Maxwell}  (F)= - \frac 14 F^{mn} F_{mn} = \hf \big( \vec{E}^2 - \vec{B}^2 \big)~,
\qquad F_{mn } = \pa_m  A_n - \pa_n  A_m\, .
\eea
The Bianchi identity and the equation of motion \cite{Minkowski}  are
\bea
\pa_m \widetilde{F}^{mn} = 0\, , \qquad 
\pa_m F^{mn} = 0 \, ,
\label{freeMaxwell}
\eea
with $\widetilde{F}^{mn} :=\hf  \ve^{mn r s} F_{rs} $  the Hodge dual of $F$. 
Since both differential equations have the same functional form, one may consider electric-magnetic duality rotations
\bea
F+\ri \widetilde F \to \re^{-\ri \l} \big( F+\ri \widetilde F \big) \quad \Longleftrightarrow \quad 
\vec{E} +\ri \vec{B} \to \re^{-\ri \l} \big( \vec{E} +\ri \vec{B} \big) ~, \qquad 
\l \in {\mathbb R}\, .
\eea
 These duality transformations change
the Lagrangian $L_{\rm Maxwell} (F)$, but the energy-momentum tensor 
\bea
T_{mn} = \hf \big( F+\ri \widetilde F\big)_{mr}  \big( F-\ri \widetilde F\big)_{ns} \eta^{rs}   
= F_{mr} F_{ns} \eta^{rs} - \frac 14 \eta_{mn} F^{rs} F_{rs} 
\label{EMT}
\eea
remains invariant.

A nonlinear extension of duality transformations was put forward by Schr\"odinger
ninety years ago \cite{Schrodinger:1935oqa}.
He studied the model for nonlinear electrodynamics proposed in 1934 by Born and Infeld \cite{BI}
\bea
L_{\rm BI} (F) &=& \frac{1}{{g}^2} \Big\{\;
1 - \sqrt{- \det (\eta_{mn} + {g} F_{mn} )} 
\Big\}  = -\frac{1}{4} F^{mn}F_{mn} + \cO(F^4) \non \\
&=&  \frac{1}{g^2} \Big\{
1 - \sqrt{1 +g^2 (\vec{B}^2 {- \vec{E}^2}) -g^4 (\vec{E}\cdot \vec{B} )^2 } 
\Big\} 
\label{BI1}
\eea
with $g$ the coupling constant. 
Born and Infeld were not guided by considerations of  duality invariance.
Their theory was designed to provide a solution to the problem of infinite self-energy of a point charge 
in Maxwell electrodynamics.\footnote{At the heart of the solution is the existence of an upper bound on values of the electric field strength in the Born-Infeld theory, which is $|\vec{E}| < g^{-1}$.}
However it was Schr\"odinger who observed that the Born-Infeld theory possesses  
a reformulation with manifest $\sU(1)$ duality invariance.
In a modern setting, the $\sU(1) $ duality-invariance of the Born-Infeld theory was first described by Bialynicki-Birula \cite{B-B}.

Born and Infeld viewed their model as a new fundamental theory of the electromagnetic field. As is well-known, their
great expectations have never come true. However the Born-Infeld action re-appeared in the spotlight of theoretical physics in the 1980s as a low-energy effective action in open string theory \cite{Fradkin:1985qd} and the world-volume action of D-branes 
\cite{Leigh:1989jq}.

Long before the Born-Infeld theory resurfaced in string theory \cite{Fradkin:1985qd, Leigh:1989jq},
it had been observed that ungauged extended supergravity theories in four dimensions exhibit electric-magnetic duality symmetry \cite{FSZ, Cremmer:1977tt, CJ}. These observations motivated Gaillard and Zumino to develop the general theory of duality invariance for Abelian vector fields non-minimally coupled to scalar and spinor matter fields \cite{GZ1, Zumino}. In particular, they demonstrated that the maximal duality group for a system of  $n$ interacting field strengths is the {\it compact} group $\sU(n)$. In the presence of scalars, the duality symmetry can be enhanced to the {\it non-compact} real symplectic group $\sSp(2n, {\mathbb R})$. The Gaillard-Zumino formalism is  very powerful, and arguably applicable to more general dynamical systems than those considered in \cite{GZ1, Zumino}. Ref. \cite{GZ1} determined the most general nonlinear $\s$-model form of the scalar field sector in any duality-invariant theory assuming its Lagrangian to be at most quadratic in the field strengths, as is typical for extended supergravity theories. However, it turns out that this restriction on the gauge field sector can be relaxed. In 1995 Gibbons and Rasheed found the general structure of models for $\sU(1)$ duality-invariant nonlinear electrodynamics \cite{GR1}
(including the Born-Infeld action) and demonstrated that the coupling of such a theory
to the dilaton and axion is completely fixed by the requirement of $\sSL(2, {\mathbb R}) \cong \sSp(2, {\mathbb R})$ duality invariance \cite{GR2}. Two years later, Gaillard and Zumino explained how one could have derived the results of \cite{GR1,GR2} by applying the formalism developed in \cite{GZ1}. Since the general structure of self-dual nonlinear electrodynamics and the 
corresponding $\sSL(2, {\mathbb R}) $ coupling to the dilaton and axion was developed in \cite{GZ1, Zumino, GR1,GR2,GZ2,GZ3},
it is natural to refer to the corresponding techniques as the Gaillard-Zumino-Gibbons-Rasheed  (GZGR) formalism.

The GZGR formalism admits a natural extension to higher dimensions \cite{GR1, Tanii,ArakiT,BMZ, ABMZ} (see also \cite{KT2,AFZ,Tanii2} for a review). In four dimensions, this setting has been generalised to  
$\sU(1)$ duality-invariant models for $\cN=1$ and $\cN=2$ supersymmetric nonlinear electrodynamics, both in 
the globally \cite{KT1,KT2} and locally \cite{KMcC,KMcC2,K12} supersymmetric cases. 
The formulation for self-dual supersymmetric nonlinear electrodynamics has been further generalised to 
the general formalism of duality rotations for $\cal N$-extended (super)conformal gauge multiplets 
of arbitrary (super)spin, with ${\cal N} \geq 0$ \cite{KR21-2, Kuzenko:2023ebe}.

In 1981, Gaillard and Zumino \cite{GZ1} made the following comment: ``It appears that the duality invariance of supergravity theories is implied by supersymmetry, a fact which still remains very mysterious.''  Some twenty years later, it was pointed out \cite{KT2}  that ``self-duality turns out to be intimately connected with spontaneous breaking of supersymmetry (for still not completely understood reasons).'' The latter phenomenon was perhaps the main motivation to study supersymmetric self-dual systems at the turn of the millennium. The intimate connections between nonlinear self-duality and supersymmetry include the following:
\begin{itemize}
\item
In the case of partial spontaneous $\cN=2 \to \cN=1$ supersymmetry breaking, 
the Maxwell-Goldstone multiplet \cite{BG,RT}
(coinciding with the $\cN=1$ supersymmetric Born-Infeld
action \cite{CF}) and the tensor Goldstone multiplet
\cite{BG2,RT} were shown  in \cite{KT1,KT2} to be invariant under supersymmetric $\sU(1)$ duality rotations.The  Maxwell-Goldstone multiplet for partial $\cN=2 \to \cN=1$ supersymmetry breaking has also been extended \cite{KT-M16} to the following maximally supersymmetric backgrounds: (i)  ${\mathbb R} \times S^3$; (ii) ${\rm AdS}_3 \times {\mathbb R}$; and (iii) a supersymmetric plane wave.\footnote{There exist only five maximally supersymmetric backgrounds in four-dimensional  $\cN=1$ off-shell supergravity \cite{FS}: (i) ${\mathbb M}^4$; (ii) AdS$_4$; (iii) ${\mathbb R} \times S^3$; (iv) ${\rm AdS}_3 \times {\mathbb R}$ (or its covering ${\rm AdS}_3 \times {\mathbb R}$); and (v) a  pp-wave spacetime isometric to the Nappi-Witten group NW$_4$ \cite{NappiW}. The Maxwell-Goldstone multiplet models
for partial $\cN=2 \to \cN=1$ supersymmetry breaking are known for all of them except for AdS$_4$.}
This theory possesses $\sU(1)$ duality invariance.

 \item
 Extending the earlier incomplete proposal of \cite{Ketov}, 
 it was suggested in \cite{KT2} that the Maxwell-Goldstone multiplet for 
 partial  $\cN=4 \to \cN=2$ supersymmetry breakdown (proposed to be the $\cN=2$ supersymmetric Born-Infeld action) is a unique $\cN=2$ vector multiplet theory with the following properties: (i) it possesses $\sU(1)$ duality invariance; and (ii) it is invariant under a nonlinearly realised central charge bosonic symmetry. Within the
perturbative approach to constructing the $\cN=2$ supersymmetric Born-Infeld action 
elaborated in  \cite{KT2}, the uniqueness of the action was demonstrated to order $W^{10}$ in powers of the chiral superfield strength $W$.
A year later, a powerful formalism of nonlinear realisations for
the partial  $\cN=4 \to \cN=2$ supersymmetry breaking was developed \cite{BIK1}
which supported the uniqueness of the $\cN=2$ supersymmetric Born-Infeld action 
and reproduced \cite{BIK2} the perturbative results of \cite{KT2}.
Further progress towards the construction of the $\cN=2$ supersymmetric Born-Infeld action has been achieved in 
\cite{BCFKR, Ivanov:2013maa}.

\item For a large family of  $\sU(1)$ duality-invariant models for $\cN=1$ supersymmetric nonlinear electrodynamics \cite{KT1}, it was demonstrated \cite{KMcC2} that the component fermionic action, which is obtained by switching off the bosonic fields, is equivalent (modulo a nonlinear field redefinition) to the Akulov-Volkov action for the Goldstino \cite{VA,AV}.

\end{itemize}

Two theoretical developments  have proved of primary importance for the construction
of self-dual models for $\cal N$-extended (super)conformal gauge multiplets 
of arbitrary (super)spin \cite{KR21-2, Kuzenko:2023ebe}. These are: (i) the Ivanov-Zupnik (IZ) 
auxiliary-field formulation for self-dual nonlinear electrodynamics \cite{IZ_N3, IZ1, IZ2}; and (ii) the ModMax theory
\cite{BLST}. 
\begin{itemize} 
\item
The IZ approach is a powerful formalism to generate self-dual models for nonlinear electrodynamics. This formalism has been extended to the $\cN=1$ and $\cN=2$ supersymmetric cases \cite{K13, ILZ}. 
Some time ago there was a revival of interest in the duality-invariant dynamical systems
\cite{BN,CKR,Chemissany:2011yv,BCFKR} inspired by the desire to achieve a better understanding
of   the UV properties of extended supergravity theories. 
The authors of \cite{BN,CKR,Chemissany:2011yv}  put forward the so-called ``twisted self-duality constraint'' 
as a systematic procedure to generate duality-invariant theories. 
However, it has been demonstrated \cite{IZ3} that the non-supersymmetric construction of  
\cite{BN,CKR,Chemissany:2011yv} naturally originates within the more general approach 
previously developed in \cite{IZ1,IZ2}. Specifically, the twisted self-duality constraint 
corresponds to an equation of motion in the approach of \cite{IZ1,IZ2}.

\item
The ModMax theory is a unique $\sU(1)$ duality-invariant and conformal model for nonlinear electrodynamics constructed by Bandos, Lechner, Sorokin and Townsend.  It is a one-parameter deformation of Maxwell's theory, which is why it was called the modified Maxwell theory. The ModMax theory has been generalised to the $\cN=1$ supersymmetric case \cite{BLST2,K21} and conformal higher-spin fields  \cite{KR21-2}. 
There also exists a supersymmetric nonlinear $\s$-model analogue of
the ModMax theory \cite{Kuzenko:2023ysh} known as the MadMax $\sigma$-model. 

\end{itemize}

This paper is a review of the general formalism of nonlinear self-duality for $\cal N$-extended (super)conformal gauge multiplets of arbitrary (super)spin in four dimensions, with ${\cal N} \geq 0$. In fact, self-dual models for an Abelian  vector field (${\cal N}=0$) and for ${\cal N}=1$ and ${\cal N}=2$ vector supermultiplets are naturally formulated on general (super)gravity backgrounds. In this paper, all models for self-dual nonlinear electrodynamics (including the higher-derivative deformations of the ModMax theory) are formulated on an arbitrary gravitational background. In the supersymmetric case, our discussion is restricted to conformally flat backgrounds in order to keep the length of the review to be under one hundred pages. Being mainly a review of the known results, this work contains a few original points, which are:
\begin{itemize} 
\item A unified (super)conformal approach to formulate $\sU(1)$ duality invariance for arbitrary spin, superspin and supersymmetry type.

\item The demonstration that every model for self-dual nonlinear electrodynamics admits a higher-spin extension.

\item The composite primary field defined in \eqref{CompositePrimary} is derived in this work for the first time.
The significance of this composite field is that: (i) it offers a manifestly conformal and $\sSL(2,{\mathbb R})$ invariant formulation for the higher-derivative nonlinear sigma model \eqref{dilaton-axionHD} that describes the dynamics of the 
dilation and axion fields taking their values in $\sSL(2,{\mathbb R})/ \sSO(2)$; and (ii) it has an application in the context of Weyl anomalies, as will be discussed elsewhere.

\end{itemize}

This paper is organised as follows. Section \ref{Section2}  contains a pedagogical review of conformal gravity as the gauge theory of the conformal group in $D$ dimensions. The language of conformal geometry reviewed in Section \ref{Section2} then will be used throughout Sections \ref{section3} to \ref{section6} to describe conformal gauge fields and the $\sU(1)$ duality-invariant models of such fields. Section \ref{section3} discusses conformal fields of arbitrary spin. Section \ref{section4} provides a modern review of models for self-dual nonlinear electrodynamics and their fundamental properties.  Section \ref{section5} is devoted to $\sU(1)$ duality-invariant systems for real conformal gauge fields, while Section \ref{section6} extends the discussion to the case of complex conformal gauge fields such as the conformal gravitino. $\cN$-extended superconformal gauge multiplets are discussed in Section 
\ref{section7}. Self-dual models for superconformal gauge multiplets are reviewed in \ref{section8}. Section \ref{section9} is devoted to self-dual models for the $\cN=2$ superconformal gravitino multiplet. Discussion and conclusions are given in Section \ref{section10}.

The main body of the paper is accompanied by four technical appendices. Appendix \ref{appendixA}  derives the commutation relations for the conformal algebra, $\mathfrak{so}(D,2)$, by making use of the conformal Killing vector fields on Minkowski space ${\mathbb M}^D$. Appendix \ref{appendixB} discusses an algebraic derivation of several conformal differential operators as well as of primary composite fields. Appendix \ref{appendixC} contains the graded commutation relations for the $\cN$-extended superconformal algebra. Finally, degauging of the $\cN$-extended  conformal superspace with flat connection is reviewed in Appendix \ref{appendixD}.

Throughout this paper we often make use of the convention whereby indices denoted by the same symbol are to be symmetrised over, e.g. 
\begin{align}
	U_{\a(m)} V_{\a(n)} = U_{(\a_1 . . .\a_m} V_{\a_{m+1} . . . \a_{m+n})} =\frac{1}{(m+n)!}\big(U_{\a_1 . . .\a_m} V_{\a_{m+1} . . . \a_{m+n}}+\cdots\big)~, \label{convention}
\end{align}
with a similar convention for dotted spinor indices.


\section{Conformal geometry} \label{Section2} 

This section is devoted to a brief review of conformal gravity as the gauge theory 
of the conformal group $\sO(D,2)/{\mathbb Z}_2$. This approach was pioneered in four dimensions by Kaku, Townsend and van Nieuwenhuizen \cite{KTvN, KTvN2}, 
as a generalisation of the MacDowell-Mansouri  construction \cite{MacDowell:1977jt}
of anti-de Sitter (super)gravity as the gauge theory of the anti-de Sitter (super)group. 
The formalism of conformal geometry, which is reviewed below, is a natural development of the ideas put forward 
in \cite{KTvN}.  In fact this formalism was introduced as a byproduct of the $\cN=1$ and $\cN=2$ conformal superspace formulations in four dimensions developed by Butter \cite{ButterN=1,ButterN=2}, and subsequently generalised to $\cN=3$ \cite{KR23} and $\cN=4$ \cite{ButterN=4} conformal supergravity theories in four dimensions, 
$\cN$-extended conformal supergravity in three dimensions \cite{BKNT-M1,BKNT-M2,KNT-M}, conformal supergravity in five dimensions \cite{BKNT-M15}, and $(1,0)$ \cite{BKNT} and $(2,0)$ \cite{Kennedy:2025nzm} conformal supergravity theories in six dimensions. 
 Here we closely follow the presentations given in  \cite{BKNT-M1,BKNT}.\footnote{One may compare this formalism with the Weyl-covariant tensor calculus developed by Boulanger \cite{Boulanger:2004eh}.}

It should be pointed out that there exists a different approach to formulate conformal  geometry. 
It  was developed by mathematicians and is often referred to as 
 tractor calculus \cite{BEG,Gover}, with its roots going back to the work of Thomas
\cite{Thomas}.  The two approaches to conformal geometry are essentially equivalent
and complementary.

Conformal gravity in $D>2$ dimensions is obtained by gauging the conformal
algebra, $\mathfrak{so}(D,2)$, which is spanned by the translation ($P_a$), Lorentz ($M_{ab}$), dilatation ($\mathbb{D}$) and special conformal  ($K^a$) generators. Their non-vanishing commutation relations\footnote{For completeness, the derivation of these commutation relations is provided in Appendix \ref{appendixA}.}  are
\begin{subequations}
	\label{ConAlg.1}
	\begin{align}
		[M_{ab} , M_{cd}] &= 2 \eta_{c[a} M_{b] d} - 2 \eta_{d [a} M_{b] c} ~, \\
		[M_{ab} , P_c ] &= 2 \eta_{c [a} P_{b]} ~, \quad [\mathbb D, P_a] = P_a ~, \\
		[M_{ab} , K_c] &= 2 \eta_{c[a} K_{b]}  ~, \quad  [\mathbb D, K_a] = - K_a ~, \\
		[K_a , P_b] &= 2 \eta_{ab} \mathbb D + 2 M_{ab} ~,
	\end{align}
\end{subequations}
with $\eta_{ab}$ the mostly plus Minkowski metric.
It is convenient to think of $\mathfrak{so}(D,2)$ as the direct sum of two Lie subalgebras 
generated by the operators
\begin{align}
	P_a \quad \& 
	\quad X_{\underline a} = (M_{ab},\mathbb{D},K^a)~,
	\label{generators}
\end{align}
respectively. Then, the conformal algebra \eqref{ConAlg.1} may be rewritten as follows
\bsubeq \label{structure}
\begin{align}
	[X_{\underline{a}} , X_{\underline{b}} ] &= -f_{\underline{a} \underline{b}}{}^{\underline{c}} X_{\underline{c}} \ , \\
	[X_{\underline{a}} , P_{{b}} ] &= -f_{\underline{a} { {b}}}{}^{\underline{c}} X_{\underline{c}}
	- f_{\underline{a} { {b}}}{}^{ {c}} P_{ {c}} \label{ConAlg.2}
	\ ,
\end{align}
\esubeq
where the structure constants can be read off from \eqref{ConAlg.1}.


\subsection{Gauging the conformal algebra}
\label{Section2.1}

Let $\mathcal{M}^D$ be a curved $D$-dimensional spacetime  parametrised by the local coordinates $x^m$. 
To gauge the conformal
algebra $\mathfrak{so}(d,2)$,
we associate a connection one-form with each generator in \eqref{generators}, 
\begin{subequations} 
\bea
P_a & ~\to ~ &e^a = \rd x^m e_m{}^a~,  \\
X_{\underline{a}} = (M_{ab},\mathbb{D},K^a)&~\to ~&\o^{\underline{a}} = (\hat{\o}^{ab},\mathfrak{b},\mathfrak{f}_a)=\rd x^{m} \o_m{}^{\underline{a}}~, 
\eea 
\end{subequations}
The vielbein one-forms $e^a = \rd x^m e_m{}^a (x)$ are assumed to form a basis of the cotangent space $T^*_p \cM^D$
at each point $p \in \cM^D$,
 $e:={\rm det}(e_m{}^a) \neq 0$, hence there exists a unique inverse vielbein 
 \begin{align}
e_a = e_a{}^m (x)\pa_m~, \qquad 	e_a{}^m e_m{}^b = \d_a{}^b~, \qquad e_m{}^a e_a{}^n=\d_m{}^n~.
\end{align}
The vector fields $e_a$
constitute a basis for the tangent space $T_p\cM^D$ at each
point $ p\in \mathcal{M}^D$. 
The inverse vielbein is  used to express the connection  $\omega^{\underline{a}} $ as
\bea
\omega^{\underline{a}} =e^b\omega_b{}^{\underline{a}}~, \qquad 
\omega_b{}^{\underline{a}}:=e_b{}^m\omega_m{}^{\underline{a}}~. 
\eea

Our next step is to introduce a conformally covariant derivative 
\begin{subequations} \label{GaugeCovDer}
\bea 
\nabla = e^a \nabla_a = \rd - \hf \hat{\omega}^{bc} M_{bc} - \mathfrak{b} \, \mathbb D - \mathfrak{f}^a K_a ~.
\eea
where the first-order operator $\nabla_a$  replaces the translation generator $P_a$, 
\bea
P_a~\to ~\nabla_a =  e_a{}^m \partial_m  - \o_a{}^{\underline b} X_{\underline b} = e_a{}^m \partial_m - \hf \hat{\omega}_a{}^{bc} M_{bc} - {\mathfrak b}_a \mathbb D - \mathfrak{f}_{ab} K^b ~.
\eea
\end{subequations}
It is postulated that a commutator $[X_{\underline{a}} , \nabla_{{b}} ] $ is obtained from \eqref{ConAlg.2} by replacing $P_b \to \nabla_b$, 
\begin{align}
	[X_{\underline{a}} , \nabla_{{b}} ] &= -f_{\underline{a} { {b}}}{}^{\underline{c}} X_{\underline{c}}
	- f_{\underline{a} { {b}}}{}^{ {c}} \nabla_{ {c}} 
	\ .
\end{align}
The commutation relation $[P_a, P_b]=0$ turns into
\bea
[\nabla_a , \nabla_b] = -\cT_{ab}{}^c \nabla_c - \hf \cR(M)_{ab}{}^{cd} M_{cd}
- \cR(\mathbb{D})_{ab} \mathbb D - \cR(K)_{abc} K^c \ ,
\eea
where the right-hand side is given in terms of  the {torsion tensor} $\cT_{ab}{}^c $ and 
the {curvature tensors} $\cR(M)_{ab}{}^{cd}, ~
 \cR(\mathbb{D})_{ab}$ and $   \cR(K)_{abc}$. They have the following explicit expressions:
\begin{subequations}
	\begin{align}
		\mathcal{T}_{ab}{}^c&=-\mathcal{C}_{ab}{}^{c}+2\hat{\o}_{[ab]}{}^c+2\mathfrak{b}_{[a}\delta_{b]}{}^{c}~,\\
		\mathcal{R}(M)_{ab}{}^{cd}&=\hat{R}_{ab}{}^{cd}+8\mathfrak{f}_{[a}{}^{[c}\delta_{b]}{}^{d]}~,\label{LorCur}\\ 
		\mathcal{R}(K)_{abc}&=-\mathcal{C}_{ab}{}^d\mathfrak{f}_{dc}-2\hat{\o}_{[a|c|}{}^{d}\mathfrak{f}_{b]d}-2{\mathfrak b}_{[a}\mathfrak{f}_{b]c}+2e_{[a}\mathfrak{f}_{b]c}~, \label{2.19cc}\\
		\mathcal{R}(\mathbb{D})_{ab}&= -\mathcal{C}_{ab}{}^{c}{\mathfrak b}_c+4\mathfrak{f}_{[ab]}+2e_{[a}{\mathfrak b}_{ b]}~.
	\end{align}
\end{subequations}
Here we have introduced the anholonomy coefficients $\mathcal{C}_{ab}{}^{c}$ defined by 
\begin{align}
	[e_a,e_b] = \mathcal{C}_{ab}{}^{c} e_c ~,
\end{align}
and the standard curvature tensor\footnote{Owing to its dependence on the dilatation connection, this curvature tensor does not satisfy the Bianchi identity $R_{[abc]d}=0$ unless $\mathfrak{b}_a=0$, see the following subsection.}  
\begin{align}
\hat{R}_{ab}{}^{cd}&=-\mathcal{C}_{ab}{}^{f}\hat{\o}_{f}{}^{cd}+2e_{[a}\hat{\o}_{b]}{}^{cd}
		-2\hat{\o}_{[a}{}^{cf}\hat{\o}_{b]f}{}^{d} \label{RiemannB}
\end{align}
constructed from the Lorentz connection $\hat{\omega}_a{}^{bc}$.  

We have used the term `gauge covariant derivative' for \eqref{GaugeCovDer}.
By definition, the gauge group of conformal gravity is generated by local transformations of the form
\begin{subequations}\label{CGgentransmations}
	\bea
	\delta_{\mathcal K} \nabla_a &=& [\mathcal{K},\nabla_a] \ , \\
	\mathcal{K} &=& \xi^b \nabla_b +  \L^{\underline{b}} X_{\underline{b}}
	=  \xi^b \nabla_b + \hf K^{bc} M_{bc} + \s \mathbb{D} + \L_b K^b ~,
	\eea
provided we interpret
\be
\nabla_a \xi^b := e_a \xi^b + \omega_a{}^{\underline{c}} \xi^d f_{d \underline{c}}{}^{b} \ , \quad 
\nabla_a \L^{\underline{b}} := e_a \L^{\underline{b}} + \omega_a{}^{\underline{c}} \xi^d f_{d\underline{c}}{}^{\underline{b}}
+ \omega_a{}^{\underline{c}} \L^{\underline{d}} f_{\underline{d} \underline{c}}{}^{\underline{b}} \ ,
\ee
where the structure constants are defined by \eqref{structure}.
These gauge transformations act on a conformal tensor field $\mathcal{U}$ (with its indices suppressed) as 
\bea 
\label{CGgenMattertfs}
\d_{\mathcal K} \mathcal{U} = {\mathcal K} \mathcal{U} ~.
\eea
\end{subequations}
Of course, it is necessary to have a realisation of the conformal algebra on $\cU$. 
Such a realisation exists for conformal primary fields.

We say that $\mathcal{U}$ is a primary field of dimension $\D$ if the following conditions hold:

	\begin{subequations}
	(i) $\cU$ is annihilated by the special
conformal generator, 
	\begin{align}
		K^a \mathcal{U} = 0;
	\end{align}
	
	(ii) $\cU$ is an eigenvector of the dilatation generator, $\mathbb D$, with eigenvalue $\D$
	\begin{align}
		\mathbb D \cU = \D \cU~.
	\end{align}
\end{subequations}
Thus $\nabla_{a}\,\cU= \big(e_{a}{}^{m}\partial_m-\frac{1}{2} \hat{\omega}_{a}{}^{bc}M_{bc}-\D\mathfrak{b}_a\big) \cU$. 


\subsection{Conformally covariant constraints} \label{Section2.2}

In order for the above geometry to describe conformal gravity, it is necessary to impose certain covariant constraints such that the only independent geometric field, modulo gauge freedom, is the vielbein. Such constraints are:
\begin{subequations}
	\label{CC}
	\begin{align}
		\cT_{ab}{}^c &=0 ~, \label{CCa}\\
		\eta^{bd} \cR(M)_{abcd} &=0~. \label{CCb}
	\end{align}
\end{subequations}
The first constraint determines $\hat{\o}_{a}{}^{bc}$ in terms of the vielbein and dilatation connection  $\mathfrak{b}_a$, 
\begin{align}
  \hat{\omega}_{abc}=\o_{abc}-2\eta_{a[b}\mathfrak{b}_{c]}~,
  \quad 
  \o_{abc}
 =\frac{1}{2}\big(\mathcal{C}_{abc}-\mathcal{C}_{acb}-\mathcal{C}_{bca}\big)~,
 \label{LorentzConnection}
  \end{align}
where  $\o_{abc} \equiv \o_{abc} (e)$
 is the  torsion-free Lorentz connection. 
Constraint \eqref{CCa} is conformal since $K_a \cT_{bc}{}^d =0$, see \cite{BKNT-M1} for the technical details.
Constraint \eqref{CCb} 
determines the special conformal connection
$\mathfrak{f}_{a}{}^b$ in terms of $e^a$ and $\mathfrak{b}_a$,
 \begin{align}
 \mathfrak{f}_{ab}=-\frac{1}{2(D-2)}\hat{R}_{ab}+\frac{1}{4(D-1)(D-2)}\eta_{ab}\hat{R}~, 
 \end{align}
 where $\hat{R}_{ab}=\eta^{cd}\hat{R}_{acbd}$ is the (non-symmetric) Ricci tensor and $\hat{R}=\eta^{ab}\hat{R}_{ab}$ is the scalar curvature. 
Constraint \eqref{CCb} proves to be conformal, see   \cite{BKNT-M1} for the technical details.

Upon imposing the constraints \eqref{CC} we stay with two independent gauge fields, the vielbein $e^a$ and dilatation connection $\mathfrak{b}$. However, the latter  is a purely gauge degree of freedom since
\bea
\d_K(\L)\, \mathfrak{b}_a = -2 \L_a \ .
\non
\eea
Thus we can choose the gauge condition 
\bea
\mathfrak{b}_a=0~,
\label{b-gauge}
\eea
which fixes the special conformal gauge freedom.


\subsection{Conformal gravity in $D>3$ dimensions} \label{Section2.3}

Making use of the constraints \eqref{CC} and analysing 
the Bianchi identity
\begin{align}
0=\big[\nabla_a,[\nabla_b,\nabla_c]\big]+\big[\nabla_b,[\nabla_c,\nabla_a]\big]+\big[\nabla_c,[\nabla_a,\nabla_b]\big]~,
\end{align}  
one observes  that the  covariant derivatives obey the algebra 
\begin{align}
[\nabla_a,\nabla_b]=-\frac{1}{2}C_{abcd}M^{cd}-\frac{1}{2(D-3)}\nabla^dC_{abcd}K^c~.
\label{Cotton}
\end{align}
Here $C_{abcd}$ is
the Weyl tensor satisfying the algebraic properties\footnote{
The symmetry property $C_{abcd} = C_{cdab}$ is not independent and follows from 
the others.} 
\bea
\eta^{bc} C_{abcd} =0\ , \quad
 C_{abcd} = C_{[ab][cd]} \ , \quad C_{[abc]d} = 0
\eea
and the Bianchi identity
\bea \nabla_{[a} C_{bc]}{}^{de} = - \frac{2}{D - 3} \nabla_{f} C_{[ab}{}^{f [d} \d_{c]}^{e]}  \ .
\eea
The Weyl tensor proves to be primary, $K^e  C_{abcd} =0$. 
Therefore, it is independent of  the dilatation connection $\mathfrak{b}$. 
In other words, when the explicit expression for the Lorentz connection is used, 
all dependence on $\mathfrak{b}_a$ drops out of the Weyl tensor.

 It follows from \eqref{Cotton} that all curvatures are expressed in terms of the Weyl tensor. 
Spacetime is conformally flat if $C_{abcd} =0$. 


\subsection{Degauging to Lorentzian geometry} \label{Section2.4}

In the gauge \eqref{b-gauge},  the special conformal gauge freedom is fixed, and  the conformally covariant 
derivative takes the form
\bea
\nabla_a = \cD_a - \mathfrak{f}_a{}^b K_b \ , \quad \cD_a := e_a - \hf \omega_a{}^{bc} M_{bc} \ ,
\eea 
with $\cD_a$ the torsion-free Lorentz covariant derivative,
\begin{align}
	[\cD_a , \cD_b] 
	= - \hf R_{ab}{}^{cd} M_{cd} ~.
\end{align}
The special conformal connection takes the form
\bea 
\mathfrak{f}_{ab} = - \frac{1}{2 (D - 2)} R_{ab} + \frac{1}{4 (D - 1) (D - 2)} \eta_{ab} R  = - \hf P_{ab}~,
\eea
where
$R_{ac} := \eta^{bd} R_{abcd} $ is the Ricci tensor, $ R := \eta^{ab} R_{ab} $
the scalar curvature and $P_{ab}$  the {Schouten tensor}.
The Riemann tensor is
\bea  
 R_{ab}{}^{cd} = C_{ab}{}^{cd} +4 \d_{[a}^{[c} P_{b]}{}^{d]}  \ .
\eea

In the gauge \eqref{b-gauge},  two types of residual gauge transformations remain: 
\begin{itemize}
\item
Combined general coordinate and local Lorentz transformation 
\bea
	\delta_{\mathcal K} \cD_a = [\mathcal{K},\cD_a] \ , \qquad
	\mathcal{K} = \xi^b \cD_b + \hf K^{bc} M_{bc} 
\eea
acting on a tensor field $\cU$ (with indices suppressed) as
	\begin{align}
		\d_\cK \cU = \cK \cU ~.
	\end{align}	
\item 
Weyl transformation 
\begin{align}
	\mathcal{K}(\s) = \s \mathbb{D} + \frac{1}{2} \nabla_b \s K^b \quad \Longrightarrow \quad \d_{\mathcal{K}(\s)} \mathfrak{b}_a = 0
\end{align}
acting on a {primary field} $\cU$ of dimension $\D$ as 
{($\d_\s \equiv \d_{\mathcal{K}(\s) }$)}
\bea
\d_\s \cU = \D \s \cU~.
\eea
The Weyl transformation of the covariant derivative 
\begin{align}	
\d_\s \cD_a = \s \cD_a + \cD^b \s M_{ba} ~\implies ~
		\d_\s C_{abcd} = 2 \s C_{abcd} ~. 
		\label{WeylTr}
	\end{align}
\end{itemize}


\subsection{Conformal action principle}\label{Section2.5}

In order to formulate conformal field theories, an action principle is required. It is
\bea
\label{action}
S = \int\text{d}^{D}x \, e \, {L} ~,
\eea
where the Lagrangian $L$ is a real primary scalar field of dimension $D$, 
\bea
 K^a L =0~, \qquad {\mathbb D} L = D L~.
\eea
It is a short exercise to demonstrate that $S$ is invariant under 
the gauge transformation \eqref{CGgentransmations}.

As a well-known  example, we consider the action for conformal gravity in four dimensions 
\begin{align}
	\label{CGaction}
	S_\text{C.G.} = \int\text{d}^{4}x \, e \, C^{abcd} C_{abcd}~,
\end{align}
which is evidently conformal.

In six dimensions there are three invariants for conformal gravity \cite{BPB,DS,KMM}. They are generated by two cubic primaries
\begin{subequations}\label{CCC}
\bea
L_{C^3}^{(1)} &:=& C_{abcd} C^{aefd} C_{e}{}^{bc}{}_{f} \ ,  \\
L_{C^3}^{(2)} &:=& C_{abcd} C^{cdef} C_{ef}{}^{ab} \ ,
\eea
\end{subequations}
 and one  $C\Box C$ primary field \cite{BKNT}
 \bea
L_{C \Box C} := C^{abcd} \Box_c C_{abcd} 
+ \hf \nabla_e C_{abcd} \nabla^e C^{abcd}
+ \frac{8}{9} \nabla^d C_{abcd} \nabla_e C^{abce} \ , \qquad \Box_c = \nabla^a \nabla_a~.
\label{CBoxC}
\eea
The latter can be recast in a different form using the identity \cite{BKNT}
\begin{align}
L_{C \Box C}
&= \frac{1}{6} C^{abcd} \Box_c C_{abcd}
+ \hf \nabla_e \Big( C_{abcd} \nabla^e C^{abcd} 
+ \frac{16}{9} C^{abce} \nabla^d C_{abcd}
\Big) - \frac{4}{3} L_{C^3}^{(1)} + \frac{1}{3} L_{C^3}^{(2)}~.
\end{align}


\subsection{Conformal compensators}\label{Section2.6}

All gravity-matter theories can be reformulated to possess Weyl invariance  \cite{Deser70, Zumino70}.
This  is achieved by coupling the gravity and matter fields to a {conformal compensator} $\J$, which is a nowhere vanishing scalar field varying by the law
\begin{align}
	\d_\s \J = \s \D \J~, \qquad \D\neq 0~.
\end{align}
under the Weyl transformation \eqref{WeylTr}.
The Weyl gauge freedom can be used to impose the condition $\J =1$ (or, more generally, $\J =\rm const$), 
and then we are back to the original theory.

As an example, we consider the  pure gravity action with  a cosmological term
\bea
S_{\rm GR}= -\frac{1}{2\k^2} \int \rd^D x \,e \,  R   -\frac{\L}{\k^2} \int \rd^D x \,e 
\label{GR1}
\eea
The Weyl-invariant extension of this action is 
 \bea
S_{\rm GR}=\hf \int \rd^D x \,e \,\Big\{ \cD^a \J \cD_a \J -\frac{1}{4} \frac{D-2}{D-1} 
R \J^2  -{\l} \J^{2D/(D-2)} 
\Big\}~,
\label{GR2}
\eea
where the conformal compensator  $\J$ is chosen to have dimension $\D=\hf (D-2)$.
Applying a finite Weyl transformation allows us to choose the gauge condition
\bea 
\J = \frac{2}{\k}  \sqrt{\frac{D-1}{D-2}}~,
\eea
in which the action \eqref{GR2} turns into \eqref{GR1}. The relationship  between $\l$ and $\L$ follows.  

In  the manifestly conformal setup, one may describe non-conformal theories by introducing dependence of the action on $\J$, whose conformal transformation law is characterised by the properties
\begin{align}
	K^a \J= 0 ~, \qquad \mathbb{D} \J = \D \J~, \qquad \Delta \neq 0~.
	\label{compensator}
\end{align} 
In this setting, the action \eqref{GR2} turns into 
 \bea
S_{\rm GR}=- \hf \int \rd^D x \,e \,\Big\{ \J \nabla^a  \nabla_a \J +{\l} \J^{2D/(D-2)} 
\Big\}~.
\label{GR23}
\eea

As another example, we consider the Born-Infeld theory 
in four dimensions, eq. \eqref{BI1}. Its conformal reformulation is 
\bea
L_{\rm BI} (F;\J) = \J^4 \Big\{
1 - \sqrt{- \det (\eta_{ab} + \J^{-2} F_{ab} )} 
\Big\} ~, 
\eea
where the dimension of $\J$ is chosen to be $\D=1$.

It should be emphasised that truly conformal theories, such as the four-dimensional conformal gravity \eqref{CGaction}, do not depend on any compensator.


\subsection{Conformal gravity in $D=3$ dimensions} \label{Section2.7}

For completeness, we also  include a brief discussion of conformal geometry in three dimensions. 
Analysing the Bianchi identity, $ \nabla_{[a} \nabla_b \nabla_{c ]} = 0 $, in the $D=3$ case, 
one observes that the conformal covariant derivatives obey the algebra 
\bea
[\nabla_a , \nabla_b] = \hf W_{ab}{}^c K_c \ , 
\eea
where $W_{ab}{}^c$ is a conformal primary, $K_d  W_{ab}{}^c = 0$, and therefore, independent of  
the dilatation connection $\mathfrak{b}$. One can show that 
\begin{align}
 W_{abc} = -4 \cD_{[a} \mathfrak{f}_{b] c} = 2\cD_{[a} R_{b] c} + \frac{1}{2} \eta_{c [a} \cD_{b]} R \ ,
\end{align}
and thus $W_{abc}$ is the Cotton tensor.  
The algebraic properties of the Cotton tensor are:
\bea
 W_{ab} := \hf \ve_a{}^{cd} W_{cdb} \ , \quad  W_{ab} =  W_{ba} \ , \quad  W^a{}_a = 0 \ .
\eea
The Cotton tensor also satisfies divergenceless condition 
\bea
\nabla^b  W_{ab} = 0 \ .
\eea

 Three-dimensional spacetime is conformally flat iff $W_{ab}=0$.


\section{Conformal fields of arbitrary spin} \label{section3}

In the remainder of this work our discussion is restricted to the $D=4$ case, and we make use of the two-component spinor formalism in the conventions of \cite{BK}, which are similar to those of \cite{WB}.  
In particular, given a one-form $h_a$, it can equivalently be described in terms of  a valence
$(1 , 1)$ spinor $h_{\a\ad }$  defined by 
\begin{subequations}
\bea
h_a \to h_{\a\ad} &=& (\s^b)_{\a\ad} h_b ~ \Longleftrightarrow ~
h_a = - \hf (\tilde \s_a)^{\bd \b} h_{\b \bd}~, \\
\s_{a} &=& ({\mathbbm 1}_{2}\,, \vec{\s} ) = \big((\s_{a})_{\a\ad}\big)\,, \quad 
\tilde{\s}_{a} = ({\mathbbm 1}_{2}\,, -\vec{\s}) =\big((\tilde{\s}_{a})^{\ad\a} \big)\, , 
\eea
\end{subequations}
with $(\tilde{\s}_{a})^{\ad\a} = \ve^{\a\b} \ve^{\ad \bd} (\s_{a})_{\b\bd}$.
Given a symmetric and traceless tensor field $h_{a(s)} := h_{a_1 \dots a_s} $, $h_{a_1 \dots a_{s-2} bc} \eta^{bc}=0$,
it is equivalently described by a 
spinor field of valence $(s,s)$, 
$h_{\a(s)\ad(s)} := h_{(\a_1 \dots \a_s) (\ad_1 \dots \ad_s)} $ defined by 
\begin{align}
h_{a(s)} \to  h_{\a_1\dots\a_s\ad_1\dots\ad_s}=(\s^{a_1})_{\a_1\ad_1}\cdots (\s^{a_s})_{\a_s\ad_s}h_{a_1\dots a_s}=h_{\a(s)\ad(s)}
\end{align}
The matrices  $\s_{ab} = -\frac{1}{4} (\s_a \tilde{\s}_b - \s_b \tilde{\s}_a)$ and 
 $\tilde{\s}_{ab} = -\frac{1}{4} (\tilde{\s}_a {\s}_b - \tilde{\s}_b {\s}_a)$ are used to provide an equivalent description of a two-form 
 $F_{ab}=-F_{ba}$ in terms of symmetric rank-two spinors $F_{\a\b}=F_{\b\a}$ and $ \bar F_{\ad\bd} = \bar F_{\bd \ad}$, 
 \bea
 F_\a{}^\b = \hf F_{ab} (\s^{ab})_\a{}^\b~,\quad \bar F^\ad{}_\bd = - \hf F_{ab} (\tilde{\s}^{ab})^\ad{}_\bd~\Longleftrightarrow~ 
 F^{ab} = (\sigma^{ab})_{\alpha \beta} F^{\alpha \beta} - (\tilde\sigma^{ab})_{\ad \bd} \bar F^{\ad \bd}~.
\label{two-form_two-spinor}
 \eea

In the two-component spinor setting the algebra of $D=4$ conformally covariant derivatives \eqref{Cotton} turns into 
\begin{align}
\big[\nabla_{\a\ad},\nabla_{\b\bd} \big]=&-\big(\ve_{\ad\bd}C_{\a\b\g\d}M^{\g\d}+\ve_{\a\b}\bar{C}_{\ad\bd\gd\dd}\bar{M}^{\gd\dd}\big) 
\non \\
& -\frac{1}{4}\big(\ve_{\ad\bd}\nabla^{\d\gd}C_{\a\b\d}{}^{\g}+\ve_{\a\b}\nabla^{\g\dd}\bar{C}_{\ad\bd\dd}{}^{\gd}\big)K_{\g\gd}~.
\label{D=4CommRel}
\end{align}
Here $C_{\a\b\g\d}$ and $\bar{C}_{\ad\bd\gd\dd}$ are the self-dual and anti-self-dual parts of the Weyl tensor $C_{abcd}$
defined as follows
\begin{subequations}\label{56.98}
\begin{align}
C_{\a\b\g\d}&=\frac{1}{2}(\s^{ab})_{\a\b}(\s^{cd})_{\g\d}C_{abcd}=C_{(\a\b\g\d)}~,\\
\bar{C}_{\ad\bd\gd\dd}&=\frac{1}{2}(\ts^{ab})_{\ad\bd}(\ts^{cd})_{\gd\dd}C_{abcd}=\bar{C}_{(\ad\bd\gd\dd)}~,\\
C_{\a\b\g\d\ad\bd\gd\dd}=(\s^a)_{\a\ad}(\s^b&)_{\b\bd}(\s^c)_{\g\gd}(\s^d)_{\d\dd}C_{abcd}=2\ve_{\ad\bd}\ve_{\gd\dd}C_{\a\b\g\d}+2\ve_{\a\b}\ve_{\g\d}\bar{C}_{\ad\bd\gd\dd}~.
\end{align}
 \end{subequations}
Both $C_{\a\b\g\d}$ and $\bar{C}_{\ad\bd\gd\dd}$ are primary fields of dimension $+2$.
The commutation relation \eqref{D=4CommRel} should be accompanied by the relations 
\bea
\big[\mathbb{D},\nabla_{\aa} \big]=\nabla_{\aa}~,\qquad 
\big[K_{\a\ad},\nabla_{\b\bd}\big] = 4\big(\ve_{\ad\bd}M_{\a\b}+\ve_{\a\b}\bar{M}_{\ad\bd}-\ve_{\a\b}\ve_{\ad\bd}\mathbb{D}\big)~.
\eea
The Lorentz generators  $M_{ab} =- M_{ba}$ can be turned into operators carrying spinor indices, $M_{\a\b} =M_{\b\a}$ 
and $\bar M_{\ad\bd} =\bar M_{\bd\ad}$, which are defined according to \eqref{two-form_two-spinor}.
The Lorentz generators 
act on vectors and two-component spinors as follows:
\bea
M_{ab} V_{c} = 2 \eta_{c[a} V_{b]} ~, \qquad 
M_{\a \b} \j_{\g} = \ve_{\g (\a} \j_{\b)} ~, \qquad \bar{M}_{\ad \bd} \bar \j_{\gd} = \ve_{\gd ( \ad} \bar \j_{\bd )} ~.
\eea

The Weyl tensor obeys the Bianchi identity (see, e.g., \cite{KP})
\begin{align}
 B_{\a(2) \ad(2)}:=\nabla^{\b_1}{}_{(\ad_1}\nabla^{\b_2}{}_{\ad_2)}
  C_{\a(2) \b(2)}
  =\nabla_{(\a_1}{}^{\bd_1}  \nabla_{\a_2)}{}^{\bd_2}
 \bar{C}_{\ad(2) \bd(2) }=\bar{B}_{\a(2) \ad(2)}~,
\end{align}
where $B_{\a(2)\ad(2)}$ is called the Bach tensor. The equation of motion for the conformal gravity model is 
\bea
 B_{\a(2) \ad(2)} =0~.
 \eea
Such spacetimes are called Bach-flat.


\subsection{Real conformal fields} \label{section3.1}

This review is devoted to duality-invariant models for conformal gauge fields. In general, we will deal with real and complex primary fields for which their dimension is determined by their spin. Our discussion in this subsection follows \cite{KP}.

Given a positive integer $s$, a real conformal spin-$s$ field $h_{\a(s) \ad(s)}=h_{(\a_1 \dots \a_s)( \ad_1 \dots \ad_s) }$
 is a real primary field\footnote{In vector notation, this gauge field is equivalently realised as a totally symmetric and traceless rank-$s$ tensor field $h_{a(s)}$ defined by  $(-2)^{s} h_{a(s)} := (\tilde \s_{a_1})^{\a_1 \ad_1} \dots (\tilde \s_{a_s})^{\a_s \ad_s} h_{\a(s) \ad(s)}$.} of dimension $2-s$,
\begin{align}
	K_{\bb} h_{\a(s) \ad(s)} = 0 ~, \qquad \mathbb{D} h_{\a(s) \ad(s)} = (2-s)h_{\a(s) \ad(s)} ~.
\end{align}
The dimension of $h_{\a(s) \ad(s)} $ is fixed by requiring a gauge variation of the form 
\bea
\d_\z h_{\a(s) \ad(s)} = \nabla_{(\a_1 (\ad_1} \z_{\a_2 \dots \a_s) \ad_2 \dots \ad_s)}
\label{gauge-tr1}
\eea
to be primary provided the transformation parameter $\z_{\a(s-1) \ad(s-1)} $ is also primary.

From $h_{\a(s) \ad(s)}$ one may construct the following higher-derivative descendant
\bea
\label{FS1}
\mathbb{C}_{\a(2s)}  := \nabla_{(\a_1}{}^{\bd_1} \dots \nabla_{\a_s}{}^{\bd_s} h_{\a_{s+1} \dots \a_{2s}) \bd(s)} 
\eea
and its conjugate $\bar{\mathbb{C}}_{\ad(2s)} $.
The crucial property of  $\mathbb{C}_{\a(2s)}  $ is that it is primary,
\begin{align}
	K_\bb \mathbb{C}_{\a(2s)} = 0 ~, \qquad \mathbb{D} \mathbb{C}_{\a(2s)} = 2~.
\end{align}
Since the dimension of $\mathbb{C}_{\a(2s)}  $ is equal to two, the quadratic action functional
\bea 
S^{(s)}_{\rm free}[\mathbb{C},\bar{\mathbb C}] = \frac{(-1)^s  }{2}  \int \rd^4x \, e \, 
\Big \{ \mathbb{C}^{\a(2s)} \mathbb{C}_{\a(2s)} +  \bar{\mathbb{C}}^{\ad(2s)} \bar{\mathbb{C}}_{\ad(2s)}   \Big \}
\label{action1}
\eea
is conformal in an arbitrary gravitational background. Choosing an opposite relative sign in the integrand \eqref{action1} would lead to a total derivative  in any conformally flat background 
\bea
C_{abcd} = 0 \quad \implies \quad 
\int \rd^4x \, e \, 
\Big \{ \mathbb{C}^{\a(2s)} \mathbb{C}_{\a(2s)} -  \bar{\mathbb{C}}^{\ad(2s)} \bar{\mathbb{C}}_{\ad(2s)}   \Big \} =0~.
\eea

Associated with  $\mathbb{C}_{\a(2s)} $ and $\bar{\mathbb{C}}_{\ad(2s)} $ 
are the primary dimension-$(2+s)$ descendants 
\begin{subequations}
\label{LinBach1}
\begin{align}
{\mathbb{B}}_{\a(s)\ad(s)}&=\nabla_{(\ad_1}{}^{\b_1}\cdots\nabla_{\ad_s)}{}^{\b_s}{\mathbb{C}}_{\a_1\dots\a_s \b_1\dots\b_s}
\bar{\mathbb{B}}_{\a(s)\ad(s)}&=\nabla_{(\a_1}{}^{\bd_1}\cdots\nabla_{\a_s)}{}^{\bd_s}
\bar{\mathbb{C}}_{\ad_1\dots\ad_s\bd_1\dots\bd_s}~.
\end{align} 
\end{subequations}
It was shown in \cite{KP} that the first term in the action \eqref{action1} can be rewritten, modulo a total derivative, to the form 
\bea
\int \rd^4x \, e \,  \mathbb{C}^{\a(2s)} \mathbb{C}_{\a(2s)} = \int \rd^4x \, e \, h^{\a(s) \ad(s)} {\mathbb{B}}_{\a(s)\ad(s)}~.
\eea
Thus the equation of motion for the model \eqref{action1} is 
\bea
{\mathbb{B}}_{\a(s)\ad(s)} + \bar{\mathbb{B}}_{\a(s)\ad(s)} =0~.
\eea 

For a generic background, the local transformation \eqref{gauge-tr1} 
leaves the field strength $\mathbb{C}_{\a(2s)}$ invariant only in the  $s=1$ case.
In this case the field strength $F_{ab} = \nabla_a h_b - \nabla_b h_a $ varies as 
\begin{align}
\d_\z F_{ab} &= 
[\nabla_a, \nabla_b] \z = -\frac{1}{2}C_{abcd}M^{cd} \z =0  ~.
\end{align}
The Bianchi identity 
\bea
 \nabla_{[a} F_{bc]}  =  \big[\nabla_{[a} , \nabla_b \big] h_{c]} = - C_{[abc]}{}^d h_d =0
 \label{BI-Maxwell}
\eea
holds in an arbitrary curved background.
The field strength $F_{ab}$ is equivalently described by two symmetric rank-2 spinors $F_{\a\b}$ and $\bar{F}_{\ad \bd}$ 
defined by \eqref{two-form_two-spinor}, 
\bea
F_{\a\b} = + \hf {\nabla_{(\a}{}^\gd h_{\b) \gd}} = \hf {\mathbb C}_{\a\b}~,
\eea
in terms of which the above Bianchi looks like 
\bea
\nabla_\ad{}^\b F_{\a \b} - \nabla_\a{}^\bd \bar F_{\ad \bd}=0~.
\label{BI-ED}
\eea

In the $s>1$ case,  $\mathbb{C}_{\a(2s)} $  is gauge invariant only if the background is conformally flat, 
\begin{align}
	\label{GTWeyl}
	C_{abcd} = 0 \quad \implies \quad \d_\z \mathbb{C}_{\a(2s)} = 0~.
\end{align}
For example, the variation of $\mathbb{C}_{\a(4)} $ is given by
\begin{align}
\delta_{\l}\mathbb{C}_{\a(4)}=\frac{1}{2}C_{\a(4)}\nabla^{\b\bd}\z_{\b\bd}-\z_{\b\bd}\nabla^{\b\bd}C_{\a(4)}-2C^{\b}{}_{(\a_1\a_2\a_3}\nabla_{\a_4)}{}^{\bd}\z_{\b\bd}~.
\end{align}
Due to \eqref{GTWeyl}, the action \eqref{action1} is invariant under the gauge transformation \eqref{gauge-tr1} in any conformally flat background. For such backgrounds, the following properties hold:
\begin{subequations}
\bea
\bar{\mathbb{B}}_{\a(s)\ad(s)}  &=& {\mathbb{B}}_{\a(s)\ad(s)}~, \\
\nabla^{\b\bd} {\mathbb{B}}_{\b\a(s-1)\bd\ad(s-1)} &=& 0~.
\eea
\end{subequations}
The former property is the higher-spin analogue of \eqref{BI-ED}. Unlike \eqref{BI-ED}, it holds only if $C_{abcd} = 0 $.
The latter relation expresses the gauge invariance of the action \eqref{action1}.

It should be pointed out that 
$\mathbb{C}_{\a(4)}$ is the linearised Weyl tensor in a conformally flat background, 
and $S^{(2)}_{\rm free}[\mathbb{C},\bar{\mathbb C}]$ is the linearised action for conformal supergravity.  
For such backgrounds, we will refer to  $\mathbb{C}_{\a(2s)}$ and its conjugate for $s>2$ as the linearised spin-$s$ Weyl tensor, and the primary field $\bar{\mathbb{B}}_{\a(s)\ad(s)}  = {\mathbb{B}}_{\a(s)\ad(s)}$ will be called the linearised spin-$s$ Bach tensor. 


\subsection{Complex conformal fields}\label{section3.2}

Particles  of  half-integer spin are described by complex fields. 
In general, all fields transforming in the $\sSL(2,{\mathbb C})$ representations $\big(m/2, n/2)$, with $m>n$, are complex.
Given two positive integers $m>n \geq1$, we are interested in 
a primary field $\phi_{\a(m) \ad(n)}$ of dimension $2-\hf(m+n)$,
\begin{align}
	K_\bb \phi_{\a(m) \ad(n)} = 0 ~, \qquad \mathbb{D} \phi_{\a(m) \ad(n)} = \Big(2 - \hf(m+n)\Big) \phi_{\a(m)\ad(n)} ~.
\end{align}
The conformal properties of $\phi_{\a(m) \ad(n)}$ are fixed by requiring a gauge variation of the form 
\bea
\label{ComplexGT}
\d_\ell \phi_{\a(m)\ad(n)} = \nabla_{(\a_1 (\ad_1} \ell_{\a_2 \dots \a_{m}) \ad_2 \dots \ad_{n})} ~.
\eea
to be primary provided the transformation parameter $\ell_{\a(m) \ad(n)}$ is also primary.
It should be pointed out that the $m=n=s$ case, which we considered earlier, may be consistently incorporated by restricting the field  to be real, $\overline{\phi_{\a(s) \ad(s)}} = \phi_{\a(s) \ad(s)}$. In Minkowski space, bosonic conformal higher-spin fields of any symmetry were studied by Vasiliev \cite{Vasiliev}.

Associated with  $\phi_{\a(m)\ad(n)}$ and its conjugate $\bar{\phi}_{\a(n) \ad(m)}$ are the following conformally primary descendants\footnote{It is evident that, for the special case $m=n=s$, these descendants coincide; $\hat{\mathbb{C}}^{[0]}_{\a(2s)} = \check{\mathbb{C}}^{[0]}_{\a(2s)}$.}
\begin{subequations}
	\label{ComplexFS}
\begin{align}
	\hat{\mathbb{C}}^{[\D]}_{\a(m+n)} &= \nabla_{(\a_1}{}^{\bd_1} \dots \nabla_{\a_n}{}^{\bd_n} \phi_{\a_{n+1} \dots \a_{m+n}) \bd(n)} ~, \\
	\check{\mathbb{C}}^{[\D]}_{\a(m+n)} &= \nabla_{(\a_1}{}^{\bd_1} \dots \nabla_{\a_m}{}^{\bd_m} \bar{\phi}_{\a_{m +1} \dots \a_{m+n}) \bd(m)} ~,
\end{align}
\end{subequations}
where we have introduced the label $\D = m-n$.\footnote{We have appended this label to the field strengths \eqref{ComplexFS} to eliminate ambiguities. In particular, its role may be appreciated by noting that the field strengths for the gauge fields $\phi_{\a(2) \ad(2)}$ and $\phi_{\a(3) \ad}$ are of identical tensor types.} 
In the 
$m=n\equiv s$ case, the field strengths \eqref{ComplexFS} coincide, $\hat{\mathbb{C}}^{[0]}_{\a(2s)} = \check{\mathbb{C}}^{[0]}_{\a(2s)}$. We emphasise that these descendants are primary in an arbitrary background, 
\begin{subequations} \label{dimensions}
	\bea 
	K_\bb \hat{\mathbb{C}}^{[\D]}_{\a(m+n)} &=&0~, \qquad 
	\mathbb{D} \hat{\mathbb{C}}^{[\D]}_{\a(m+n)} = \Big(2 - \frac \D 2\Big) \hat{\mathbb{C}}^{[\D]}_{\a(m+n)}~,\\
	K_\bb\check{{\mathbb{C}}}^{[\D]}_{\a(m+n)}&=&0~,\qquad \mathbb{D}\check{{\mathbb{C}}}^{[\D]}_{\a(m+n)}
	=\Big(2+\frac{\D}{2}\Big)\check{{\mathbb{C}}}^{[\D]}_{\a(m+n)}~.
	\eea
\end{subequations} 
It follows from \eqref{dimensions} that the dimensions of the field strengths \eqref{ComplexFS} are determined by $\D$, which explains why the field strengths carry the label $\D$.

The conformal properties of $\hat{\mathbb{C}}^{[\D]}_{\a(m+n)} $ and $\check{\mathbb{C}}^{[\D]}_{\a(m+n)} $
imply that the free action 
\bea
\label{Free}
S^{(m,n)}_{\rm free}[\hat{\mathbb{C}},\check{\mathbb{C}}]
= {\ri^{m+n}}\int \rd^4 x \, e\,  \hat{ {\mathbb{C}}}^{[\D] \a(m+n)}\check{{\mathbb{C}}}^{[\D]}_{\a(m+n)} 
+{\rm c.c.} ~,
\eea
is conformal. 
It holds that 
\bea
C_{abcd} = 0 \quad \implies \quad 
 {\ri^{m+n+1}}\int \rd^4 x \, e\,  \hat{ {\mathbb{C}}}^{[\D] \a(m+n)}\check{{\mathbb{C}}}^{[\D]}_{\a(m+n)} 
+{\rm c.c.} =0~.
\eea
For conformally flat backgrounds, we will refer to  $\hat{\mathbb{C}}^{[\D]}_{\a(m+n)} $ and $\check{\mathbb{C}}^{[\D]}_{\a(m+n)} $
as (linearised) higher-spin Weyl tensors.

In a general curved space, one may construct the following primary descendants from the higher-spin Weyl tensors,
\begin{subequations}\label{4.122}
\begin{align}
\hat{\mathbb{B}}_{\a(n)\bd(m)}&=\nabla_{(\bd_1}{}^{\g_1}\cdots\nabla_{\bd_m)}{}^{\g_m}\hat{\mathbb{C}}^{[\D]}_{\a_1\dots\a_n\g_1\dots\g_m}\label{4.122a}~,\\
\check{\mathbb{B}}_{\a(m)\bd(n)}&=\nabla_{(\bd_1}{}^{\g_1}\cdots\nabla_{\bd_n)}{}^{\g_n}\check{\mathbb{C}}^{[\D]}_{\a_1\dots\a_m\g_1\dots\g_n}~. \label{4.122b}
\end{align} 
\end{subequations}
Both \eqref{4.122a} and \eqref{4.122b} have dimension $2+\frac 12 (m+n)$. The proof that they are primary is similar to that of the higher-spin Weyl tensors.
The primary fields \eqref{4.122a} and \eqref{4.122b} originate from
two alternative expressions for one and the same conformal invariant 
\begin{align} 
\int\text{d}^4x\, e \, \hat{\mathbb{C}}^{[\D]\a(m+n)}\check{\mathbb{C}}^{[\D]}_{\a(m+n)}  
= \int\text{d}^4x\, e \, \phi^{\a(m)\bd(n)}\check{\mathbb{B}}_{\a(m)\bd(n)} 
= \int\text{d}^4x\, e \, \bar{\phi}^{\a(n)\bd(m)}\hat{\mathbb{B}}_{\a(n)\bd(m)}~.
\end{align}

In a general gravitational background, the primary fields \eqref{ComplexFS} are not invariant under the local transformation \eqref{ComplexGT}. As an example, it is instructive to consider the so-called conformal gravitino
described by a prepotential $\f_{\a(2)\ad} $ and its conjugate.\footnote{The conformal gravitino  model can be extracted, e.g., from the action for $\cN=1$ conformal supergravity  \cite{KTvN,KTvN2} by linearising it around a Bach-flat background, see \cite{KP} for the technical details.}  
Associated with the gravitino are the two primary field strengths \cite{KP}
 \begin{align}
 \label{GravitinoFS}
 \hat{\mathbb{C}}_{\a(3)} \equiv  \hat{\mathbb{C}}^{[1]}_{\a(3)}
 =\nabla_{(\a_1}{}^{\bd}\phi_{\a_2\a_3)\bd}~,\qquad 
 \check{\mathbb{C}}_{\a(3)} \equiv \check{\mathbb{C}}^{[1]}_{\a(3)} 
 =\nabla_{(\a_1}{}^{\bd_1}\nabla_{\a_2}{}^{\bd_2}\bar{\phi}_{\a_3)\bd(2)}~,
 \end{align}
 and their conjugates, which are primary fields of dimensions $+3/2$ and $+5/2$ respectively. Under the gauge transformation 
  \begin{align}\label{GravitinoGaugeTr}
 \delta_{\z}\phi_{\a(2)\ad}=\nabla_{(\a_1\ad}\z_{\a_2)}~,
 \end{align}
  their variations are given by 
 \begin{align}
 \d_{\z}\hat{\mathbb{C}}_{\a(3)}=C_{\a(3)\d}\z^{\d}~,\qquad 
 \d_{\z}\check{\mathbb{C}}_{\a(3)}=\frac{1}{2}C_{\a(3)\d}\nabla^{\d\dd}\bar{\z}_{\dd}
 -\bar{\z}_{\dd}\nabla^{\d\dd}C_{\a(3)\d}~.
 \end{align}

For any conformally flat background it may be shown that the descendants \eqref{ComplexFS}
are inert under the gauge transformations \eqref{ComplexGT}
\begin{align}
C_{abcd} = 0 \quad \implies \quad 	\d_\ell \hat{\mathbb{C}}^{[\D]}_{\a(m+n)} = \d_\ell \check{\mathbb{C}}^{[\D]}_{\a(m+n)} = 0~,
\end{align}
and therefore the conformal action \eqref{Free} is gauge invariant.
For such backgrounds
the field strengths \eqref{ComplexFS} are related via the Bianchi identity
\be
\label{ComplexBI}
\nabla^{\b_1}{}_{(\ad_1} \dots \nabla^{\b_m}{}_{\ad_m)} \hat{\mathbb{C}}^{[\D]}_{\a(n) \b(m)} 
= \nabla_{(\a_1}{}^{\bd_1} \dots \nabla_{\a_n)}{}^{\bd_n} \overline{\check{\mathbb{C}}}_{\ad(m) \bd(n)}^{[\D]} ~.
\ee


\section{Self-dual nonlinear electrodynamics} \label{section4}

As follows from the discussion in section \ref{section3.1}, the electromagnetic field $h_{\a\ad}$ is special in  the family of conformal spin-$s$ gauge fields $h_{\a(s) \ad(s)} $ in the sense that the primary descendant \eqref{FS1} is invariant under the local transformation \eqref{gauge-tr1} on an arbitrary gravitational background only in the $s=1$ case. This means that any model for the electromagnetic field in curved space with its action $S$ being a functional of the field strength, $S=S[F_{ab}]$,  is gauge-invariant.
We assume $S[F_{ab}]$ to be conformal, which is always possible to achieve by coupling $h_a$ to a (nowhere vanishing) conformal compensator $\J$ with the conformal properties \eqref{compensator}.
Strictly speaking, the action functional also depends on the gauge fields of conformal gravity
and the compensator, $S[F; \nabla, \J]$. This dependence will only be explicitly indicated  when necessary.


\subsection{Self-duality equation}\label{section4.1}

To formulate the equation of motion for the model $S[F]$, we introduce 
\bea
\widetilde G^{ab}[F] := \hf  \ve^{abcd}\, G_{cd}[F] =2 \frac{\delta S[F]}{\delta F_{ab}}~,
\eea
where the functional derivative of the action with respect to $F$ is defined by 
\bea
\d S =  \int \rd^4x\, e \, \d F_{a b} \,\frac{\delta S[F]}{\delta F_{ab}}~.
\eea
By construction, $G_{ab}[F]$ is a primary field of dimension $+2$.
Now the Bianchi identity \eqref{BI-Maxwell} and the equation of motion for $S[F]$ are given by
\bea
\nabla_a \widetilde F^{ab}=0~, \qquad \nabla_a \widetilde G^{ab}[F]=0~. 
\label{BI+EoM}
\eea

Since (i) $F_{ab}$ and $G_{ab}$ have the same conformal properties, and (ii) the equations \eqref{BI+EoM}  have the same functional form, one may consider duality transformations
\bea
  \left( \begin{array}{c} G'[F']  \\  F'  \end{array} \right)
= M
\left( \begin{array}{c} G[F] \\ F  \end{array} \right) ~, \qquad 
M
 \in
\sGL(2, {\mathbb R})~,
\label{GL-duality}
\eea
such that the transformed quantities $F'$ and $G'$
also satisfy the equations \eqref{BI+EoM}, where
\bea
\widetilde{G}'\equiv \widetilde{G}'^{ab} [F'] = 2  \frac{\d S'[F']}{\d F'_{ab}}~.
\eea
The transformed action, $S'[F]$, always exists. To demonstrate this, it suffices to consider the infinitesimal transformation
\bea
M = {\mathbbm 1} + \left( \begin{array}{cc} a~& ~b \\ c~ & ~d \end{array} \right) 
\label{DualTr1}
\eea
with infinitesimal matrix elements $a,b, c, d$.
In this case one finds\footnote{We  often use the notation $F \cdot G= F^{ab} G_{ab}$
implying $F \cdot \widetilde{G}= \widetilde{F}\cdot G$ and
$ \widetilde{F} \cdot \widetilde{G} = -F \cdot G$.}
\bea
\D S := S'[F] - S[F] &=& (a+d) S[F] -  d \int \rd^4 x\, e\, 
F_{a b} \,\frac{\delta S[F]}{\delta F_{ab}}
\non \\
&&+ \frac{1}{4} \int \rd^4 x\, e\, \left \{b \,\widetilde{F} \cdot F
-
c \,\widetilde{G} \cdot G \right\}~.
\label{DualTr2}
\eea

The above consideration becomes nontrivial  if 
the model under consideration is {\it self-dual}, which means
\bea
S'[F] =  S[F]~.
\eea
The contributions in the first and second lines of \eqref{DualTr2} should cancel out independently of each other.\footnote{In particular, in a parity invariant theory these contributions involve functional structures of opposite parity.}
In order for the contributions in the first line of \eqref{DualTr2} to vanish, we must require one of the two options:
\begin{subequations}
\bea
a&=&d=0 ~;  \label{DualTr3.a} \\
a&=&d \neq 0 \quad   \implies  \quad \int \rd^4 x\, e\, 
F_{a b} \,\frac{\delta S[F]}{\delta F_{ab}} =2S[F]~. \label{DualTr3.b} 
\eea
\end{subequations}
Now let us turn to the contributions in the second lines of \eqref{DualTr2}. 
Maxwell's theory is assumed to belong to the family of self-dual models, 
which implies the following condition on the parameters in \eqref{DualTr1}:
\bea
b=-c \quad & \implies & \quad \int \rd^4 x\, e\, \left \{\widetilde{F} \cdot F 
+ \widetilde{G} \cdot G\right\}=0~.
\label{DualTr3.c}
\eea

Option \eqref{DualTr3.b} corresponds to a scale transformation
\bea
  \left( \begin{array}{c} G'[F']  \\  F'  \end{array} \right)
= \re^\t
\left( \begin{array}{c} G[F] \\ F  \end{array} \right) ~, \qquad 
\t
 \in {\mathbb R}~.
\label{DualTr4}
\eea
Such a symmetry transformation exists in the case of conformal theories, including Maxwell's theory and its one-parameter deformation \eqref{ModMax} known as the ModMax theory.
Scale transformations \eqref{DualTr4} do not preserve the energy-momentum tensor in the Maxwell case \eqref{EMT}, 
and should be discarded. 
As a result, we stay with the conditions  \eqref{DualTr3.a} and  \eqref{DualTr3.c}.
We see that the requirement of self-duality leads to  the following fundamental properties.
\begin{itemize}
\item Only $\sU(1)$ duality rotations can be consistently
defined in the non-conformal case, 
\bea
  \left( \begin{array}{c}  G'[F'] \\ F'  \end{array} \right)
=  \left( \begin{array}{cr} \cos \l & ~
-\sin \l \\ \sin \l  &  ~\cos \l \end{array} \right) \;
\left( \begin{array}{c}  G[F] \\ F  \end{array} \right) ~.
\label{U(1)-duality}
\eea
\item The action is a solution of the {\it self-duality equation} \cite{KT2}
\bea
\int \rd^4 x\, e\,  \left(\widetilde{G}^{ab}[F] {G}_{ab}[F] + \widetilde{F}^{ab}  {F}_{ab} \right)= 0~,
\label{GZ1}
\eea
which must hold for an unconstrained two-form $F_{ab}$. 
\end{itemize}

It is instructive to give an alternative derivation of (\ref{GZ1}) by directly analysing the $\sU(1)$ duality transformations 
\eqref{U(1)-duality}.
For an infinitesimal duality rotation, we have
\bea
\widetilde{G}'_{ab}[F'] = \widetilde{G}_{ab}[F] - \l \widetilde{F}_{ab}
= \widetilde{G}_{ab}[F]
+ 2 \frac{\d}{\d F^{ab}} \left(
- \frac{1}{4}\l  \int \rd^4 x\, e\, \widetilde{F} \cdot {F} \right)~,
\label{var-1}
\eea
where we have used the infinitesimal version of \eqref{U(1)-duality}.
On the other hand, from the definition of
$\widetilde{G}'[F']$ it follows that
\bea
\widetilde{G}' [F'] = 2  \frac{\d S[F']}{\d F'}
= 2\left( \frac{\d }{\d F'} S[F]
+ \frac{\d}{\d F} \d S \right)~,
\label{tilde-G'-def}
\eea
where $\d S = S[F'] - S[F]$.
Since $F' = F + \l G[F]$, one can express
$\d / \d F'$ on the right-hand side of \eqref{tilde-G'-def}
via $\d / \d F$ to result with
\bea
\widetilde{G}'_{ab}[F'] = \widetilde{G}_{ab}[F]
+ 2\frac{\d}{\d F^{ab}}
\left( \d S - \frac{1}{4} \l  \int \rd^4 x\, e\, 
\widetilde{G}[F] \cdot {G} [F]
\right)~.
\label{var-2}
\eea
Comparing the equations  \eqref{var-1} and \eqref{var-2} gives
\bea
\d S = \frac{1}{4} \l  \int \rd^4 x\, e\,
\left( \widetilde{G}[F] \cdot {G}[F] - \widetilde{F} \cdot {F} \right)~.
\label{var-3}
\eea
On the other hand, the action can be varied directly to give
\bea
\d S =  \int \rd^4 x\, e\,  \frac{\d S[F]}{\d F^{ab}} \d F^{ab}
=\hf \l  \int \rd^4 x\, e\, \widetilde{G}[F]\cdot G[F].
\label{var-4}
\eea
This is consistent with eq. (\ref{var-3}) provided the self-duality equation (\ref{GZ1}) holds.

The most well-known family of self-dual theories are 
$\sU(1)$ duality-invariant models for nonlinear electrodynamics
\bea
S[F] = \int \rd^4 x\, e\, L(F) \quad \implies \quad 
\widetilde G^{ab}[F]=2 \frac{\pa L(F)}{\pa F_{ab}} \equiv \widetilde G^{ab}(F)~.
\label{SDwithoutHD}
\eea
For such models the self-duality equation \eqref{GZ1} turns into the equation discovered in \cite{B-B,GR1, GZ2}
\bea
\widetilde{G}^{ab}(F) {G}_{ab}(F) + \widetilde{F}^{ab}  {F}_{ab} =0~.
\label{GZ2}
\eea
The term `self-duality equation' was introduced by Gaillard and Zumino \cite{GZ2,GZ3}.

In the case of $\sU(1)$ duality-invariant theories with higher derivatives, one must use the integral form of the self-duality equation \eqref{GZ1}.\footnote{Duality-invariant theories with higher derivatives naturally occur in 
$\cN=2$ supersymmetry \cite{KT1, KT2}.}
Further aspects of duality-invariant theories with higher derivatives were studied, e.g., in\cite{AFZ,Chemissany:2011yv,AF,AFT}.


\subsection{Self-duality equation in spinor notation}\label{section4.2}

Every real two-form $F_{ab}=-F_{ba}$ can be equivalently described in terms of a symmetric rank-two spinor $F_{\a\b}=F_{\b\a}$ and its conjugate $ \bar F_{\ad\bd} = \bar F_{\bd \ad}$, which are defined by the rule 
\bea
(\s^a)_{\a\ad} (\s^b)_{\b\bd} F_{ab} = 2 \ve_{\ad\bd} F_{\a\b}+ 2 \ve_{\a\b} \bar F_{\ad\bd}~,
\eea
such that the following identities hold:
\begin{subequations}
\bea
\widetilde{F}_{\a\b} &=& -\ri F_{\a\b}~, \\
\hf F^{ab}H_{ab} &=& F^{\a\b} H_{\a\b} + \bar F^{\ad \bd } \bar H_{\ad\bd}~.
\eea
\end{subequations}
Here we recast the results of the previous subsection in  spinor notation.

The action $S[F_{ab}]$ considered in the previous subsection can equivalently be viewed as a functional of $F_{\a\b}$ and its conjugate, $S[F_{\a\b} , \bar F_{\ad\bd} ]$. We define
\bea
\d S =  \int \rd^4x\, e \, \d F_{\a \b} \,\frac{\delta S[F]}{\delta F_{\a \b}} +{\rm c.c.}~,
\eea
and therefore 
\bea
\frac{\delta S[F]}{\delta F_{\a \b}} = -\ri G^{\a\b}[F]~.
\eea
The Bianchi identity and equation of motion \eqref{BI+EoM} take the form
\bea
\nabla^\b{}_\ad F_{\a\b}= \nabla_\a{}^\bd \bar F_{\ad \bd}~, \qquad 
\nabla^\b{}_\ad G_{\a\b}[F]= \nabla_\a{}^\bd \bar G_{\ad \bd}[F]~.
\eea
Finally the self-duality equation \eqref{GZ1} turns into
\bea
{\rm Im} \int \rd^4 x\, e\, \left( {G}^{\a\b}[F] {G}_{\a\b}[F] + F^{\a\b}  {F}_{\a\b} \right)= 0~.
\eea

 Given a model for nonlinear electrodynamics, its Lagrangian $L(F_{ab})$ is a Lorentz scalar 
that can be realised as a real function of one complex variable, 
\begin{subequations}
\bea
L(F_{ab})= L(\o , \bar{\o} )~, \qquad 
\o = \a + {\rm i} \, \b = F^{\a\b}F_{\a\b}~, 
\eea
where
\bea
\a := \frac{1}{4} \, F^{ab} F_{ab}~,  \qquad \b := \frac{1}{4} \, F^{ab} \widetilde{F}_{ab} 
\eea
\end{subequations}
are the independent invariants of the electromagnetic field \cite{Minkowski}. 
The self-duality equation 
\bea
{\rm Im}  \Big( {G}^{\a\b}[F] {G}_{\a\b}[F] + F^{\a\b}  {F}_{\a\b} \Big)= 0
\eea
takes the form 
\bea
{\rm Im}\,  \left\{ \o - 4\, \o\,
\left( \frac{\pa L}{\pa \o} \right)^2 \right\} = 0~.
\eea
Making the Ansatz 
\bea
L(\o, \bar \o)  = -\hf \, \Big( \o + \bar{\o} \Big) +
\o \, \bar{\o} \; \L (\o, \bar{\o} )~,
\eea
the self-duality equation turns into \cite{KT2}
\bea
{\rm Im}\, \bigg\{ \frac{\pa (\o \, \L) }{\pa \o}
- \bar{\o}\,
\left( \frac{\pa (\o \, \L )  }{\pa \o} \right)^2 \bigg\} = 0~.
\label{SDE-spinor}
\eea
For theories possessing a weak-field limit, such as the Born-Infeld model, $\L(\o,\bar \o)$ is a real analytic function in a neighbourhood of $\o=0$.
In this case the general solution of \eqref{SDE-spinor} has the form \cite{KT2}
\bea
\L (\o , \bar \o ) &=& \sum_{n=0}^{\infty} ~
\sum_{p+q =n} \l_{p,q} \; \o^p {\bar \o}^q~,
\qquad  \l_{p,q}=\l_{q,p} \in {\mathbb R}
\label{GenSol}
\eea
where 
the self-duality equation
uniquely fixes the level-$n$ coefficients
$\l_{p,q}$ with $p \neq q$ through those at lower levels,
while $\l_{r,r}$ remain undetermined. This means that a general solution of the self-duality equation
involves an arbitrary real analytic function of one real argument, 
\bea
\sum_{n=0}^{\infty}  \l_{n,n}  |\o|^n ~.
 \eea
  The same conclusion was reached in \cite{GR1, Hatsuda:1999ys} by analysing a different form of \eqref{GZ2}.

 In general, given a model for nonlinear electrodynamics $ L(\omega , \bar{\omega} )$, the theory is parity invariant if 
 $ L(\omega , \bar{\omega} ) =  L(\bar{\omega} , {\omega} )$. It follows from \eqref{GenSol} that self-dual nonlinear electrodynamics is parity invariant.
 
Two comments are in order. Firstly, in the literature one finds alternative forms of the self-duality
equation  \cite{GR1,GZ3} but it is the equation  \eqref{SDE-spinor} 
 which turns out to be most convenient for supersymmetric generalisations \cite{KT1,KT2}.
Secondly, let  $L(\o , \bar{\o} )$ be a general model for nonlinear electrodynamics. Such a theory is parity invariant iff $L(\o , \bar{\o} )
= L( \bar{\o}, \o )$. It follows from \eqref{GenSol} that every self-dual nonlinear electrodynamics is parity invariant \cite{KT2}.

The well-known solutions of the self-duality equation are:
\begin{itemize} 
\item
 the Maxwell theory 
\bea
L_{\rm Maxwell}(\o, \bar \o)  = -\hf \, \big( \o + \bar{\o} \big) ~;
\eea
\item
 the Born-Infeld theory 
\bea
L_{\rm BI} (\o, \bar \o; \J) 
=
  \J^4 \left\{ 
1 - \sqrt{1 + \J^{-4} (\o + \bar \o )
+\frac 14 \J^{-8} (\o - \bar \o )^2 } 
\right\} ~.
\label{BI-spinor}
\eea
\end{itemize}
Other explicit solutions of the self-duality equations were given in \cite{Hatsuda:1999ys}.


\subsection{ModMax theory} \label{section4.3}

Until the year 2020, it was tacitly assumed that physically interesting solutions of the self-duality equation \eqref{GZ2}
should possess the weak-field limit, see e.g. \cite{Hatsuda:1999ys},
and therefore the self-interaction $\L(\o,\bar \o)$ must be a real analytic function in a neighbourhood of $\o=0$. Omitting the latter requirement, new solutions of the self-duality equation become possible.
A remarkable self-dual theory was discovered by Bandos, Lechner, Sorokin and Townsend in \cite{BLST}.
Its Lagrangian has the form
\bea
 L_{\rm MM}(\o, \bar \o) = - \hf \big( {\o} + {\bar \o}\big)\cosh \g  
+{\sqrt{\o\bar \o} } \, {\sinh \g } ~,
\label{ModMax}
\eea
with $\g$ the coupling constant restricted to be non-negative, $ \g\geq 0$, to avoid superluminal propagation. This theory is conformal, since there is no dependence on the compensator, and reduces to Maxwell's theory for $\g=0$. It is a one-parameter deformation of Maxwell theory, which is why \eqref{ModMax} was called the ModMax (modified Maxwell) theory. 
The ModMax proves to be a unique self-dual and conformal model for nonlinear electrodynamics.

ModMax theory \eqref{ModMax} is characterised by the self-coupling \cite{K21}
\bea
\L_{\rm MM}(\o, \bar \o) = \frac{\sinh \g }{\sqrt{\o\bar \o} } 
- \hf  \Big( \frac{1}{\o} + \frac{1}{\bar \o}\Big)(\cosh \g -1)
~,
\label{LambdaMM}
\eea
which is evidently not of the form \eqref{GenSol}. 

The uniqueness of the ModMax theory is lost if one allows for higher-derivative conformal and duality-invariant deformations
\cite{Kuzenko:2024zra}.


\subsection{Fundamental properties of self-dual theories} \label{section4.4}

Self-dual theories possess several fundamental properties established in \cite{GZ1,GR2,GZ3}. 

\subsubsection{Duality invariance of the energy-momentum tensor}\label{section4.4.1}

The energy-momentum tensor of every $\sU(1)$ duality-invariant theory is duality invariant \cite{GZ1}. 
In general, given a self-dual theory with action $S[F]$,  
an observable $\U[F]$ is said to be duality invariant if it does not change under the duality transformations,
\bea
 \int \rd^4 x\, e\, 
G_{a b} \,\frac{\delta \U[F]}{\delta F_{ab}} =0~.
\eea
An example of a duality-invariant observable is 
\bea
S[F] - \frac14  \int \rd^4 x\, e\, F  \cdot \tilde{G}[F]~.
\label{duality-invariant-observable}
\eea

The duality invariance of the energy-momentum tensor follows from the fact that the vielbein is duality invariant.  
More generally, given a self-dual theory with its action functional $S[F; g] $ depending on a duality-invariant parameter $g$ (or a field), 
the following observable $\U[F; g]:= \pa S[F; g] \pa g$ is duality invariant.\footnote{Partial derivative $\pa/ \pa g$ should be replaced with a functional derivative if $g$ is a field.} 
Indeed, applying an infinitesimal duality transformation to $\U[F;g]$ gives
\bea
\d \, \frac{\pa }{\pa g} S = \frac{\pa}{\pa g}  \d S
= \hf \l\, \frac{\pa}{\pa g} \int \rd^4 x\, e\,   \tilde{G}\cdot G 
= \hf\l\, \frac{\pa}{\pa g} \int \rd^4 x\, e\,  \left( \tilde{G}\cdot G + \tilde{F}\cdot F
\right) =0~,
\eea
since $F$ is $g$-independent.

\subsubsection{$\sSL(2,{\mathbb R})$ duality in the presence of dilaton and axion} \label{section4.4.2} 

Given a $\sU(1)$ duality-invariant theory $S[F]$, the duality group can be enhanced to the non-compact group 
$\sSL(2,{\mathbb R})$ by coupling the field strength to the dilaton $\vf$ and axion $\mathfrak a$ fields 
\cite{GZ1,GR2} taking their values in $\sSL(2,{\mathbb R})/ \sSO(2)$. 
This is achieved by replacing\footnote{Our discussion here is restricted to  self-dual theories of the form \eqref{SDwithoutHD}.} 
\bea
L(F) ~  \to ~ 
L(F, \t, \bar \t )=L( {\rm e}^{-\vf /2}
\, F) + \frac{1}{4} 
{\mathfrak a}\,
F \cdot \tilde{F}~,
\qquad \t= {\mathfrak a} + {\rm i} \,{\rm e}^{-\vf}~.
\label{NLEDLagrangian-axion-dilaton}
\eea
The fields  $\vf$ and $\mathfrak a$ are defined to be primary and dimensionless, and therefore the Lagrangian $L(F, \t, \bar \t )$ is conformally primary and of dimension $+4$.
The duality group $\sSL(2,{\mathbb R})$
acts on the fields by  transformations of the form 
\bea
  \left( \begin{array}{c} G'  \\  F'  \end{array} \right)
=  \left( \begin{array}{cc} a& b \\ c & d \end{array} \right) 
\left( \begin{array}{c} G \\ F  \end{array} \right) , \qquad
\t' = \frac{a\t +b}{c\t+d}~, \qquad
\left( \begin{array}{cc} a& b \\ c & d \end{array} \right) \in
\sSL(2, {\mathbb R})
\eea

In addition to the Lagrangian \eqref{NLEDLagrangian-axion-dilaton}, which describes the electromagnetic field coupled to the dilaton and axion, a kinetic term for these scalar fields should also be included. There are two possible case. 
\begin{itemize} 

\item The model for nonlinear electrodynamics is not conformal. This means that its Lagrangian, denoted above $L(F)$,
 explicitly depends on the compensator $\Psi$,  $L(F) \equiv L(F;\Psi)$. Then, the $\sSL(2,{\mathbb R})$-invariant 
Lagrangian for the dilaton and axion is 
\bea
L(\tau, \bar \tau; \Psi) =- \frac{\Psi^2 }{ ({\rm Im}\, \tau)^2} \nabla^a \tau \nabla_a \bar \tau~,
\eea
where 
\bea 
\rd s^2 = \frac{1}{ ({\rm Im}\, \tau)^2} \rd \tau \rd \bar \tau = 2 g_{\tau \bar \tau} \rd \tau \rd \bar \tau
\label{metric}
\eea
 is the K\"ahler metric on the Poincar\'e upper half-plane. 
 
\item Nonlinear electrodynamics is described by the ModMax theory \eqref{ModMax}, with Maxwell's theory corresponding to 
$\gamma =0$. 
Then, it is natural to choose the purely dilaton-axion action to be conformal and $\sSL(2,{\mathbb R})$ invariant.  
In general, its Lagrangian should have a higher-derivative form 
\begin{align}
L(\tau, \bar \tau) =& - \frac{1}{ 2 ({\rm Im}\, \tau)^2} \Big\{
\hf  \D \tau \D \bar \tau 
+ \Big( \nabla^a\tau \hat{\nabla}_a\D \bar \tau + \nabla^a \bar \tau \hat{\nabla}_a \D \tau \Big)
+ \hat{\nabla}^a \nabla^b \tau \hat{\nabla}_a \nabla_b \bar \tau\Big\} \non \\
&+ \frac{1}{12 ( {\rm Im}\, \t)^4} 
\Big\{ \a \nabla^a \t \nabla_a \t \nabla^b \bar \t  \nabla_b \bar \t 
+\b \nabla^a \t \nabla_a \bar  \t \nabla^b  \t  \nabla_b \bar \t \Big\}~,
\label{dilaton-axionHD}
\end{align}
with $\alpha$ and $\beta$ being dimensionless coupling constants. 
Here we have denoted\footnote{The Christoffel symbols for the K\"ahler metric \eqref{metric} are 
 $\Gamma^\tau{}_{\tau\tau} = {\rm i} ( {\rm Im}\,\tau)^{-1}$ and   
 $\Gamma^{\bar \tau}{}_{\bar \tau \bar \tau} = -{\rm i} ( {\rm Im} \,  \tau)^{-1}$.  }
 \begin{subequations}
 \bea
 \hat{\nabla}_a \nabla_b \tau &:=&  \Big(\nabla_a + \frac{\rm i}{{\rm Im}\, \tau} \nabla_a\tau \Big) \nabla_b \tau~, 
 \qquad 
 \D \tau :=  \hat{\nabla}^a \nabla_a \tau~, 
 \\
\hat{\nabla}_a\D \tau &:=& \Big(\nabla_a + \frac{\rm i}{{\rm Im}\, \tau} \nabla_a\tau \Big) 
\D \tau~.
 \eea
 \end{subequations}
It follows from the discussion in Appendix \ref{appendixB} that the Lagrangian \eqref{dilaton-axionHD} is conformal  primary.
\end{itemize}

As demonstrated by Osborn \cite{Osborn}, and re-derived later in \cite{BPT} in the framework of induced ${\cal N}=4$ conformal supergravity, the higher-derivative model \eqref{dilaton-axionHD} originates as the one-loop logarithmically divergent quantum correction 
in Maxwell's theory coupled to the dilaton and axion fields
\bea
L(F, \t, \bar \t )=-\frac 14  {\rm e}^{-\vf } F^{ab}F_{ab} + \frac{1}{4} 
{\mathfrak a}\, F \cdot \tilde{F}~.
\eea
This simple model may be generalised to a system of $n$ Abelian vector fields coupled to $\frac 12 n(n+1)$ complex scalars parametrising the Hermitian symmetric space
$\mathsf{Sp}(2n, {\mathbb R})/ \mathsf{U}(n)$. Such a theory is conformal invariant and possesses the maximal non-compact duality group $\mathsf{Sp}(2n, {\mathbb R})$. The corresponding induced action, obtained by integrating out the vector fields, was computed in 
\cite{Grasso:2023hmv}. The induced action, which determines the logarithmically divergent part of the one-loop effective action, is conformal  and $\mathsf{Sp}(2n, {\mathbb R})$ invariant.


\subsubsection{Self-duality under Legendre transformation} \label{section4.4.3}

Every $\sU(1)$ duality-invariant theory is self-dual under a Legendre transformation \cite{GZ3}. Let us recall the definition of a dual formulation for an Abelian gauge theory with action $S[F]$. We can associate with $S[F]$ an equivalent first-order model defined by 
\bea
S[F, F_{\rm D}] = S[F] -\hf \int \rd^4 x\, e\, 
F \cdot \widetilde{F}_{\rm D}~,
\qquad  F_{\rm D}{}^{ab} = \nabla^a h_{\rm D}{}^b
- \nabla^b h_{\rm D}{}^a~,
\label{parent-NLED}
\eea
in which $F_{ab} $ is an unconstrained two-form (auxiliary field), and $h_{\rm D}{}^a$ the dual gauge field. 
Varying $S[F, F_{\rm D}] $ with respect to the dual gauge field leads to  the Bianchi identity $\nabla_a \widetilde F^{ab}=0$, 
and then $S[F, F_{\rm D}] $ reduces to the original action $S[F]$. On the other hand, we can, in principle,  
eliminate $F_{ab} $ using its equation of motions $G(F) = F_{\rm D}$
to yield 
\bea
S_{\rm D} [F_{\rm D}]
:=  \Big( S[F]
-\hf \int \rd^4 x\, e\, F\cdot \widetilde F_{\rm D} \Big)\, \Big|_{ F=F(F_{\rm D})} ~.
\eea
This relation defines the dual formulation, $ S_{\rm D} [F]$, of the theory. One often refers to \eqref{parent-NLED} as the parent action, 
see, e.g., \cite{Hjelmeland:1997eg}.

So far, the action $S[F]$ has been arbitrary. In the case that $S[F]$ is a solution of the self-duality equation \eqref{GZ1}, 
it holds that 
 \bea
S_{\rm D} [F] = S[F]~.
\eea 
This result can be proved by making use of the duality-invariant observable \eqref{duality-invariant-observable}. 
Its duality invariance means that 
\bea
S[F] - \frac14  \int \rd^4 x\, e\, F  \cdot \tilde{G}[F] = S[F'] - \frac14  \int \rd^4 x\, e\, F'  \cdot \tilde{G}[F']~,
\eea
for any duality rotation \eqref{U(1)-duality}. For a finite  rotation \eqref{U(1)-duality} by $\lambda = \pi/2$, this relation reads
\bea
S[F] - \frac12  \int \rd^4 x\, e\, F  \cdot \tilde{F}_{\rm D} = S[F_{\rm D}] ~, \qquad F_{\rm D} \equiv G[F]~.
\eea


\subsection{The Ivanov-Zupnik formulation} \label{section4.5}

A natural framework to generate $\sU(1)$ duality-invariant models for nonlinear electrodynamics is the Ivanov-Zupnik (IZ) approach developed in \cite{IZ_N3,IZ1,IZ2}.
In the case of theories without higher derivatives, it is a reformulation of the GZGR formalism  \cite{GZ1,GR1,GR2,GZ2,GZ3} which is obtained by replacing $L(F_{ab}) \to \mathfrak{L}(F_{ab} , V_{ab})$, where  $V_{ab}=-V_{ba}$ is an auxiliary unconstrained 
two-form. More generally, the IZ reformulation is obtained by replacing the action functional $S[F]$ with the following
\bea
\mathfrak{S}[F, V] = \int \rd^4 x \,e \,\left\{ \frac{1}{4} F^{ab}F_{ab} +\hf V^{ab}V_{ab} 
 - V^{ab}F_{ab} \right\}+ \mathfrak{S}_{\rm int} [V]~,
\label{AuxAction}
 \eea
such that imposing the equation of motion 
\bea
\frac{\d}{\d V_{ab} }  \mathfrak{S}[F, V] =0~,
\label{AuxEoM}
\eea
reduces the action \eqref{AuxAction} to $S[F]$. 
For theories without higher derivatives, 
\bea
 \mathfrak{S}^{\rm int} [V] =  \int \rd^4 x \,e \,\mathfrak{L}_{\rm int} (V_{ab}) ~,
 \label{no-higher-derivative}
\eea
the equation of motion \eqref{AuxEoM} is algebraic. 

It may be shown that the self-duality equation \eqref{GZ1} is equivalent to\footnote{This condition has a natural generalisation to $4n$ dimensions \cite{Kuzenko:2019nlm}.} 
\bea
\int \rd^4 x \,e \, \widetilde{V}_{ab}  \frac{\d}{\d V_{ab} }  \mathfrak{S}_{\rm int}[ V] =0~.
\label{invariance1}
\eea
Introducing (anti) self-dual components of $V$ defined by 
\bea
V^\pm_{ab} = \hf \Big( V_{a b}  
\pm \ri \widetilde V_{a b} \Big) ~, \quad 
\widetilde V^\pm = \mp\ri V^\pm ~,\quad V = V^+ +V^-~,
\non
\eea
equation \eqref{invariance1} turns into 
\bea
\int \rd^4 x \,e \, \Big( 
 V^+_{a b} \frac{\d}{\d V^+_{ab} } 
 - V^-_{ab} \frac{\d}{\d V^-_{ab} } \Big)\mathfrak{S}_{\rm int}[ V^+, V^-] 
=0~.
\label{invariance2}
 \eea
This means 
that 
the self-interaction
is invariant under rigid $\sU(1)$ phase transformations, 
\bea
\mathfrak{S}_{\rm int}[ \re^{\ri \vf}  V^+, \re^{-\ri \vf} V^-]  = \mathfrak{S}_{\rm int}[ V^+, V^-]  ~, \qquad 
\vf \in {\mathbb R}~.
\label{invariance3}
\eea
We see that the condition of $\sU(1) $ duality invariance is equivalent, within the IZ formulation, to the $\sU(1)$ invariance of 
the self-interaction $ \mathfrak{S}_{\rm int}[ V_+,V_-]  $.  For theories without higher derivatives, eq. 
\eqref{no-higher-derivative}, condition \eqref{invariance3} is equivalent to the $\sU(1)$ invariance of $\mathfrak{L}_{\rm int} (V^\pm)$,
\bea
\mathfrak{L}_{\rm int}( \re^{\ri \vf}  V^+, \re^{-\ri \vf} V^-)  = \mathfrak{L}_{\rm int}( V^+, V^-) ~\implies ~
 \mathfrak{L}_{\rm int} (V^+, V^-) = f ( V^+ \cdot V^+ V^- \cdot V^-)~, 
 \label{invariance4}
 \eea
 with $f(x)$ a real function of a real variable. As follows from our consideration, associated with such a function $f(x)$ is a
 $\sU(1)$ duality-invariant theory. Thus the IZ formulation provides a powerful formalism to generate self-dual models for nonlinear electrodynamics. 

Since $\mathfrak{L}_{\rm int} (V_{ab}) $ is a Lorentz scalar, it can be recast as a function of two invariants of $V_{ab}$, 
\bea
\mathfrak{L}_{\rm int} (V_{ab}) = \mathfrak{L}_{\rm int} (\n, \bar \n)~, \qquad \n:=V^{\a\b}V_{\a\b}
\eea
where the symmetric rank-two spinor $V_{\a\b}$ and its conjugate $\bar V_{\ad\bd}$ are defined as usual, 
\bea
(\s^a)_{\a\ad} (\s^b)_{\b\bd} V^+_{ab} = 2 \ve_{\ad\bd} V_{\a\b}~,\qquad
(\s^a)_{\a\ad} (\s^b)_{\b\bd} V^-_{ab} =  2 \ve_{\a\b} \bar V_{\ad\bd}~,
\eea
The condition of $\sU(1)$ duality invariance, eq. \eqref{invariance4}, becomes 
\bea
\mathfrak{L}_{\rm int} (V_{ab}) = \mathfrak{L}_{\rm int} (\n \bar \n) ~.
\eea
In the case of the ModMax theory \eqref{ModMax}, it holds that \cite{K21}
\bea
\mathfrak{L}_{\rm int, MM} = \k \sqrt{\n \bar \n}~,
\label{MM-self-coupling}
\eea
where the coupling constant $\k$ is related to $\g$ by the rule
\bea
\sinh \g = \frac{\k}{1-(\k/2)^2} ~.
\eea


\subsection{Higher-derivative deformations of the ModMax theory}

As an application of the IZ approach, we discuss, following \cite{Kuzenko:2024zra},
those higher-derivative deformations of the ModMax theory which may contribute to a low-energy effective action of the theory. 
An important insight is obtained by considering the in-out vacuum amplitude for the ModMax theory 
\bea
Z = \int [\mathfrak{D} A_a] [\mathfrak{D} V_{ab} ]  \,\d\big[ \nabla_a A^a - \x\big ] {\rm Det}(\nabla^2 ) \,
\exp \left\{ \frac{\ri} {\hbar} \mathfrak{S}_{\rm MM} [F,V] \right\}~,
\label{in-out}
\eea
where $A_a$ is the gauge potential, 
$F_{ab} = \nabla_a A_b - \nabla_b A_a$, and $\x(x)$ is a background scalar field.\footnote{The  in-out vacuum amplitude is independent of $\x(x)$, in accordance with \cite{Faddeev:1967fc}.}   
In accordance with \eqref{AuxAction} and \eqref{MM-self-coupling},
the functional $\hbar^{-1} \mathfrak{S}_{\rm MM}[F,V]$ is invariant under 
re-scalings
\bea
\hbar \to \l^2 \hbar ~, \qquad F_{ab}(x) \to \l F_{ab}(x)~, \qquad V_{ab}(x) \to \l V_{ab}(x)~.
\label{scale}
\eea
Formally, the effective action is expected to possess such a scale symmetry.  
Thus it is natural to assume that (a local part of) the effective action has the form
\begin{subequations} \label{LEEA}
\bea
\label{2.11a}
\G_{\rm MM} [F,V] = \mathfrak{S}_{\rm MM} [F,V] + \sum_{n=1}^{\infty} \hbar^n \G^{(n)} [V]
\eea
and possesses the following properties: 
 (i) $\hbar^{-1} \G_{\rm MM}[F,V]$ is 
invariant under \eqref{scale}; (ii) each functional $\G^{(n)} [V] $ is Weyl invariant;
and (iii) each functional $\G^{(n)} [V]$ obeys the condition \eqref{invariance1}.\footnote{These properties imply, in particular, that the ModMax coupling \eqref{MM-self-coupling} cannot be generated as a loop quantum correction.} 
A solution to these requirements is given by 
\bea
\G^{(n)} [V] = g_n \int \rd^4 x \,e \, 
\frac{ \big[ \square_c (\n\bar \n)^{ 1/8} \big]^{2n} } {  (\n\bar \n)^{ (3n-2)/4}}~, 
\eea
\end{subequations}
where $g_n$ is a dimensionless numerical factor, and $\square_c:= \nabla^a\nabla_a$ is the conformal d'Alembertian.\footnote{In four dimensions, the operator $\square_c$ is conformal when acting on the space of primary dimension-one scalar fields, see Appendix \ref{appendixB}.}

Our next step is to eliminate the auxiliary two-form $V_{ab}$ by solving its equation of motion \eqref{AuxEoM}. 
In the spinor notation, this equation is equivalent to 
\begin{align}
	V_{\a \b} = F_{\a \b} - \hf \frac{\d}{\d V^{\a \b}} 
		 \mathfrak{S}^{\rm int} [V]~,
	 \label{AuxEoM2}
\end{align}
and its conjugate. The latter equations can be solved in perturbation theory, say, within the loop expansion. 
For simplicity, our analysis will be restricted to the one-loop deformation in \eqref{LEEA}, and we also set $\hbar =1$.
Thus our model is
\begin{align}
 \mathfrak{L}^{\rm int}_\text{MM,def} = \k \sqrt{\n \bar \n} 
+ g \frac{ \big[ \square_c (\n\bar \n)^{ 1/8} \big]^2 } {  (\n\bar \n)^{ 1/4}}~, \label{2.9a} 
\end{align}
where $g\equiv g_1$.
The equation of motion \eqref{AuxEoM2} takes the form
\begin{align}
		 V_{\a \b} = F_{\a \b} - V_{\a \b} \bigg \{ \frac{\k}{2} \Big ( \frac{\bar{\n}}{\n} \Big )^{\frac 1 2} 
		+ \frac{g \bar{\nu}}{4} \bigg [ (\nu \bar{\nu})^{-\frac 7 8} \square_c \bigg ( \frac{\square_c(\n \bar{\n})^{\frac 1 8}}{(\n \bar{\n})^{\frac 1 4}}\bigg )
		- (\nu \bar{\nu})^{- \frac 5 4} (\square_c (\nu \bar{\nu})^{\frac 1 8} )^2
		\bigg ]
		\bigg \} ~
\end{align}
Eliminating the auxiliary fields gives 
\begin{align}
		L &= L_\text{MM} + g \O^{-\frac12} \big(\square_c \O^{\frac14}\big)^2 
		+ \frac{g^2\O^{- \frac 32}}{4  (1-(\k/2)^2)(1+(\k/2)^2)^2} \Big ( \Box_c (\O^{-\hf} \Box_c \O^{\frac 1 4}) - \O^{-\frac 3 4} (\Box_c \O^{\frac 1 4})^2 \Big )^2  \non \\
		&\phantom{=} \qquad \times  \bigg \{\Big(3-12(\k/2)^2 +20(\k/2)^4\Big) (\o +\bar{\o}) - 4(\k/2)\Big(2+\k/2-5(\k/2)^2 +2(\k/2)^3 \non \\
		& \qquad \qquad \qquad \quad  +9(\k/2)^4 +(\k/2)^5 \Big) \O  \bigg \}+\mathcal{O}(g^3)~,
\label{QA}
\end{align}
where we have defined
\begin{align}
	\O = \frac{\big( 1 + (\k/2)^2 \big) (\o \bar{\o})^{\frac 1 2} - (\k/2) (\o + \bar{\o})}{\big( 1 - (\k/2)^2\big)^2} = 
		\hf (\cosh \g + 1)
	 \frac{\partial L_\text{MM}}{\partial \g}~.
\label{Omega}
\end{align}

What is the significance of the composite field $\O$ appearing in \eqref{QA}? Let $\mathfrak{S} [F,V; g]$ be the action corresponding to the self-coupling \eqref{2.9a}, and $S[F;g]$ denote the $\sU(1)$ duality-invariant model which is obtained upon elimination of the auxiliary field 
$V_{ab}$. Since the parameter $g$ is inert under the $\sU(1)$ duality transformations, 
the functional 
\bea
\label{3.9}
\U(g) := \frac{\pa}{\pa g} S[F;g]
\eea
is duality invariant for any value of $g$, in accordance with our discussion in section \ref{section4.4.1}.
In particular, $\U(g=0)$ is 
a duality-invariant functional in the ModMax theory.  As demonstrated in 
\cite{Ferko:2023wyi}, any two duality-invariant local observables $\cO_1 (F; g)$ and 
$\cO_2(F; g) $, which originate in self-dual nonlinear electrodynamics {\it without higher derivatives},
are functionally dependent. In particular,
every duality-invariant scalar observable $\cO(F;\g)$ in the ModMax theory can be expressed as a function of $\O$. However, this is no longer the case if we allow for functionals involving derivatives of the field strength $F_{ab}$.
Let us consider an infinitesimal duality transformation in the ModMax theory 
\begin{align} 
\d_\lambda F_{\a\b } = \ri \lambda F_{\a\b} \left( \cosh \g -  \sqrt{ \frac{\bar \o}{\o} }\sinh \g \right)
~ \implies ~ \d_\lambda \o = 2\ri \lambda \Big( \o \cosh \g - \sqrt{\o\bar \o} \sinh \g \Big)~.
\end{align} 
Introducing 
\bea
I :=  \sqrt{\o} (1+\cosh \g) - \sqrt{\bar \o} \sinh \g ~\implies ~ I \bar I = 4\O~, 
\eea
we observe that
\bea
\d_\lambda I = \ri \lambda I~.
\eea
This result immediately implies that $\O$ is duality invariant. Moreover, it also implies the existence of new primary and duality-invariant observables, such as $I (\square_c \sqrt{\bar I})^2$, which are functionally independent of $\O$.

\subsection{Flows in the space of self-dual theories} \label{section4.6}

A few years ago, a remarkable result was established \cite{Ferko:2023wyi}
for arbitrary self-dual theories of the form \eqref{SDwithoutHD}.
Given a one-parameter family of $\sU(1)$ duality-invariant theories, $L(F;g)$, with $g$ a duality-invariant parameter, 
the Lagrangian obeys a $T\bar T$-like flow equation
\bea
\frac{\pa }{\pa g}  L  = \mathfrak{F} ( T_{ab} )~,
\eea
for some function $\mathfrak{F} $ of the energy-momentum tensor $T_{ab}$.
This theorem extends  several explicit examples considered earlier in the literature in the context of $T\bar T$ deformations \cite{Conti:2018jho, Babaei-Aghbolagh:2022uij, Ferko:2022iru}.
The quoted result of \cite{Ferko:2023wyi}
has been extended to nonlinear chiral theories in six dimensions \cite{Ferko:2024zth}. 


\section{Duality-invariant models for conformal gauge fields} \label{section5}

We are prepared to introduce $\sU(1)$ duality-invariant models for conformal spin-$s$ gauge fields with $s>1$ \cite{KR21-2}, as a generalisation of self-dual nonlinear electrodynamics. For the real conformal higher-spin (CHS) field $h_{\a(s) \ad(s)}$, the analogue of Maxwell's field strength is the linearised spin-$s$ Weyl tensor which is described by $\mathbb{C}_{\a(2s)}$, eq. \eqref{FS1}, 
and its conjugate $\bar{\mathbb{C}}_{\ad(2s)}$. Unlike Maxwell's field strength, these primary fields are invariant under the gauge transformation \eqref{gauge-tr1} only in conformally flat spacetimes, eq. \eqref{GTWeyl}.
 In what follows we restrict our attention only to conformally flat geometries.
 As discussed in Section \ref{section3.1}, 
 $\mathbb{C}_{\a(2s)}$ obeys the Bianchi identity
\bea
\label{CHS-Bianchi}
\nabla^{\b_1}{}_{(\ad_1} \dots \nabla^{\b_s}{}_{\ad_s)} \mathbb{C}_{\a(s) \b(s)} 
=
\nabla_{(\a_1}{}^{\bd_1} \dots \nabla_{\a_s)}{}^{\bd_s} \bar{\mathbb{C}}_{\ad(s) \bd(s)}  ~.
\eea

Let us consider a dynamical system describing the propagation of $h_{\a(s) \ad(s)}$ on a conformally flat spacetime. 
Its action functional $S^{(s)}[\mathbb{C},\bar{\mathbb C}]$ is assumed to depend only on the field strength $\mathbb{C}_{\a(2s)}$ and its conjugate 
$\bar{\mathbb{C} }_{\ad (2s) }$, hence it is manifestly gauge-invariant.
Next, we assume that  $S^{(s)}[\mathbb{C},\bar{\mathbb C}]$ is extended to a functional of an unconstrained field $\mathbb{C}_{\a(2s)}$ and its conjugate. We introduce the following primary field of dimension $+2$
\be
\ri \mathbb{M}_{\a(2s)} := \frac{\d S^{(s)}[\mathbb{C},\bar{\mathbb{C}}]}{\d \mathbb{C}^{\a(2s)}} ~,
\ee
where the functional derivative with respect to $\mathbb{C}^{\a(2s)}$ is defined by the rule
\be
\d S^{(s)}[\mathbb{C},\bar{\mathbb{C}}] = \int \rd^4x\, e \, \d \mathbb{C}^{\a(2s)} \frac{\d S^{(s)}[\mathbb{C},\bar{\mathbb{C}}]}{\d \mathbb{C}^{\a(2s)}}  + \text{c.c.}
\ee
Varying $S^{(s)}[\mathbb C , \bar{\mathbb C} ]$ with respect to $h_{\a(s) \ad(s)}$ leads to
\be
\label{CHS-EoM}
\nabla^{\b_1}{}_{(\ad_1} \dots \nabla^{\b_s}{}_{\ad_s)} \mathbb{M}_{\a(s) \b(s)} 
=
\nabla_{(\a_1}{}^{\bd_1} \dots \nabla_{\a_s)}{}^{\bd_s} \bar{\mathbb{M}}_{\ad(s) \bd(s)}  ~.
\ee

\subsection{Self-duality equation}\label{section5.1}

A crucial feature of our analysis above is that the functional form of the equation of motion \eqref{CHS-EoM} mirrors that of the Bianchi identity \eqref{CHS-Bianchi}. Consequently,
the union of equations  \eqref{CHS-Bianchi} and  \eqref{CHS-EoM}
is invariant under infinitesimal  $\sSO(2)\cong \sU(1)$ duality transformations:
\bea
\label{DR-spin-s}
\d_\l \mathbb{C}_{\a(2s)} = \l \mathbb{M}_{\a(2s)} ~, \qquad \d_\l \mathbb{M}_{\a(2s)} = - \l \mathbb{C}_{\a(2s)} ~,
\eea
where $\l$ is a constant, real parameter. One may then obtain two equivalent expressions for the variation of 
$S^{(s)}[\mathbb C , \bar{\mathbb C}]$ with respect to \eqref{DR-spin-s}
\bea
\d_\l S^{(s)}[\mathbb C,\bar{\mathbb C}] &=& \frac{\ri \l}{4} \int \rd^4x \, e \, \Big \{ \mathbb C^{\a(2s)} \mathbb C_{\a(2s)}
- \mathbb M^{\a(2s)} \mathbb M_{\a(2s)} \Big \} + \text{c.c.} \non\\
&=& -\frac{\ri \l}{2} \int \rd^4x \, e \, \mathbb M^{\a(2s)} \mathbb M_{\a(2s)}+ \text{c.c.}~,
\label{U1variation}
\eea
as a generalisation of similar derivations in nonlinear electrodynamics 
\cite{GZ2,GZ3,KT2}, see Section \ref{section4.1}.
This implies the self-duality equation \cite{KR21-2}
\bea
\label{SDE-spin-s}
\text{Im} \int \rd^4x \, e \,\Big \{ 
\mathbb C^{\a(2s)} \mathbb C_{\a(2s)}
+ \mathbb M^{\a(2s)} \mathbb M_{\a(2s)}
\Big \}  = 0 ~,
\eea
which must hold for an unconstrained field $\mathbb{C}_{\a(2s)}$ and its conjugate. 

The simplest solution of the self-duality equation \eqref{SDE-spin-s} is the
free CHS model
\begin{align}
	\label{FreeCHSBosonic}
	S^{(s)}_\text{Free}[\mathbb{C}, \bar{\mathbb{C}}] = \frac{(-1)^s}{2} \int \rd^4x \, e \, \Big \{ 
	\mathbb C^{\a(2s)} \mathbb C_{\a(2s)} + \bar{\mathbb C}^{\ad(2s)} \bar{\mathbb C}_{\ad(2s)} \Big \} ~,
\end{align}
which was introduced 
in \cite{FT,FL,FL2} in the case of Minkowski space and extended to arbitrary conformally flat backgrounds in \cite{KP}.

 For models without higher derivatives, 
 \bea
 S^{(s)}[\mathbb C , \bar{\mathbb C}] = \int \rd^4x \, e \,L^{(s)}(\mathbb C , \bar{\mathbb C})~,
 \eea
 the self-duality equation \eqref{SDE-spin-s} is equivalent to 
 \bea
\text{Im} \,\Big \{ 
\mathbb C^{\a(2s)} \mathbb C_{\a(2s)}
+ \mathbb M^{\a(2s)} \mathbb M_{\a(2s)}
\Big \}  = 0 ~.
\label{SDE-spin-s-local}
\eea


\subsection{Examples of self-dual nonlinear theories}\label{section5.2}
 
 Now we consider several nonlinear solutions of the self-duality equation \eqref{SDE-spin-s-local}.
 
Our first example is the following $\sU(1)$ duality-invariant model \cite{KR21-2}
\bea
\label{HSBI}
L^{(s)}_\text{BI}(\mathbb C,\bar{\mathbb C}) =  \J^4 \left\{ 1 - 
\sqrt{1 - (-1)^s \frac{{\mathbb C^2 + \bar{\mathbb C}^2}}{\J^4}+  \frac{(\mathbb C^2 - \bar{\mathbb C}^2)^2}{4 \J^8} 
} \right\} ~, \quad \mathbb C^2 := \mathbb C^{\a(2s)} \mathbb C_{\a(2s)}~.
\eea
This is a higher-spin generalisation of the Born-Infeld theory \eqref{BI-spinor}.

Another example is provided by the following self-dual and conformal theory \cite{KR21-2}
\begin{align}
L^{(s)}_{\rm MM}(\mathbb{C},\bar{\mathbb C}) = \frac{(-1)^s}{2} 
\big( \mathbb{C}^2 + \bar{\mathbb{C}}^2 \big) \cosh \, \g 
+ \sqrt{\mathbb{C}^2 \bar{\mathbb{C}}^2}\, \sinh \, \g 
~, \quad \g \in \mathbb{R},
\label{HSMM}
\end{align}
This is a higher-spin generalisation of the ModMax theory \eqref{ModMax}.

Now let us look for a solution to \eqref{SDE-spin-s-local} of the form 
\begin{subequations}
\bea
L^{(s)}(\mathbb{C},\bar{\mathbb C}) := L (\mathfrak S, \mathfrak P)~,
\label{HSSDmodel}
\eea
where we have denoted 
\bea
\mathfrak S = \hf (-1)^s \big( \mathbb{C}^2 + \bar{\mathbb{C}}^2 \big)~,
\qquad \mathfrak P := \frac{\ri}{2} \big( \mathbb{C}^2 - \bar{\mathbb{C}}^2 \big)~.
\eea
\end{subequations}
For such a model, the self-duality equation \eqref{SDE-spin-s-local} turns into
\begin{equation}
    \mathfrak P (L^2_{\mathfrak S} - L^2_{\mathfrak P}-1)=2\mathfrak S L_{\mathfrak S} L_{\mathfrak P}~, \label{selfdual2}
\end{equation}
with $L_{\mathfrak S} = \pa L/\pa \mathfrak S$ and  $L_{\mathfrak P} = \pa L/\pa \mathfrak P$.
Eq. \eqref{selfdual2} proves to be the self-duality equation for nonlinear electrodynamics written in the form given for the first time by Bialynicki-Birula \cite{B-B}. The two invariants of the electromagnetic field \cite{Minkowski} were defined in \cite{B-B} as 
\bea
S = - \hf (\o +\bar \o) = - \frac{1}{4}  F^{ab} F_{ab}~, \qquad P= \frac{\ri}{2} (\o - \bar \o) = - \frac{1}{4} \, F^{ab} \widetilde{F}_{ab} ~,
\eea
and this parametrisation has become popular since the work of Bandos, Lechner, Sorokin and Townsend in \cite{BLST}.
Self-duality equation \eqref{GZ2} on the Lagrangian $L(F_{ab}) = L(S,P)$ is 
\begin{equation}
    P (L^2_{S} - L^2_{P}-1)=2 S L_{S} L_{P}~.
     \label{selfdual3}
\end{equation}

The above discussion implies that every model for self-dual nonlinear electrodynamics admits a higher-spin extension.\footnote{There is a higher-dimensional analogue of this result. Specifically, every model for self-dual nonlinear electrodynamics in four dimensions has a $\mathsf{U}(1)$ duality-invariant extension to $4p>4$ dimensions \cite{Kuzenko:2026kvr}.}
In particular, we can introduce a higher-spin generalisation of the algorithm proposed in \cite{Murcia:2025psi} to generate $\sU(1)$ duality-invariant models. Specifically, if the Lagrangian \eqref{HSSDmodel} is a solution of \eqref{selfdual2}, then the following model 
\bea
\hat L^{(s)}(\mathbb{C},\bar{\mathbb C}) := L (\O, \mathfrak P)~, \qquad
\O = \frac{(-1)^s}{2} 
\big( \mathbb{C}^2 + \bar{\mathbb{C}}^2 \big) \cosh \, \g + \sqrt{\mathbb{C}^2 \bar{\mathbb{C}}^2}\, \sinh \, \g 
\eea
 is also a solution of the self-duality equation \eqref{selfdual2}. For example, applying this algorithm to the higher-spin Born-Infeld action \eqref{HSBI} gives \cite{KR21-2}
\begin{align}
	\label{HSBIgen}
	L^{(s)}_\text{BIgen}(\mathbb{C},\bar{\mathbb C} ) =  
		\J^4 \bigg \{ 1 - \bigg (  1 &- \frac{2}{\J^4} \bigg [ \frac{(-1)^s}{2} \big(\mathbb{C}^2 + \bar{\mathbb C}^2 \big) \,\cosh \g 
	+ \sqrt{\mathbb{C}^2 \bar{\mathbb C}^2}  \,\sinh \g
	\bigg ] \non \\
	& 
		+ \frac{({\mathbb C}^2 - \bar{\mathbb C}^2)^2}{4 \J^8} \bigg)^{\frac 1 2} \bigg \}~.
	\end{align}
For $s=1$ this model coincides with that introduced in \cite{Bandos:2020hgy}.


\subsection{Self-duality under Legendre transformation}\label{section5.3}

In the case of nonlinear (supersymmetric) electrodynamics, $\sU(1)$ duality invariance implies self-duality under Legendre transformations, see section \ref{section4.4.3}.
This remarkable property proves to extend to the higher-spin case, as will be shown below.

We start by describing a Legendre transformation for a generic theory with action 
$S^{(s)}[\mathbb{C},\bar{\mathbb C}]$.
For this we introduce the parent action
\begin{align}
	\label{parent}
	S^{(s)} [\mathbb C,\bar{\mathbb C},
	\mathbb C^{\rm D},\bar{\mathbb C}^{\rm D}] = S^{(s)}[\mathbb C,\bar{\mathbb C}] + \int \rd^4x \, e \, \Big ( \frac \ri 2 \mathbb C^{\a(2s)} \mathbb C^{\rm D}_{\a(2s)} + \text{c.c.} \Big ) ~.
\end{align}
Here $\mathbb{C}_{\a(2s)}$ is an unconstrained field, and $\mathbb C^{\rm D}_{\a(2s)}$ has the form
\begin{align}
	\mathbb C^{\rm D}_{\a(2s)} = \nabla_{(\a_1}{}^{\bd_1} \dots \nabla_{\a_s}{}^{\bd_s} h^{\rm D}_{\a_{s+1} \dots \a_{2s}) \bd(s)} ~,
\end{align}
where $h^{\rm D}_{\a(s) \ad(s)}$ is a Lagrange multiplier field (a primary field  of dimension $2-s$). 
Indeed, upon varying \eqref{parent} with respect to 
$h^{\rm D}_{\a(s) \ad(s)}$ 
one obtains the Bianchi identity \eqref{CHS-Bianchi}, 
and its general solution is given by eq. \eqref{FS1}, for some real field $h_{\a(s)\ad(s)}$.\footnote{This result may be established by making use of the spin projection operators \cite{KP,GGRS, Hutchings:2024qqf}, see also an alternative proof given below.}  
As a result the second term in \eqref{parent} becomes a total derivative, and we end up with the original action 
$S^{(s)}[\mathbb{C},\bar{\mathbb C}]$.
Alternatively, if we first vary  \eqref{parent} with respect to $\mathbb{C}^{\a(2s)}$, the equation of motion is
\begin{align}
	\mathbb{M}_{\a(2s)} = - \mathbb{C}^{\rm D}_{\a(2s)}~,
\end{align}
which we may solve to express $\mathbb{C}_{\a(2s)}$ as a function of 
$\mathbb{C}^{\rm D}_{\a(2s)}$ and its conjugate. 
Inserting this solution into \eqref{parent}, we obtain the dual model
\begin{align}
	\label{dualmodel}
	S^{(s)}_{\rm D}[\mathbb C^{\rm D},\bar{\mathbb C}^{\rm D}] 
	:= \Big [ S^{(s)}[\mathbb C,\bar{\mathbb C}] +  \int \rd^4x \, e \, \Big ( \frac \ri 2 \mathbb C^{\a(2s)} \mathbb C^{\rm D}_{\a(2s)} + \text{c.c.} \Big ) \Big ]\Big |_{\mathbb{C} = \mathbb{C}( \mathbb{C}^{\rm D},  \bar{\mathbb{C}}^{\rm D})}~.
\end{align}

Now, given an action $S^{(s)}[\mathbb C , \bar{\mathbb C} ]$ obeying the self-duality equation\eqref{SDE-spin-s}, our aim is to show that the following property holds
\begin{align}
	\label{legendre}
	S^{(s)}_{\rm D}[\mathbb C,\bar{\mathbb C}] = S^{(s)} [\mathbb C,\bar{\mathbb C}]~,
\end{align}
which means that the corresponding Lagrangian is invariant under Legendre transformations. A routine calculation, being analogous to the one given in Section \ref{section4.4.3}, allows one to show that the following functional
\begin{align}
	\label{invariant}
	S^{(s)}[\mathbb C,\bar{\mathbb C}] +  \int \rd^4x \, e \, \Big ( \frac \ri 4 \mathbb C^{\a(2s)} \mathbb M_{\a(2s)} + \text{c.c.} \Big )
\end{align}
is invariant under infinitesimal duality rotations \eqref{DR-spin-s}. 
Indeed, varying the two terms in {invariant} gives 
\begin{subequations}
\begin{align}
\delta_\lambda S^{(s)}[\mathbb C,\bar{\mathbb C}] 
&= -\frac{\ri \l}{2} \int \rd^4x \, e \, \mathbb M^{\a(2s)} \mathbb M_{\a(2s)}+ \text{c.c.}~, \\
\delta_\lambda \int \rd^4x \, e \, \Big ( \frac \ri 4 \mathbb C^{\a(2s)} \mathbb M_{\a(2s)} + \text{c.c.} \Big )
&=\frac{\ri}{4} \lambda
 \int \rd^4x \, e \, \Big ( \mathbb M^{\a(2s)} \mathbb M_{\a(2s)} - \mathbb C^{\a(2s)} \mathbb C_{\a(2s)} \Big )+ \text{c.c.} 
\end{align}
\end{subequations}
The sum of these variations is proportional to the left-hand side of the self-duality equation  \eqref{SDE-spin-s}.
Thus, the functional \eqref{invariant} is invariant under the infinitesimal transformation \eqref{DR-spin-s}. 
The latter may be exponentiated to obtain the finite $\sU(1)$ duality transformations
\begin{subequations}
\begin{align}
	\mathbb{M}'_{\a(2s)} &= - \text{sin} \l \, \mathbb{C}_{\a(2s)} + \text{cos} \l \, \mathbb{M}_{\a(2s)}  ~,\\
	\mathbb{C}'_{\a(2s)} &= \text{cos} \l \, \mathbb{C}_{\a(2s)} + \text{sin} \l \, \mathbb{M}_{\a(2s)} ~.
\end{align}
\end{subequations}
Performing such a transformation with $\l = \frac \pi 2$ on \eqref{invariant} yields 
\begin{align}
	S^{(s)}[\mathbb C, \bar{\mathbb C}] 
	= S^{(s)}[\mathbb C^{\rm D}, \bar{\mathbb C}^{\rm D}] 
	- \int \rd^4x \, e \, \Big( \frac \ri 2 \mathbb{C}^{\a(2s)} \mathbb{C}^{\rm D}_{\a(2s)} + \text{c.c.} \Big) ~. 
\end{align}
Upon inserting this expression into \eqref{dualmodel}, we obtain \eqref{legendre}, which completes the proof.

In conclusion, we would like to give a simple proof of the fact that the general solution of the Bianchi identity \eqref{CHS-Bianchi}
in Minkowski space  is given by eq. \eqref{FS1}, for some real field $h_{\a(s)\ad(s)}$. 
Let $\mathbb{C}_{\a(2s)} $ be a field subject to the equation 
\bea
\label{Bianchi-flat}
\pa^{\b_1}{}_{(\ad_1} \dots \pa^{\b_s}{}_{\ad_s)} \mathbb{C}_{\a(s) \b(s)} 
=
\pa_{(\a_1}{}^{\bd_1} \dots \pa_{\a_s)}{}^{\bd_s} \bar{\mathbb{C}}_{\ad(s) \bd(s)}  ~.
\eea
Introduce its descendant $h^{\perp}_{\a(s) \ad(s)} $ defined by
\begin{subequations}
\bea
h^{\perp}_{\a(s) \ad(s)} := \pa^{\b_1}{}_{\ad_1} \dots \pa^{\b_s}{}_{\ad_s} \mathbb{C}_{\a(s) \b(s)} =
\pa^{\b_1}{}_{(\ad_1} \dots \pa^{\b_s}{}_{\ad_s)} \mathbb{C}_{\a(s) \b(s)} ~,
\label{perp}
\eea
which is transverse, 
\bea
\pa^{\b\bd} h^{\perp}_{\b\a(s-1) \bd \ad(s-1)} =0~.
\eea
\end{subequations}
The Bianchi identity \eqref{Bianchi-flat} tells us that $h^{\perp}_{\a(s) \ad(s)} $ is real, 
$\overline{h^{\perp}_{\a(s) \ad(s)}  }= h^{\perp}_{\a(s) \ad(s)} $.
Now we can express $\mathbb{C}_{\a(2s)} $ in terms of \eqref{perp},
\bea
\mathbb{C}_{\a(2s)} =  \Box^{-s} \pa_{(\a_1}{}^{\bd_1} \dots \pa_{\a_s}{}^{\bd_s} 
h^{\perp}_{\a_{s+1} \dots \a_{2s}) \bd(s)} 
=\pa_{(\a_1}{}^{\bd_1} \dots \pa_{\a_s}{}^{\bd_s} 
h_{\a_{s+1} \dots \a_{2s}) \bd(s)} ~,
\eea
where $\Box = \pa^a \pa_a$.
In the final relation the real field $h_{\a(s)\ad(s)}$ is not assumed to be transverse.
This field differs from $\Box^{-s} h^{\perp}_{\a (s) \ad(s)} $ by a finite gauge transformation,
\bea
h_{\a(s) \ad(s)} = \Box^{-s} h^{\perp}_{\a (s) \ad(s)} + \pa_{(\a_1 (\ad_1} \z_{\a_2 \dots \a_s) \ad_2 \dots \ad_s)}~,
\eea
with a real gauge parameter $\z_{\a(s-1) \ad(s-1)}$.  


\subsection{General properties of self-dual higher-spin theories}\label{section5.4}

Self-duality under the Legendre transformation studied in Section \ref{section4.4},
 is one of the fundamental properties  of $\sU(1)$ duality-invariant models for the spin-1 field. We have just demonstrated that a similar property holds for $\sU(1)$ duality-invariant higher-spin models. It is natural to wonder whether other properties of self-dual nonlinear electrodynamics extend to higher spins. 
The $\sU(1)$ duality-invariant higher-spin models introduced in Section \ref{section5.2} may be coupled to the dilaton and axion in such a way that the duality group gets enhanced to the non-compact group $\sSL(2,{\mathbb R})$. Thus the results of section \ref{section4.4.2} admit a higher-spin generalisation.

Suppose we are given a self-dual theory such that its action $S^{(s)}[\mathbb{C},\bar{\mathbb C}; g]$ depends on a duality-invariant 
parameter $g$. Then, repeating the arguments given in Section \ref{section4.4.1}, one observes that  $\pa S /\pa g$ is a duality-invariant observable. Thus one of the properties of self-dual nonlinear electrodynamics reviewed in Section \ref{section4.4.1} has a natural higher-spin extension. Unfortunately, the most important property -- duality invariance of the energy-momentum tensor, does not extend to the higher-spin self-dual theories. The point is that for $s>1$ all self-dual theories are formulated on a conformally flat spacetime. However, the energy-momentum tensor is obtained by varying the action with respect to the vielbein. Once the vielbein is given an arbitrary infinitesimal displacement, the resulting spacetime is no longer conformally flat. 
There is a different derivation of this negative result.  As discussed earlier, the linearised spin-$s$ Weyl tensor  $\mathbb{C}_{\a(2s)}$, eq. \eqref{FS1}, and its conjugate $\bar{\mathbb{C}}_{\ad(2s)}$ are primary fields on an arbitrary curved background. 
However, for $s>1$ they are invariant under the gauge transformation \eqref{gauge-tr1} only on conformally flat spacetimes, eq. \eqref{GTWeyl}. Given a self-dual theory $S^{(s)}[\mathbb{C},\bar{\mathbb C}; \J]$, with $\J$ the conformal compensator,  its action can be uniquely lifted to a generic curved background to remain conformal, and then the energy-momentum tensor can be computed using the standard procedure. However, the resulting energy-momentum tensor $T_{ab}$ will be neither gauge invariant nor expressible entirely in terms $\mathbb{C}_{\a(2s)}$ and $\bar{\mathbb{C}}_{\ad(2s)}$, even in the case of the free model \eqref{action1}.  Since \eqref{action1} is conformal, the corresponding energy-momentum tensor $T_{\a(2) \bd (2) }$ is a primary field of dimension $+4$. There is now way to express $T_{\a(2) \bd (2) }$ is terms of the  primary dimension-2 fields
$\mathbb{C}_{\a(2s)}$ and $\bar{\mathbb{C}}_{\ad(2s)}$ for $s>1$.


\subsection{Auxiliary variable formulation} \label{section5.5}

As a generalisation of the Ivanov-Zupnik approach \cite{IZ_N3,IZ1,IZ2}, 
here we will introduce a formalism to generate duality-invariant models that makes use of auxiliary variables. 

Consider the following action functional
\begin{align}
	\label{auxiliaryCHS}
	S^{(s)}[\mathbb{C},\bar{\mathbb C}, \r, \bar \r] &= (-1)^s \int \rd^4x 
	\, e \, \Big \{ 2 \r \mathbb{C} - 
	\r^2- \frac{1}{2}\mathbb{C}^2 \Big \} + \text{c.c.} + \mathfrak{S}^{(s)}_{\text{int}} [\r , \bar{\r}] ~.
\end{align}
Here we have introduced the auxiliary variable $\r_{\a(2s)}$
which is chosen to be an unconstrained primary dimension-2 field.
The functional 
$\mathfrak{S}^{(s)}_{\text{int}} [\r , \bar{\r}]$, by definition, contains cubic and higher powers of $\r_{\a(2s)}$ and its conjugate. The equation of motion for $\r^{\a(2s)}$ is 
\be
\r_{\a(2s)} = \mathbb{C}_{\a(2s)} + \frac{(-1)^s}{2} \frac{\d {S}^{(s)}_{\text{int}} [\r , \bar{\r}]}{\d \r^{\a(2s)}}~.
\ee
This equation allows one to express $\r_{\a(2s)}$ as a functional of $\mathbb{C}_{\a(2s)}$ and its conjugate. This means that \eqref{auxiliaryCHS} is equivalent to a CHS theory with action
\begin{align}
	\label{dualTheory}
	S^{(s)}[\mathbb{C},\bar{\mathbb C}] &= \frac{(-1)^s}{2} \int \rd^4x 
	\, e \, \mathbb{C}^2 + \text{c.c.} + {S}^{(s)}_{\text{int}} [\mathbb{C} , \bar{\mathbb C}] ~.
\end{align}
Thus, \eqref{auxiliaryCHS} and \eqref{dualTheory} provide two equivalent realisations of the same model.

The power of this formulation is most evident when $S^{(s)}[\mathbb{C},\bar{\mathbb C}]$ satisfies
the self-duality equation \eqref{SDE-spin-s}.   
A routine computation reveals that this constraint is equivalent to
\be
\label{dualCondition}
\text{Im} \int \rd^4x \, e \, \r^{\a(2s)} \frac{\d \mathfrak{S}^{(s)}_{\text{int}} [\r , \bar{\r}]}{\d \r^{\a(2s)}} = 0 ~.
\ee
Thus, self-duality of the action \eqref{auxiliaryCHS} is equivalent to the requirement that $\mathfrak{S}^{(s)}_{\text{int}}[\r,\bar{\r}]$ is invariant under rigid $\sU(1)$ phase transformations
\be
\mathfrak{S}^{(s)}_{\text{int}} [\re^{\ri \varphi} \r , \re^{- i \varphi} \bar{\r}] 
= \mathfrak{S}^{(s)}_{\text{int}} [\r , \bar{\r}] ~, \quad \varphi \in \mathbb{R} ~.
\label{dualCondition2}
\ee

For instance we can consider the model 
\bea 
\mathfrak{S}^{(s)}_{\text{int}} [\r , \bar{\r} ; \Psi]  = 
\int \rd^4 x\, e\,\J^4 {\mathfrak F} \Big( 
\frac{ {\r}^2 \bar{{\r}}^2 }{\J^8} \Big)~,
\eea
where ${\mathfrak F}(x) $ is a real analytic function of a real variable. However, such models are not conformal if the action does depend on $\J$. The condition of conformal invariance imposes additional nontrivial restrictions. 


\subsection{Conformal $\sU(1)$ duality-invariant models} \label{section5.6}

The higher-spin extension of the ModMax theory given by eq. \eqref{HSMM} is conformal. Unlike the original spin-1 theory 
\eqref{ModMax}, it is no longer unique. There exist other self-dual conformal spin-$s$ models. 
In order to construct such more general models, it is advantageous to make use of the auxiliary variable formulation described above. 

We introduce algebraic invariants of the symmetric rank-$(2s)$ spinor $\r_{\a(2s)} $, which has the same algebraic properties as  the linearised spin-$s$ Weyl tensor $\mathbb{C}_{\a(2s)}$:
\bea
\r^2 = (-1)^s \r_{\a(s)}{}^{\b(s)} \r_{\b(s)}{}^{\a(s)} ~, \quad 
\r^3 = \r_{\a(s)}{}^{\b(s)} \r_{\b(s)}{}^{\g(s)} \r_{\g(s)}{}^{\a(s)}~, \quad \dots
\label{invariants}
\eea
If $s$ is odd, all invariants $\r^{2n +1}$, with $n$ a non-negative integer, vanish.

For simplicity, we restrict our analysis to the conformal graviton, $s=2$. 
In this case the family of invariants \eqref{invariants} contains only two functionally independent invariants \cite{PenroseR}, $\r^2 $ and $\r^3$. In particular, one may show that 
\bea
s=2: \qquad \r^4 = \hf (\r^2)^2~.
\eea
Now we choose the self-interaction in \eqref{auxiliaryCHS} to be of the form 
\bea
\label{HSconformalAction}
\mathfrak{S}^{(2)}_{\text{int}} [\r , \bar{\r}] 
=\int \rd^4 x\, e\, \Big\{ \b \big( \r^2 \bar \r^2\big)^{\hf} + \k \big( \r^3 \bar \r^3\big)^{\frac 13} \Big\}~,
\eea
where $\b$ and $\k$ are real coupling constants. The resulting model is clearly 
conformal and $\sU(1)$ duality-invariant. For $\k \neq 0$, elimination of the auxiliary variables $\r_{\a(4)}$ and $\bar \r_{\ad(4)}$ does not result 
in a simple action like  \eqref{HSMM}.
In particular, such an elimination, to quadratic order in the couplings, yields the following self-dual model
\bea
\label{confgravitonNLAction}
S^{(2)}[\mathbb{C},\bar{\mathbb{C}}] &=& \int \rd^4 x\, e\, \bigg\{ 
\frac 1 2 \Big (1 + \frac 1 2 \b^2 \Big) (\mathbb{C}^2 +\bar{\mathbb{C}}^2) 
+ \b (\mathbb{C}^2 \bar{\mathbb{C}}^2)^{\frac 1 2} + \kappa (\mathbb{C}^3 \bar{\mathbb{C}}^3)^{\frac 1 3} \non \\
&& \quad + \frac{1}{2} \b \k \frac{(\mathbb{C}^3)^2 \bar{\mathbb{C}}^2 + (\bar{\mathbb{C}}^3)^2 {\mathbb{C}}^2 }{(\mathbb{C}^3 \bar{\mathbb{C}}^3)^{\frac 2 3} (\mathbb{C}^2 \bar{\mathbb{C}}^2)^{\frac 1 2}}
+ \frac{1}{12} \kappa^2 \frac{(\mathbb{C}^2)^2 + (\bar{\mathbb{C}}^2)^2}{(\mathbb{C}^3 \bar{\mathbb{C}}^3)^{\frac 1 3}} \non \\
&& \quad - \frac{1}{24} \kappa^2 \frac{(\mathbb{C}^3)^2 (\bar{\mathbb{C}}^2)^2 + (\bar{\mathbb{C}}^3)^2 ({\mathbb{C}}^2)^2}{(\mathbb{C}^3 \bar{\mathbb{C}}^3)^{\frac 4 3}} + \dots \bigg \} ~.
\eea
The ellipsis in \eqref{confgravitonNLAction} denotes additional contributions to the full nonlinear theory which are cubic or higher order in the coupling constants.
We emphasise that for the special case $\k=0$ the above action yields \eqref{HSMM}
upon making the identification
\begin{align}
	\cosh \, \g = \frac{1 + (\b/2)^2}{1-(\b/2)^2} \quad \Longleftrightarrow \quad \sinh \, \g = \frac{\b}{1 - (\b/2)^2} ~.
\end{align}
With $\k \neq 0$ the action is ill-defined unless $\mathbb{C}^2 \neq $ and $\mathbb{C}^3 \neq 0$.

For $s> 2$ the number of functionally independent  invariants of $\r_{\a(2s)}$ can be shown to be $2(s-1)$, see e.g. \cite{Cederwall:2025ywy}, and therefore one can define families of conformal $\sU(1)$ duality-invariant models.


\section{Self-dual models for complex conformal gauge fields} \label{section6}

This section is devoted to a review of the formalism of $\sU(1)$ duality rotations for complex conformal gauge fields developed in
\cite{Kuzenko:2023ebe}. The simplest example of such a dynamical system is the free conformal gravitino model \cite{KP} is a conformally flat background. It is described by a prepotential $\f_{\a(2)\ad} $ and its conjugate, and the corresponding superconformal action is
\bea
\label{gravitino}
S^{(2,1)}_{\rm free}[\hat{\mathbb{C}},\check{\mathbb{C}}]
= {\ri^{3}}\int \rd^4 x \, e\,  \hat{ {\mathbb{C}}}^{ \a(3)}\check{{\mathbb{C}}}_{\a(3)} 
+{\rm c.c.} ~,
\eea
where the gravitino field strengths are given by eq. \eqref{GravitinoFS}.

\subsection{$\sU(1)$ duality rotations for conformal gauge fields} \label{section6.1}

We now consider a general dynamical system describing the propagation of a complex conformal field $\phi_{\a(m) \ad(n)}$, with $m>n>0$, in a conformally flat spacetime. 
The corresponding action functional, which we denote $S^{(m,n)}[\hat{\mathbb{C}},\check{\mathbb{C}}]$, is assumed to depend only on the gauge-invariant primary field strengths $\hat{\mathbb{C}}^{[\D]}_{\a(m+n)}$ and  $\check{\mathbb{C}}^{[\D]}_{\a(m+n)}$, defined in \eqref{ComplexFS}, and their conjugates. 
We recall that these field strength obey the Bianchi identity \eqref{ComplexBI}.

Considering $S^{(m,n)}[\hat{\mathbb{C}},\check{\mathbb{C}}]$ as a functional of the unconstrained primary fields $\hat{\mathbb{C}}^{[\D]}_{\a(m+n)}$, $\check{\mathbb{C}}^{[\D]}_{\a(m+n)}$ and their conjugates, we may introduce the primary fields
\be
\ri^{m+n+1} \hat{\mathbb{M}}^{[\D]}_{\a(m+n)} :=  \frac{\d S^{(m,n)}[\hat{\mathbb{C}},\check{\mathbb{C}}]}{\d \check{\mathbb{C}}^{{[\D]} \a(m+n)}} ~, 
\qquad \ri^{m+n+1} \check{\mathbb{M}}^{[\D]}_{\a(m+n)} :=  \frac{\d S^{(m,n)}[\hat{\mathbb{C}},\check{\mathbb{C}}]}{\d \hat{\mathbb{C}}^{{[\D]} \a(m+n)}}~,
\label{CFD}
\ee
where the variational derivative is defined in the following way
\begin{align}
	\d S^{(m,n)}[\hat{\mathbb{C}},\check{\mathbb{C}}] &= \int \rd^4x\,e\,
	\Big \{ \d \hat{\mathbb{C}}^{{[\D]} \a(m+n)} \frac{\d S^{(m,n)}[\hat{\mathbb{C}},\check{\mathbb{C}}]}{\d \hat{\mathbb{C}}^{{[\D]} \a(m+n)}} \non \\
	&\qquad \qquad \qquad \qquad \qquad \quad
	+ \d \check{\mathbb{C}}^{{[\D]},\a(m+n)} \frac{\d S^{(m,n)}[\hat{\mathbb{C}},\check{\mathbb{C}}]}{\d \check{\mathbb{C}}^{{[\D]} \a(m+n)}} \Big \} + \text{c.c.}
\end{align}
The conformal properties of $\hat{\mathbb{M}}^{[\D]}_{\a(m+n)}$ and  $\check{\mathbb{M}}^{[\D]}_{\a(m+n)}$ coincide with those of 
$\hat{\mathbb{C}}^{[\D]}_{\a(m+n)}$ and  $\check{\mathbb{C}}^{[\D]}_{\a(m+n)}$, respectively, 
\begin{subequations} \label{Mdimensions}
	\bea 
	K_\bb \hat{\mathbb{M}}^{[\D]}_{\a(m+n)} &=&0~, \qquad 
	\mathbb{D} \hat{\mathbb{M}}^{[\D]}_{\a(m+n)} = \Big(2 - \frac \D 2\Big) \hat{\mathbb{M}}^{[\D]}_{\a(m+n)}~,\\
	K_\bb\check{{\mathbb{M}}}^{[\D]}_{\a(m+n)}&=&0~,\qquad \mathbb{D}\check{{\mathbb{M}}}^{[\D]}_{\a(m+n)}
	=\Big(2+\frac{\D}{2}\Big)\check{{\mathbb{M}}}^{[\D]}_{\a(m+n)}~,
	\eea
\end{subequations} 
compare with \eqref{dimensions}.

Varying $S^{(m,n)}[\hat{\mathbb{C}},\check{\mathbb{C}}]$ with respect to $\bar{\phi}_{\a(n) \ad(m)}$ yields the following equation of motion
\bea
\label{ComplexEoM}
\nabla^{\b_1}{}_{(\ad_1} \dots \nabla^{\b_m}{}_{\ad_m)} \hat{\mathbb{M}}^{[\D]}_{\a(n) \b(m)} 
= \nabla_{(\a_1}{}^{\bd_1} \dots \nabla_{\a_n)}{}^{\bd_n} \overline{\check{\mathbb{M}}}^{[\D]}_{\ad(m) \bd(n)} ~,
\eea
which is of the same functional form as the Bianchi identity \eqref{ComplexBI}.

It is clear from the discussion above that the system of equations \eqref{ComplexBI} and \eqref{ComplexEoM} is invariant under infinitesimal $\sU(1)$ duality rotations of the form
\begin{subequations}
	\label{ComplexU(1)}
	\begin{align}
		\d_\l \hat{\mathbb{C}}^{[\D]}_{\a(m+n)} &= \l \hat{\mathbb{M}}^{[\D]}_{\a(m+n)} ~, \quad \d_\l \hat{\mathbb{M}}^{[\D]}_{\a(m+n)} = - \l \hat{\mathbb{C}}^{[\D]}_{\a(m+n)}  ~, \\
		\d_\l\check{\mathbb{C}}^{[\D]}_{\a(m+n)} &= \l \check{\mathbb{M}}^{[\D]}_{\a(m+n)}~, \quad \d_\l \check{\mathbb{M}}^{[\D]}_{\a(m+n)} = - \l \check{\mathbb{C}}^{[\D]}_{\a(m+n)} ~,
	\end{align}
\end{subequations}
where $\ \l$ is a real parameter. One may then construct $\sU(1)$ duality-invariant nonlinear models for such fields.
They may be shown to satisfy the self-duality equation 
\cite{Kuzenko:2023ebe}
\bea
\label{ComplexSDE}
\ri^{m+n+1} \int \rd^4x \, e\,\Big \{ \hat{\mathbb{C}}^{[\D] \a(m+n)}  \check{\mathbb{C}}^{[\D]}_{\a(m+n)}
+ \hat{\mathbb{M}}^{[\D] \a(m+n)} \check{\mathbb{M}}^{[\D]}_{\a(m+n)} \Big \} + \text{c.c.}  = 0 ~,
\eea
which must hold for unconstrained fields $\hat{\mathbb{C}}^{[\D]}_{\a(m+n)}$ and $\check{\mathbb{C}}^{[\D]}_{\a(m+n)}$.
The simplest solution of this equation is the free action \eqref{Free}.


\subsection{Self-duality under Legendre transformations}
\label{section6.2}

The analysis given in Section \ref{section5.3} naturally extends to the case of complex conformal fields. 

We begin by describing the Legendre transformation of the model described by action $S^{(m,n)}[\hat{\mathbb{C}},\check{\mathbb{C}}]$. To facilitate this, we introduce the parent action
\begin{align}
	\label{ComplexLegendre}
	S^{(m,n)}[\hat{\mathbb{C}},\check{\mathbb{C}},\hat{\mathfrak{C}},\check{\mathfrak{C}}] &= S^{(m,n)}[\hat{\mathbb{C}},\check{\mathbb{C}}] +
	\Big \{
	\ri^{m+n+1} \int \rd^4 x \, e\, \Big(
	\hat{\mathbb{C}}^{[\D] \a(m+n)} \check{\mathfrak{C}}^{[\D]}_{\a(m+n)} \non \\
	& \qquad \qquad \qquad \qquad \qquad \quad + \hat{\mathfrak{C}}^{[\D] \a(m+n)} \check{\mathbb{C}}^{[\D]}_{\a(m+n)} \Big)+ \text{c.c.} 
	\Big \} ~,
\end{align}
where $\hat{\mathbb{C}}^{[\D]}_{\a(m+n)}$ and $\check{\mathbb{C}}^{[\D]}_{\a(m+n)}$ are unconstrained fields, while $\hat{\mathfrak{C}}^{[\D]}_{\a(m+n)}$ and $\check{\mathfrak{C}}^{[\D]}_{\a(m+n)}$ take the form:
\begin{subequations}
\begin{align}
	\hat{\mathfrak{C}}^{[\D]}_{\a(m+n)} &= \nabla_{(\a_1}{}^{\bd_1} \dots \nabla_{\a_n}{}^{\bd_n} \phi^{(\rm D)}_{\a_{n+1} \dots \a_{m+n}) \bd(n)} ~, \\
	\check{\mathfrak{C}}^{[\D]}_{\a(m+n)} &= \nabla_{(\a_1}{}^{\bd_1} \dots \nabla_{\a_m}{}^{\bd_m} \bar{\phi}^{(\rm D)}_{\a_{m +1} \dots \a_{m+n}) \bd(m)} ~,
\end{align}
\end{subequations}
where $\phi^{(\rm D)}_{\a(m) \ad(n)}$ is a Lagrange multiplier field. Now, if one varies eq. \eqref{ComplexLegendre} with respect to $\phi^{(\rm D)}_{\a(m) \ad(n)}$, the resulting equation of motion is exactly the Bianchi identity \eqref{ComplexBI}, whose general solution is given by \eqref{ComplexFS}. As a result, we recover the original self-dual model.

Next, we vary the parent action \eqref{ComplexLegendre} with respect to the unconstrained fields $\hat{\mathbb{C}}^{[\D]}_{\a(m+n)}$ and $\check{\mathbb{C}}^{[\D]}_{\a(m+n)}$. The resulting equations of motion are
\begin{align}
	\hat{\mathbb{M}}^{[\D]}_{\a(m+n)} = - \hat{\mathfrak{C}}^{[\D]}_{\a(m+n)}~, \qquad \check{\mathbb{M}}^{[\D]}_{\a(m+n)} = - \check{\mathfrak{C}}^{[\D]}_{\a(m+n)}~,
\end{align}
which we may solve to obtain $\hat{\mathbb{C}}^{[\D]}_{\a(m+n)} = \hat{\mathbb{C}}^{[\D]}_{\a(m+n)}(\hat{\mathfrak{C}},\check{\mathfrak{C}})$ and $\check{\mathbb{C}}^{[\D]}_{\a(m+n)} = \check{\mathbb{C}}^{[\D]}_{\a(m+n)}(\hat{\mathfrak{C}},\check{\mathfrak{C}})$. Inserting this solution into \eqref{ComplexLegendre} results in the dual action
\begin{align}
	\label{ComplexDualAction}
	S^{(m,n)}_{\mathrm{Dual}}[\hat{\mathfrak{C}},\check{\mathfrak{C}}]
	:= S^{(m,n)}[\hat{\mathbb{C}},\check{\mathbb{C}},\hat{\mathfrak{C}},\check{\mathfrak{C}}] \Big |_{
		\substack{ \hat{\mathbb{C}} = \hat{\mathbb{C}}(\hat{\mathfrak{C}},\check{\mathfrak{C}}) \\
			\check{\mathbb{C}} = \check{\mathbb{C}}(\hat{\mathfrak{C}},\check{\mathfrak{C}})}} ~.
\end{align}

Now, assuming that the action $S^{(m,n)}[\hat{\mathbb{C}},\check{\mathbb{C}}]$ satisfies the self-duality equation \eqref{ComplexSDE}, we will show that it coincides with the dual action \eqref{ComplexDualAction}
\begin{align}
	\label{SDLegendre}
	S_{\mathrm{Dual}}^{(m,n)}[\hat{\mathbb{C}},\check{\mathbb{C}}] = S^{(m,n)}[\hat{\mathbb{C}},\check{\mathbb{C}}]~.
\end{align}
A routine calculation allows one to show that the following functional is invariant under infinitesimal $\sU(1)$ rotations \eqref{ComplexU(1)}
\begin{align}
	\label{ComplexInvariant}
	S^{(m,n)}[\hat{\mathbb{C}},\check{\mathbb{C}}] +
	\Big \{
	\frac{\ri^{m+n+1}}{2} \int \rd^4 x \, e\,  \big(
	\hat{\mathbb{M}}^{[\D] \a(m+n)} \check{\mathbb{C}}_{\a(m+n)}^{[\D]}
	+ \check{\mathbb{M}}^{[\D] \a(m+n)} \hat{\mathbb{C}}^{[\D]}_{\a(m+n)} \big)+ \text{c.c.} 
	\Big \} ~.
\end{align}
Hence, it must also be preserved by the following finite duality transformations:
\begin{subequations}
	\begin{align}
		\hat{\mathbb{C}}_{\a(m+n)}^{'[\D]} &= \text{cos} \l \, \hat{\mathbb{C}}_{\a(m+n)}^{[\D]} + \text{sin} \l \, \hat{\mathbb{M}}_{\a(m+n)}^{[\D]} ~, \\
		\hat{\mathbb{M}}_{\a(m+n)}^{'[\D]} &= - \text{sin} \l \, \hat{\mathbb{C}}_{\a(m+n)}^{[\D]} + \text{cos} \l \, \hat{\mathbb{M}}_{\a(m+n)}^{[\D]}  ~, \\
		\check{\mathbb{C}}_{\a(m+n)}^{'[\D]} &= \text{cos} \l \, \check{\mathbb{C}}_{\a(m+n)}^{[\D]} + \text{sin} \l \, \check{\mathbb{M}}_{\a(m+n)}^{[\D]} ~, \\
		\check{\mathbb{M}}_{\a(m+n)}^{'[\D]} &= - \text{sin} \l \, \check{\mathbb{C}}_{\a(m+n)}^{[\D]} + \text{cos} \l \, \check{\mathbb{M}}_{\a(m+n)}^{[\D]}  ~.
	\end{align}
\end{subequations}
Performing this transformation on functional \eqref{ComplexInvariant} with $\l = \frac \pi 2$ yields
\begin{align}
	S^{(m,n)}[\hat{\mathbb{C}},\check{\mathbb{C}}] &= 
	S^{(m,n)}[\hat{\mathfrak{C}},\check{\mathfrak{C}}] -
	\Big \{
	\ri^{m+n+1} \int \rd^4 x \, e\, \Big(
	\hat{\mathbb{C}}^{[\D] \a(m+n)} \check{\mathfrak{C}}^{[\D]}_{\a(m+n)} \non \\
	& \qquad \qquad \qquad \qquad \qquad\qquad\quad
	+ \hat{\mathfrak{C}}^{[\D] \a(m+n)} \check{\mathbb{C}}_{\a(m+n)}^{[\D]} \Big) + \text{c.c.} 
	\Big \} ~.
\end{align}
Upon inserting this expression into \eqref{ComplexDualAction}, we obtain \eqref{SDLegendre}. Thus, the Lagrangian associated with any duality-invariant theory is invariant under Legendre transformations.


\section{Superconformal gauge multiplets}\label{section7}

In Sections \ref{section8} and \ref{section9}, we will review $\sU(1)$ duality-invariant systems for $\cN$-extended superconformal gauge multiplets. The present section contains the preparatory material devoted to a brief overview of the conformal superspace approach and superconformal multiplets.

The conformal superspace approach to describe $\cN \leq 3$ conformal supergravity in four dimensions was developed in  \cite{ButterN=1,ButterN=2,KR23}, and its $\cN=4$ extension has been sketched in \cite{ButterN=4}. The formulations for the $\cN=1$ and $\cN=2$ cases are reviewed in \cite{KRTM1} and \cite{KRTM2}, respectively. Beyond $\cN=4$, only the conformal superspace with a flat connection \cite{KR21} is known (see \cite{ERThesis, KKR, KKR2} for a review).

\subsection{Conformal superspace with flat connection}\label{section7.1}

Before discussing superconformal gauge multiplets, we provide a brief review of the $\cN$-extended conformal superspace with a flat connection \cite{KR21}. Our starting point will be an $\cN$-extended superspace $\cM^{4|4\cN}$, parametrised by local coordinates 
$z^{M} = (x^{m},\theta^{\m}_\imath,\bar \theta_{\dot{\mu}}^\imath)$, where $m=0, 1, 2, 3$, $\mu = 1, 2$, $\dot{\mu} = \dot{1}, \dot{2}$ and
$\imath = \underline{1}, \dots, \underline{\cN}$. We take the structure group to be the superconformal group $\sSU(2,2|\cN)$.
Its Lie superalgebra, $\mathfrak{su}(2,2|\cN)$, is spanned by the super-translation $P_A=(P_a, Q_\a^i ,\bar Q^\ad_i)$, Lorentz $M_{ab}$,  $R$-symmetry
$\mathbb{Y}$ and $\mathbb{J}^{i}{}_j$, dilatation $\mathbb{D}$,  and the special superconformal $K^A=(K^a, S^\a_i ,\bar S_\ad^i)$ generators.
The graded commutation relations for $\mathfrak{su}(2,2|\cN)$ are collected in Appendix \ref{appendixC}.

The superspace geometry is formulated in terns of conformally covariant derivatives $\nabla_A = (\nabla_a,\nabla_\a^i,\bar{\nabla}_i^\ad)$, which take the form:
\begin{align}
	\label{CFSS.1}
	\nabla_A &= E_A{}^M \partial_M - \hf \Omega_A{}^{bc} M_{bc} - \Phi_A{}^j{}_k \mathbb{J}^{k}{}_j - \ri \Phi_A \mathbb{Y}
	- B_A \mathbb{D} - \mathfrak{F}_{AB} K^B \eol
	&= E_A{}^M \partial_M - \Omega_A{}^{\b\g} M_{\b\g} - \bar{\Omega}_A{}^{\bd\gd} \bar{M}_{\bd\gd}
	- \Phi_A{}^j{}_k \mathbb{J}^{k}{}_j - \ri \Phi_A \mathbb{Y} - B_A \mathbb{D} - \mathfrak{F}_{A B} K^B ~,
\end{align} 
where $E_{A}{}^M$ denotes the inverse supervielbein while the remaining superfields are connections associated with the non-translational generators of the superconformal group. The graded commutation relations of $\nabla_A$ with the superconformal generators $X_{\underline A} = (M_{ab} ,{\mathbb D}, {\mathbb Y}, {\mathbb J}^i{}_j , K^A)$, $[\nabla_A , X_{\underline B}  \}$,
are obtained from the relations $[P_A , X_{\underline B}\}$ given in Appendix \ref{appendixC} by the replacement $P_A \to \nabla_A$.

By definition, the gauge group of conformal supergravity is generated by local transformations of the form
\begin{align}
	\label{CFSS.2}
	\nabla_A' = \re^{\mathscr{K}} \nabla_A \re^{-\mathscr{K}} ~, \qquad
	\mathscr{K} =  \xi^B \nabla_B+ \hf K^{bc} M_{bc} + \S \mathbb{D} + \ri \rho \mathbb{Y} 
	+ \chi^{i}{}_j \mathbb{J}^{j}{}_i + \L_B K^B ~ ,
\end{align}
where the gauge parameters satisfy natural reality conditions. Given a conformally covariant tensor superfield $\mathscr{U}$ (with its indices suppressed), it transforms under such transformations as follows:
\begin{align}
	\label{CFSS.3}
	\mathscr{U}' = \re^{\mathscr{K}} \mathscr{U}~.  
\end{align}

Within the conformal superspace approach to $\cN$-extended conformal supergravity with $ \cN\leq 4$ developed in 
\cite{ButterN=1,ButterN=2,KR23,ButterN=4}, the graded commutator 
$[\nabla_A,\nabla_B\}$ is expressed in terms of the corresponding super-Weyl tensor and its covariant derivatives. The super-Weyl tensor $\cW_{\a_1\dots \a_{4-\cN}}$
is covariantly chiral in the  $\cN<4$ case \cite{ButterN=1,ButterN=2,KR23}; its structure is more involved for $\cN=4$ 
\cite{ButterN=4}.
In the case of  $\cN$-extended conformal superspace with a flat connection, the graded commutator 
$[\nabla_A,\nabla_B\}$ takes the flat-superspace form 
 \cite{KR21}, which means that 
\bea
\label{CSSCFlat}
\{ \nabla_\a^i , \bar{\nabla}^{\bd}_j \} = - 2 \ri \d^i_j \nabla_\a{}^{\bd} ~,
\eea
and the other  (anti)commutators are equal to zero. 
Since the super-Weyl tensor vanishes, applying a gauge transformation \eqref{CFSS.2}
allows one  (at least locally) to turn the covariant derivatives $\nabla_A$ into 
$ D_{A} = (\partial_{a}, D^i_{\a}, \bar{D}_i^{\ad})$ corresponding to $\cN$-extended Minkowski superspace ${\mathbb M}^{4|4\cN}$. 
Upon degauging \eqref{super-degauging} (compare with the non-supersymmetric case described in Section \ref{Section2.4}), one ends up with a complicated algebra of covariant derivatives given in Appendix \ref{appendixD}.


\subsection{Primary superfields} \label{section7.2}

A tensor superfield $\mathscr{U}$ is said to be primary if it is annihilated by the special superconformal generators 
$K^A = (K^a,S^\a_i,\bar{S}_\ad^i)$, 
\begin{subequations}
\bea
K^A \mathscr{U}=0~.
\eea
 This superfield is said to have dimension (or dilatation weight) $\D_\mathscr{U}$ and $\sU(1)_R$ charge $q_\mathscr{U}$ if 
 \bea
 {\mathbb D} \mathscr{U} = \D_\mathscr{U} \mathscr{U}~, \qquad {\mathbb Y} \mathscr{U} = q_\mathscr{U} \mathscr{U}~.
 \eea
\end{subequations} 
Suppose we are given a primary chiral superfield $\F$,
\bea
K^A \F=0~, \qquad \bar \nabla^\ad_i \F=0~.
\eea
Then it follows from the anti-commutation relation \eqref{C.4d} recast in the form
\bea
\{ \bar{S}_\ad^i , \bar{\nabla}^\bd_j \} &=& \d_j^i \d^\bd_\ad \Big(2 \mathbb{D} + \mathbb{Y} \Big) + 4 \d_j^i  \bar{M}_\ad{}^\bd 
	- 4 \d_\ad^\bd  \mathbb{J}^i{}_j  ~,
\eea
that $\F$ can carry neither iso-spinor no dotted spinor indices, 
$\F = \F_{\a(n)}$. Furthermore, the $\sU(1)_R$ charge of $\F$ is determined by its dimensions as follows: 
\bea
q_\F = -2 \D_\F~.
\eea

There is a regular procedure to construct a primary chiral superfield $\F_{\a(n)}$ of dimension $\D_\F$. Let us introduce the following differential operators: 
\begin{subequations}
\begin{align}
	\nabla_{\a(\cN)} &= \frac{1}{\cN!} \ve_{i_1 \dots i_\cN} \nabla_{(\a_1}^{i_1} \dots \nabla_{\a_\cN)}^{i_\cN} ~, \qquad \bar{\nabla}^{\ad(\cN)} = \frac{1}{\cN!} \ve^{i_1 \dots i_\cN} \bar{\nabla}^{(\ad_1}_{i_1} \dots \bar{\nabla}^{\ad_\cN)}_{i_\cN}~, \\
{\nabla}^{2 \cN}&= \frac{ (-1)^{\cN (\cN+1) /2} }{2^\cN (\cN+1) } {\nabla}^{\a(\cN)} {\nabla}_{\a(\cN)} ~, \qquad	
	\bar{\nabla}^{2 \cN}= \frac{ (-1)^{\cN (\cN+1) /2} }{2^\cN (\cN+1) } \bar{\nabla}_{\ad(\cN)} \bar{\nabla}^{\ad(\cN)} ~.
\end{align}
\end{subequations}
Here the totally antisymmetric $\sSU(\cN)$ tensor $\ve_{i_1 \dots i_\cN}$ is normalised as $\ve^{1 \dots \cN} = \ve_{1 \dots \cN}  = 1$. 
Let us also consider a primary $\sSU(\cN)_R$ neutral rank-$n$ spinor superfield $\X_{\a(n) } $ with the superconformal properties 
\bea
K^B \X_{\a(n)} =0~, \quad {\mathbb D} \X_{\a(n)} = (\D_\F - \cN) \X_{\a(n)} ~, \quad 
{\mathbb Y}\X_{\a(n)} = 2\big(4 -\cN - \D_\F\big) \X_{\a(n)}~.
\eea
 Then the following descendant of $\X_{\a(n)}$,  
 \bea
 \F_{\a(n)} := \bar{\nabla}^{2 \cN}\X_{\a(n)} ~,
 \eea
is a primary chiral superfield of dimension $\D_{\a(n)}$.


\subsection{Superconformal actions} \label{section7.3}

In order for formulate superconformal field theories in $\cN$-extended conformal superspace, there exist two action principles. 
One of them is given by an integral over the full superspace, 
\bea
\cS= \int \rd^{4|4\cN} z\, E\, \cL ~, \qquad \rd^{4|4\cN} z := \rd^4 x \rd^{2\cN} \q \rd^{2\cN} \bar \q~, \qquad 
E^{-1} = {\rm Ber} \,(E_A{}^M)~,
\label{super-action}
\eea
where the Lagrangian is a primary real superfield of dimension $4-2\cN$,
\bea
K^A \cL =0~, \qquad \bar \cL =\cL~, \qquad {\mathbb D}\cL = (4-2\cN)\cL~.
\eea
Using the superconformal  properties of $\cL$, one may show that the action \eqref{super-action} is invariant under the gauge transformations \eqref{CFSS.2} and \eqref{CFSS.3}.

Another action principle makes use of a primary chiral superfield $\cL_c$ of dimension $4-\cN$, 
\bea
K^A \cL_c=0~, \qquad  \bar\nabla^\ad_i \cL_c =0~, \qquad {\mathbb D}\cL_c = (4-\cN) \cL_c~.
\eea
The associated action is given by an integral over the chiral subspace of the full superspace, 
\bea
\cS_c = \int \rd^4x \rd^{2\cN} \q \, \cE\, \cL_c~, 
\label{super-action-chiral}
\eea
with $\cE $ being a suitably chosen chiral measure. The chiral measure is uniquely fixed by requiring the following relation
between \eqref{super-action} and \eqref{super-action-chiral}:
\bea
\int \rd^{4|4\cN} z\, E\, \cL =  \int \rd^4x \rd^{2\cN} \q \, \cE\, \bar \nabla^{2\cN} \cL~.
\eea
This relation allows one to evaluate the chiral measure by making use of superspace normal coordinates following the procedure described in \cite{KT-M2009}. 

Chiral action  \eqref{super-action-chiral} may be reduced to an ordinary conformal action \eqref{action} with $D=4$ by integrating out the Grassmann variables
\bea
S_c = \int \rd^4 x \, e \, L~, \qquad L := \nabla^{2\cN} \cL_c \Big|_{\q = \bar \q =0}~.
\eea
In fact this result holds upon imposing a gauge condition that reduces the superspace gauge group to that of conformal gravity.


\subsection{$\cN$-extended superconformal gauge multiplets} \label{section7.4}

Superconformal gauge multiplets in a conformally flat background 
are described by complex tensor superfields $\U_{\a(m)\ad(n)}$, $m,n \geq 0$. They are defined modulo gauge transformations  of the form given in \cite{KR21}:
\begin{subequations}
\label{SCGT}
\begin{align}
m,n \geq 1 :& \qquad \d_\z \U_{\a(m)\ad(n)} = \nabla_\a^i {\z}_{\a(m-1) \ad(n) i} - \bar{\nabla}_{\ad i} \z_{\a(m) \ad(n-1)}{}^i  ~, \\
m \geq 1, n = 0 :& \qquad \d_\z \U_{\a(m)} = \nabla_\a^i \z_{\a(m-1) i} + \bar{\nabla}_{ij} \z_{\a(m)}{}^{ij}  ~, \\
m=n=0 :& \qquad \d_\z \U = \nabla^{ij} \z_{ij} + \bar{\nabla}_{ij} \bar{\z}^{ij}  ~, \label{SCGTc}
\end{align}
where 
 we have introduced the second-order operators
\begin{align}
	\nabla^{ij} = \nabla^{\a (i} \nabla_\a^{j)} ~, \qquad \bar{\nabla}_{ij} = \bar{\nabla}_{\ad (i} \bar{\nabla}_{j)}^\ad ~.
\end{align}
\end{subequations}

The requirement that \eqref{SCGT} is consistent with superconformal symmetry implies that $\U_{\a(m) \ad(n)}$ is primary, 
\begin{subequations}
\bea 
K^B \U_{\a(m) \ad(n)}=0~, 
\eea
and its dimension and $\sU(1)_R$ charge are as follows:
\begin{align}
	\mathbb{D} \U_{\a(m) \ad(n)} = -\hf \Big(m+n+4\cN-4\Big) \U_{\a(m) \ad(n)} ~, \quad \mathbb{Y} \U_{\a(m) \ad(n)} = (m-n) \U_{\a(m) \ad(n)} ~.
\end{align}
\end{subequations}

In the $m=n\equiv s$ case, the gauge prepotential $\U_{\a(s) \ad(s)}$ may be consistently restricted to be real, $\overline{\U_{\a(s)\ad(s)}} = \U_{\a(s) \ad(s)}$ (this reality condition will be assumed in what follows).
Then the gauge transformations \eqref{SCGT} reduce to:
\begin{subequations}
	\begin{align}
		s\geq1 :& \qquad \d_\z \U_{\a(s)\ad(s)} = \nabla_\a^i \bar{\z}_{\a(s-1) \ad(s) i} - \bar{\nabla}_{\ad i} \z_{\a(s) \ad(s-1)}{}^i  ~, \label{HSTGT}\\
		s=0 :& \qquad \d_\z \U = \nabla^{ij} \bar{\z}_{ij} + \bar{\nabla}_{ij} {\z}^{ij}  ~.
	\end{align}
\end{subequations}
It should be pointed out that the flat-superspace version of eq. \eqref{HSTGT} first appeared in \cite{HST}, as a natural generalisation of the gauge transformation for linearised $\cN=1$ conformal supergravity \cite{FZ2}.

A few comments are in order regarding the important cases $\cN \leq 2$. For $\cN=1$, transformations \eqref{SCGT} are equivalent to those given in \cite{KMT,KP,KPR}:
\begin{subequations}
\begin{align}
	m,n \geq 1 :& \qquad \d_\z \U_{\a(m)\ad(n)} = \nabla_\a {\z}_{\a(m-1) \ad(n)} - \bar{\nabla}_{\ad} \z_{\a(m) \ad(n-1)} ~, \label{N=1SCHS.a}\\
	m \geq 1, n = 0 :& \qquad \d_\z \U_{\a(m)} = \nabla_\a \z_{\a(m-1)} + \z_{\a(m)}  ~, \qquad \bar{\nabla}_\ad \z_{\a(m)} = 0~, 
	\label{N=1SCHS.b}\\
	m=n=0 :& \qquad \d_\z \U = \z + \bar{\z}  ~, \qquad \bar{\nabla}_\ad \z = 0~. \label{N=1SCHS.c}
\end{align}
\end{subequations}
Transformation law \eqref{N=1SCHS.b} with $m=1$ corresponds to the $\cN=1$ conformal gravitino multiplet model \cite{KMT}. Finally, 
the transformation law  \eqref{N=1SCHS.c} describes the $\cN=1$ vector multiplet \cite{WB}.

In the $\cN=2$ case, the transformation law \eqref{SCGTc} describes the linearised $\cN=2$ conformal supergravity multiplet \cite{KT}.


\subsection{Primary gauge-invariant field strengths} \label{section7.5}

Associated with the gauge prepotential $\U_{\a(m) \ad(n)}$  and its conjugate $\bar{\U}_{\a(n) \ad(m)}$ are the following chiral 
descendants:
\begin{subequations}
\label{SUSYFS}
\begin{align}
	\hat{\mathbb{W}}_{\a(m+n+\cN)}^{[\D]} (\U)&= \bar{\nabla}^{2\cN} (\nabla_\a{}^\bd)^n \nabla_{\a(\cN)} \U_{\a(m) \bd(n)} ~, \qquad \bar{\nabla}^\ad_i \hat{\mathbb{W}}_{\a(m+n+\cN)}^{[\D]} = 0~, \\
	\check{\mathbb{W}}^{[\D]}_{\a(m+n+\cN)} (\bar \U) &= \bar{\nabla}^{2\cN} (\nabla_{\a}{}^{\bd})^m \nabla_{\a(\cN)} \bar{\U}_{\a(n) \bd(m)} ~, \qquad \bar{\nabla}^\ad_i \check{\mathbb{W}}^{[\D]}_{\a(m+n+\cN)} = 0~,
\end{align}
\end{subequations}
with $\D = m-n$. 
It should be emphasised that, if $m=n=s$, the descendants \eqref{SUSYFS} coincide,
 $\hat{\mathbb{W}}^{[0]}_{\a(2s+\cN)} = \check{\mathbb{W}}^{[0]}_{\a(2s+\cN)}$, since $\U_{\a(s) \ad(s)}$ 
 is chosen to be real.\footnote{In the super-Poincar\'e case, practically all chiral field strengths given in our paper were introduced (perhaps in a somewhat disguised form) many years ago in \cite{SG81,GGRS}.}  

The chiral descendants \eqref{SUSYFS} have three fundamental properties. Firstly, they prove to be primary superfields, and their  
superconformal properties are:
\begin{subequations} 
	\bea 
	K^B \hat{\mathbb{W}}^{[\D]}_{\a(m+n+\cN)} &=&0~, \qquad 
	\mathbb{D} \hat{\mathbb{W}}^{[\D]}_{\a(m+n+\cN)} = \hf\Big(4 - \D -\cN\Big) \hat{\mathbb{W}}^{[\D]}_{\a(m+n+\cN)}~,\\
	K^{B} \check{\mathbb{W}}^{[\D]}_{\a(m+n+\cN)}&=&0~,\qquad \mathbb{D}\check{{\mathbb{W}}}^{[\D]}_{\a(m+n+\cN)}
	=\hf\Big(4 + \D -\cN\Big)\check{{\mathbb{W}}}^{[\D]}_{\a(m+n+\cN)}~.
	\eea
\end{subequations} 
Secondly, these descendants are invariant under the gauge transformations \eqref{SCGT},
\bea
\d_\z \hat{\mathbb{W}}^{[\D]}_{\a(m+n+\cN)} = \d_\z  \check{\mathbb{W}}^{[\D]}_{\a(m+n+\cN)} =0~,
\eea
and therefore they may be interpreted as linearised gauge-invariant field strengths.
Thirdly, 
the field strengths \eqref{ComplexFS} obey the following Bianchi identity
\bea
\label{SCHSBI}
(\nabla^\b{}_\ad)^m \nabla^{\b(\cN)} \hat{\mathbb{W}}^{[\D]}_{\a(n) \b(m+\cN)} 
= (-1)^{\cN(m+n+1)} (\nabla_\a{}^\bd)^n \bar{\nabla}^{\bd(\cN)}\bar{\check{\mathbb{W}}}^{[\D]}_{\ad(m) \bd(n+\cN)} ~.
\eea


\subsection{Superconformal higher-spin models} \label{section7.6}

The free superconformal action to describe the dynamics of $\U_{\a(m)\ad(n)}$ and its conjugate is   
\bea
\label{FreeSuperconformalAction}
\cS^{(m,n;\cN)}_{\rm Free}[\hat{\mathbb{W}},\check{\mathbb{W}}]
= {\ri^{m+n}}\int \rd^{4}x \, \rd^{2\cN} \q \, {\mathcal E} \,  \hat{ {\mathbb{W}}}^{[\D] \a(m+n+\cN)}(\U)
\check{{\mathbb{W}}}^{[\D]}_{\a(m+n+\cN)} (\bar \U) 
+{\rm c.c.}
\eea
Here ${\cE}$ is the chiral integration measure. 
The overall factor of $\ri^{m+n}$ in \eqref{FreeSuperconformalAction} has been chosen due to the identity
\bea
\ri^{m+n+1} \int \rd^4x \rd^4 \q \, {\cE} \, \hat{\mathbb{W}}^{[\D] \a(m+n+\cN)} (\U) 
\check{\mathbb{W}}^{[\D]}_{\a(m+n+\cN)}  (\bar \U) + \text{c.c.} = 0~,
\eea
which holds up to a total derivative. If $m=n=s$ and $\U_{\a(s)\ad(s)} = \bar \U_{\a(s)\ad(s)} \equiv H_{\a(s)\ad(s)}$, the right-hand side of
\eqref{FreeSuperconformalAction} should be multiplied by $1/2$.

For $\cN=1$, the flat-superspace version of \eqref{FreeSuperconformalAction} was first given\footnote{Extension of 
\eqref{FreeSuperconformalAction} for $\D=0$ to the case of $\cN=1$ anti-de Sitter superspace was also given in \cite{KMT}.}
 in \cite{KMT} and then extended to general conformally flat backgrounds in \cite{KP}. The extension to $\cN > 1$ followed shortly thereafter \cite{KR21}.  


\section{Self-dual models for superconformal gauge multiplets} \label{section8}

This section provides a brief review of the $\cN=1$ duality-invariant theories for superconformal gauge multiplets 
in  $\cN$-extended conformal superspace with flat connection described in \cite{KR21-2, Kuzenko:2023ebe}.
The corresponding formalism was modelled of the self-dual models for $\cN=1$ and $\cN=2$ vector multiplets
\cite{KT1,KT2}, including supersymmetric nonlinear electrodynamics. The latter theories have been studied over two decades, see  
\cite{KMcC, KMcC2, K12, BCFKR, K13, ILZ, CKR, IZ3,  BLST2,K21, Ferko:2022iru} for an incomplete list of references.
That is why we will not discuss them in some detail. However, we should make several related comments. 
\begin{itemize} 
\item Gauge-invariant and $\sU(1)$ duality-invariant models for $\cN=1$ and $\cN=2$ vector multiplets are defined on arbitrary supergravity backgrounds, which implies the fact that the supercurrent multiplet is duality invariant. 

\item The $\cN=1$ vector multiplet corresponds to the choice \eqref{SCGTc} or, equivalently, \eqref{N=1SCHS.c}.
The corresponding chiral field strength 
\bea
{\mathbb W}_\a = \bar \nabla^2 \nabla_\a \U 
\eea
is invariant under the gauge transformation \eqref{N=1SCHS.c} for any $\cN=1$ conformal supergravity background
\cite{ButterN=1}.

\item 
The $\cN=2$ vector multiplet may be described by a primary chiral scalar field strength $\mathbb W$ of dimension $+1$ and its conjugate $\bar{\mathbb W}$, 
\begin{subequations}\label{VMFS}
\bea
K^A {\mathbb W} =0~, \qquad \bar \nabla^\ad_i {\mathbb W}=0~, \qquad {\mathbb D} {\mathbb W} ={\mathbb W}~,
\eea
 subject to the Bianchi identity \cite{GSW}
\bea
\nabla^{ij} \mathbb{W} = \bar{\nabla}^{ij} \bar{\mathbb{W}}~.
\eea
\end{subequations}
This constraint is solved in terms  of the curved superspace analogue of Mezincescu's prepotential \cite{Mezincescu} (see also \cite{HST}),  $V_{ij}=V_{ji}$, which is a primary unconstrained real SU(2) triplet, $\overline{V_{ij}} = V^{ij}= \ve^{ik}\ve^{jl}V_{kl}$. 
The expression for $\mathbb W$ in terms of $V_{ij}$ 
was found in \cite{ButterK} 
\bea
{\mathbb W} = \frac{1}{4}\bar\nabla^4 \nabla^{ij} V_{ij}  ~, \qquad K^A V_{ij}=0~, \qquad 
{\mathbb D} V_{ij} = -2 V_{ij}~.
\eea
The field strength  is invariant under gauge transformations of the form \cite{ButterK2}
\bea
\delta_\z V^{ij} &= \nabla^{\alpha}{}_k \z_\alpha{}^{kij}
+ \bar\nabla_{\ad}{}_k \bar\z^\ad{}^{kij}, \qquad
\z_\alpha{}^{kij} = \z_\alpha{}^{(kij)}~,
\label{pre-gauge}
\eea
with $ \z_\alpha{}^{kij} $ being primary and unconstrained modulo the algebraic condition given. 
In the flat-superspace limit, the gauge transformation law \eqref{pre-gauge} reduces to that given in 
\cite{Mezincescu,HST}. It is important to emphasise that $\bm{\cW}$ is invariant under the gauge transformations \eqref{pre-gauge} in an arbitrary supergravity background.

\item 
Mezincescu's prepotential $V_{ij}$ does not belong to the family of superconformal gauge multiplets  $\U_{\a(m)\ad(n)}$ 
introduced in Section \ref{section7.4}.

\end{itemize}


\subsection{Self-duality equation}\label{section8.1}

Considering $\cS^{(m,n;\cN)}[\hat{\mathbb{W}},\check{\mathbb{W}}]$ as a functional of the chiral, but otherwise unconstrained, superfields $\hat{\mathbb{W}}^{[\D]}_{\a(m+n+\cN)}$, $\check{\mathbb{W}}^{[\D]}_{\a(m+n+\cN)}$ and their conjugates, we define the dual tensors
\begin{subequations}
\label{DualSfs}
\begin{align}
\ri^{m+n+1} \hat{\mathbb{M}}^{[\D]}_{\a(m+n+\cN)} &:=  \frac{\d \cS^{(m,n;\cN)}[\hat{\mathbb{W}},\check{\mathbb{W}}]}{\d \check{\mathbb{W}}^{[\D] \a(m+n+\cN)}} ~, \qquad \bar{\nabla}^\ad_i \hat{\mathbb{M}}^{[\D]}_{\a(m+n+\cN)} = 0~, \\
\qquad \ri^{m+n+1} \check{\mathbb{M}}^{[\D]}_{\a(m+n+\cN)} &:=  \frac{\d \cS^{(m,n;\cN)}[\hat{\mathbb{W}},\check{\mathbb{W}}]}{\d \hat{\mathbb{W}}^{[\D] \a(m+n+\cN)}}~, \qquad
\bar{\nabla}^\ad_i \check{\mathbb{M}}^{[\D]}_{\a(m+n+\cN)} = 0~,
\label{5.15}
\end{align}
\end{subequations}
where the variational derivative is defined as follows
\begin{align}
	\d \cS^{(m,n;\cN)}[\hat{\mathbb{W}},\check{\mathbb{W}}] &= \int \rd^{4}x \, \rd^{2\cN} \q \, {\mathcal E} \, 
	\Big \{ \d \hat{\mathbb{W}}^{[\D] \a(m+n+\cN)} \frac{\d \cS^{(m,n;\cN)}[\hat{\mathbb{W}},\check{\mathbb{W}}]}{\d \hat{\mathbb{W}}^{[\D] \a(m+n+\cN)}}\non  \\
	&\qquad\qquad\qquad\qquad\qquad
	+ \d \check{\mathbb{W}}^{[\D] \a(m+n+\cN)} \frac{\d \cS^{(m,n;\cN)}[\hat{\mathbb{W}},\check{\mathbb{W}}]}{\d \check{\mathbb{W}}^{[\D] \a(m+n+\cN)}} \Big \} + \text{c.c.}
\end{align}
Here ${\mathcal E}$ denotes the chiral measure. The superconformal transformation law of the dual superfields \eqref{DualSfs} are characterised by the properties: 
\begin{subequations} 
	\bea 
	K^B \hat{\mathbb{M}}^{[\D]}_{\a(m+n+\cN)} &=&0~, \qquad 
	\mathbb{D} \hat{\mathbb{M}}^{[\D]}_{\a(m+n+\cN)} = \hf\Big(4 - \D -\cN\Big) \hat{\mathbb{M}}^{[\D]}_{\a(m+n+\cN)}~, \\
	K^B \check{{\mathbb{M}}}^{[\D]}_{\a(m+n+\cN)}&=&0~,\qquad \mathbb{D}\check{{\mathbb{M}}}^{[\D]}_{\a(m+n+\cN)}
	=\hf\Big(4 + \D -\cN\Big) \check{{\mathbb{M}}}^{[\D]}_{\a(m+n+\cN)}~.
	\eea
\end{subequations} 
Additionally, varying $\cS^{(m,n;\cN)}[\hat{\mathbb{W}},\check{\mathbb{W}}]$ with respect to $\bar{\U}^{\a(n) \ad(m)}$ yields the following equation of motion
\bea
\label{SCHSEoM}
(\nabla^\b{}_\ad)^m \nabla^{\b(\cN)} \hat{\mathbb{M}}^{[\D]}_{\a(n) \b(m+\cN)} 
= (-1)^{\cN(m+n+1)} (\nabla_\a{}^\bd)^n \bar{\nabla}^{\bd(\cN)}\bar{\check{\mathbb{M}}}^{[\D]}_{\ad(m) \bd(n+\cN)} ~.
\eea

It is clear that the Bianchi identity \eqref{SCHSBI} and the equation of motion \eqref{SCHSEoM} are together invariant under the following $\sU(1)$ duality rotations:
\begin{subequations}
	\label{SCComplexU(1)}
	\begin{align}
		\d_\l \hat{\mathbb{W}}^{[\D]}_{\a(m+n+\cN)} &= \l \hat{\mathbb{M}}^{[\D]}_{\a(m+n+\cN)} ~, \quad \d_\l \hat{\mathbb{M}}^{[\D]}_{\a(m+n+\cN)} = - \l \hat{\mathbb{W}}^{[\D]}_{\a(m+n+\cN)} ~; \\
		\d_\l \check{\mathbb{W}}^{[\D]}_{\a(m+n+\cN)} &= \l \check{\mathbb{M}}^{[\D]}_{\a(m+n+\cN)} ~, \quad \d_\l \check{\mathbb{M}}^{[\D]}_{\a(m+n+\cN)} = - \l \check{\mathbb{W}}^{[\D]}_{\a(m+n+\cN)} ~.
	\end{align}
\end{subequations}
Here $\l = \bar{\l}$ is an arbitrary constant parameter. A routine analysis then leads to the self-duality equation for $\cS^{(m,n;\cN)}[\hat{\mathbb{W}},\check{\mathbb{W}}]$
\bea
\label{SDeqSUSY}
\ri^{m+n+1} \int \rd^{4}x \, \rd^{2\cN} \q \, {\mathcal E} \, \Big \{ \hat{\mathbb{W}}^{[\D] \a(m+n+\cN)}  \check{\mathbb{W}}^{[\D]}_{\a(m+n+\cN)}
+ \hat{\mathbb{M}}^{[\D] \a(m+n+\cN)} \check{\mathbb{M}}^{[\D]}_{\a(m+n+\cN)} \Big \} + \text{c.c.}  = 0 ~.~~~~~~~
\eea
We emphasise that this equation must hold for chiral, but otherwise unconstrained, superfields $\hat{\mathbb{W}}^{[\D]}_{\a(m+n+\cN)}$ and $\check{\mathbb{W}}^{[\D]}_{\a(m+n+\cN)}$.

The simplest solution of the self-duality equation \eqref{SDeqSUSY} is the free action \eqref{FreeSuperconformalAction}. 


\subsection{Self-duality under Legendre transformations} \label{section8.2}

Earlier, we extended the well-known result, in (supersymmetric)  nonlinear electrodynamics, that $\sU(1)$ duality invariance implies self-duality under Legendre transformations to the case of general conformal gauge fields. Here, we will generalise this result to the case of a superconformal gauge multiplet.

We begin by introducing the parent action
\begin{align}
	\label{SCComplexLegendre}
	\cS^{(m,n;\cN)}[\hat{\mathbb{W}},\check{\mathbb{W}},\hat{\mathfrak{W}},\check{\mathfrak{W}}] &= \cS^{(m,n;\cN)}[\hat{\mathbb{W}},\check{\mathbb{W}}] +
	\Big \{
	\ri^{m+n+1} \int \rd^4 x \, \rd^{2\cN} \q \, \mathcal{E} \, \big(
	\hat{\mathbb{W}}^{[\D] \a(m+n+\cN)} \check{\mathfrak{W}}^{[\D]}_{\a(m+n+\cN)} \non \\
	& \qquad \qquad \qquad \qquad \qquad \quad + \hat{\mathfrak{W}}^{[\D] \a(m+n+\cN)} \check{\mathbb{W}}^{[\D]}_{\a(m+n+\cN)} \big)+ \text{c.c.} 
	\Big \} ~,
\end{align}
in which $\hat{\mathbb{W}}^{[\D]}_{\a(m+n+\cN)}$ and $\check{\mathbb{W}}^{[\D]}_{\a(m+n+\cN)}$ are chiral, but otherwise unconstrained, superfields, while $\hat{\mathfrak{W}}^{[\D]}_{\a(m+n+\cN)}$ and $\check{\mathfrak{W}}^{[\D]}_{\a(m+n+\cN)}$ take the form:
\begin{subequations}
\begin{align}
	\hat{\mathfrak{W}}^{[\D]}_{\a(m+n+\cN)} &= \bar{\nabla}^{2\cN} (\nabla_\a{}^\bd)^n \nabla_{\a(\cN)} \U^{(\rm D)}_{\a(m) \bd(n)} ~, \\
	\check{\mathfrak{W}}^{[\D]}_{\a(m+n+\cN)} &= \bar{\nabla}^{2\cN} (\nabla_{\a}{}^{\bd})^m \nabla_{\a(\cN)} \bar{\U}^{(\rm D)}_{\a(n) \bd(m)} ~.
\end{align}
\end{subequations}
Here $\U^{(\rm D)}_{\a(m) \ad(n)}$ is a Lagrange multiplier superfield. If one varies the action \eqref{SCComplexLegendre} with respect to $\U^{(\rm D)}_{\a(m) \ad(n)}$, the resulting equation of motion is exactly the Bianchi identity \eqref{SCHSBI}, whose general solution is given by \eqref{SUSYFS}.\footnote{The latter claim may be proved by making use of the superspin projector
  operators \cite{GGRS, SG81}.}
Consequently, we recover the original model.

Next, varying \eqref{SCComplexLegendre} with respect to $\hat{\mathbb{W}}^{[\D]}_{\a(m+n+\cN)}$ and $\check{\mathbb{W}}^{[\D]}_{\a(m+n+\cN)}$ leads to
\begin{align}
	\hat{\mathbb{M}}^{[\D]}_{\a(m+n+\cN)} = - \hat{\mathfrak{W}}^{[\D]}_{\a(m+n+\cN)}~, \qquad \check{\mathbb{W}}^{[\D]}_{\a(m+n+\cN)} = - \check{\mathfrak{W}}^{[\D]}_{\a(m+n+\cN)}~,
\end{align}
which we may solve to obtain 
\bea
\hat{\mathbb{W}}^{[\D]}_{\a(m+n+\cN)} = \hat{\mathbb{W}}^{[\D]}_{\a(m+n+\cN)}(\hat{\mathfrak{W}},\check{\mathfrak{W}})~, \qquad 
\check{\mathbb{W}}^{[\D]}_{\a(m+n+\cN)} = \check{\mathbb{W}}^{[\D]}_{\a(m+n+\cN)}(\hat{\mathfrak{W}},\check{\mathfrak{W}})~. 
\eea
Inserting this solution into \eqref{SCComplexLegendre} leads to the dual action
\begin{align}
	\label{SCComplexDualAction}
	\cS^{(m,n;\cN)}_{\mathrm{Dual}}[\hat{\mathfrak{W}},\check{\mathfrak{W}}]
	:= \cS^{(m,n;\cN)}[\hat{\mathbb{W}},\check{\mathbb{W}},\hat{\mathfrak{W}},\check{\mathfrak{W}}] \Big |_{
		\substack{ \hat{\mathbb{W}} = \hat{\mathbb{W}}(\hat{\mathfrak{W}},\check{\mathfrak{W}}) \\
			\check{\mathbb{W}} = \check{\mathbb{W}}(\hat{\mathfrak{W}},\check{\mathfrak{W}})}} ~.
\end{align}

Now, assuming that the action $\cS^{(m,n;\cN)}[\hat{\mathbb{W}},\check{\mathbb{W}}]$ satisfies the self-duality equation \eqref{SDeqSUSY}, we will show that it coincides with the dual action \eqref{SCComplexDualAction}
\begin{align}
	\label{SCSDLegendre}
	\cS^{(m,n;\cN)}_{\mathrm{Dual}}[\hat{\mathbb{W}},\check{\mathbb{W}}] = \cS^{(m,n;\cN)}[\hat{\mathbb{W}},\check{\mathbb{W}}]~.
\end{align}
A routine calculation allows one to show that the following functional is invariant under infinitesimal $\sU(1)$ rotations \eqref{SCComplexU(1)}
\begin{align}
	\label{SCComplexInvariant}
	\cS^{(m,n;\cN)}[\hat{\mathbb{W}},\check{\mathbb{W}}] +
	\Big \{
	\frac{\ri^{m+n+1}}{2} \int \rd^4 x \, \rd^{2\cN} \q \, {\mathcal E} \, \big(&
	\hat{\mathbb{M}}^{[\D] \a(m+n+\cN)} \check{\mathbb{W}}_{\a(m+n+\cN)}^{[\D]} \non \\
	& \qquad
	+ \check{\mathbb{M}}^{[\D] \a(m+n+\cN)} \hat{\mathbb{W}}^{[\D]}_{\a(m+n+\cN)} \big)+ \text{c.c.} 
	\Big \} ~.
\end{align}
Hence, it must also be invariant under the following finite duality transformations:
\begin{subequations}
	\begin{align}
		\hat{\mathbb{W}}_{\a(m+n+\cN)}^{'[\D]} &= \text{cos} \l \, \hat{\mathbb{W}}_{\a(m+n+\cN)}^{[\D]} + \text{sin} \l \, \hat{\mathbb{M}}_{\a(m+n+\cN)}^{[\D]} ~, \\
		\hat{\mathbb{M}}_{\a(m+n+\cN)}^{'[\D]} &= - \text{sin} \l \, \hat{\mathbb{W}}_{\a(m+n+\cN)}^{[\D]} + \text{cos} \l \, \hat{\mathbb{M}}_{\a(m+n+\cN)}^{[\D]}  ~, \\
		\check{\mathbb{W}}_{\a(m+n+\cN)}^{'[\D]} &= \text{cos} \l \, \check{\mathbb{W}}_{\a(m+n+\cN)}^{[\D]} + \text{sin} \l \, \check{\mathbb{M}}_{\a(m+n+\cN)}^{[\D]} ~, \\
		\check{\mathbb{M}}_{\a(m+n+\cN)}^{'[\D]} &= - \text{sin} \l \, \check{\mathbb{W}}_{\a(m+n+\cN)}^{[\D]} + \text{cos} \l \, \check{\mathbb{M}}_{\a(m+n+\cN)}^{[\D]}  ~.
	\end{align}
\end{subequations}
Performing this transformation on \eqref{SCComplexInvariant} with $\l = \frac \pi 2$ yields
\begin{align}
	\cS^{(m,n;\cN)}[\hat{\mathbb{W}},\check{\mathbb{W}}] &= 
	\cS^{(m,n;\cN)}[\hat{\mathfrak{W}},\check{\mathfrak{W}}] -
	\Big \{
	\ri^{m+n+1} \int \rd^4 x \, \rd^{2\cN} \q \, {\mathcal E} \, \Big(
	\hat{\mathbb{W}}^{[\D] \a(m+n+\cN)} \check{\mathfrak{W}}^{[\D]}_{\a(m+n+\cN)} \non \\
	& \qquad \qquad \qquad \qquad \qquad \qquad 
	+ \hat{\mathfrak{W}}^{[\D] \a(m+n+\cN)} \check{\mathbb{W}}_{\a(m+n+\cN)}^{[\D]} \Big) + \text{c.c.} 
	\Big \} ~, 
\end{align}
and upon inserting this expression into \eqref{SCComplexDualAction}, we obtain \eqref{SCSDLegendre}.


\section{The $\cN=2$ superconformal gravitino multiplet} \label{section9}

In the $\cN=2$ case, the gauge-invariant chiral field strengths \eqref{SUSYFS} contain at least two spinor indices, 
with ${\mathbb W}_{\a\b}$ corresponding to the linearised $\cN=2$ conformal supergravity.
The index-free field strength $\mathbb W$, eq. \eqref{VMFS}, corresponds to the $\cN=2$ vector multiplet.   
There exists an $\cN=2$ superconformal gauge multiplet such that its gauge-invariant chiral fields strengths are spinor,
$\hat{\mathbb{W}}_\a $ and $ \check{\mathbb{W}}_\a $. This is the $\cN=2$ superconformal gravitino multiplet constructed
in 2023 \cite{HKR}. In the same year, $\sU(1)$ duality models for $\cN=2$ superconformal gravitino multiplet were formulated 
\cite{Kuzenko:2023ebe}.

\subsection{Superconformal gravitino multiplet}  \label{section9.1}

Superconformal gravitino multiplet is described by a primary unconstrained iso-spinor prepotential $\U_i$, and its conjugate 
$\bar \U^i$, 
defined modulo gauge transformations of the form\footnote{The prepotentials $\U_i$ and $\bar \U^i$ may be embedded into harmonic-superspace prepotentials introduced in \cite{Ivanov:2024bsb}.} 
\begin{align}
	\label{SGr.3}
	\d_{\z} \U_i = {\nabla}^{jk} \z_{ijk} + \bar{\nabla}^{\ad j} \z_{\ad ij}~.
\end{align}
Gauge transformation \eqref{SGr.3} is superconformal provided $\U_i$ is characterised by the properties
\begin{align}
	\label{SGr.4}
	K^B \U_i = 0 ~, \qquad \mathbb{D} \U_i = - \mathbb{Y} \U_i = -2 \U_i~,
\end{align}
Associated with $\U_i$ and $\bar \U_i$ are
the chiral field strengths $\hat{\mathbb{W}}_\a$, $\check{\mathbb{W}}_\a$,
which are defined as follows:
\begin{align}
	\label{SGr.1}
	\hat{\mathbb{W}}_\a (\U)= \bar{\nabla}^{4} \nabla_\a^i {\U}_i~, \qquad 
	\check{\mathbb{W}}_\a (\bar \U) = \bar{\nabla}^{4} \nabla^{ij} \nabla_{\a i} {\bar \U}_j ~.
\end{align}
The field strengths prove to be primary in general curved backgrounds:
\begin{subequations}
	\label{SGr.5}
	\begin{align}
		K^B \hat{\mathbb{W}}_\a  &= 0 ~, \qquad \mathbb{D} \hat{\mathbb{W}}_\a  = \frac{1}{2} \hat{\mathbb{W}}_\a  ~, 
			\\
		K^B \check{\mathbb{W}}_\a &= 0 ~, \qquad \mathbb{D} \check{\mathbb{W}}_\a = \frac{3}{2} \check{\mathbb{W}}_\a ~.
			\end{align}
\end{subequations}
However, the gauge transformation \eqref{SGr.3} leaves the field strengths \eqref{SGr.1} invariant only in conformally flat backgrounds
\be
\label{SGr.6}
W_{\a \b} = 0 \quad \implies \quad \d_{\z} \hat{\mathbb{W}}_{\a} = \d_{\z} \check{\mathbb{W}}_{\a} = 0~.
\ee
Here $W_{\a \b}$ denotes the $\cN=2$ super-Weyl tensor. Such a geometry will be assumed in what follows. It is important to note that the field strengths \eqref{SGr.1} satisfy the Bianchi identity
\begin{align}
	\label{SGr.7}
	\nabla^\a_j \nabla^{ij} \hat{\mathbb W}_\a = \bar{\nabla}^{\ad i} \bar{\check{\mathbb{W}}}_\ad~.
\end{align}


\subsection{$\sU(1)$ duality-invariant models}  \label{section9.2}

Here we consider a dynamical system describing the propagation of the $\cN=2$ superconformal gravitino multiplet in curved superspace.
The associated action functional $\cS[\hat{\mathbb{W}},\check{\mathbb{W}}]$ is assumed to depend on the chiral field strengths $\hat{\mathbb{W}}_\a$, $\check{\mathbb{W}}_\a$ and their conjugates.

We now view $\cS[\hat{\mathbb{W}},\check{\mathbb{W}}]$ as a functional of the unconstrained fields $\hat{\mathbb{W}}_{\a}$, $\check{\mathbb{W}}_{\a}$ and their conjugates. This allows us to introduce the dual chiral superfields
\begin{subequations}
\label{SGr.8}
\begin{align}
\hat{\mathbb{M}}_{\a} &:= \ri  \frac{\d \cS[\hat{\mathbb{W}},\check{\mathbb{W}}]}{\d \check{\mathbb{W}}^{\a}} ~, \qquad \bar{\nabla}^\ad_i \hat{\mathbb{M}}_{\a} = 0 ~, \\
\quad \check{\mathbb{M}}_{\a} &:= \ri \frac{\d \cS[\hat{\mathbb{W}},\check{\mathbb{W}}]}{\d \hat{\mathbb{W}}^{\a}}~, \qquad \bar{\nabla}^\ad_i \check{\mathbb{M}}_{\a} = 0 ~,
\end{align}
\end{subequations}
where we have made the definition
\begin{align}
	\label{SGr.9}
	\d \cS[\hat{\mathbb{W}},\check{\mathbb{W}}] = \int \rd^4x \rd^4\q \, \cE \, 
	\Bigg \{ \d \hat{\mathbb{W}}^{\a} \frac{\d \cS[\hat{\mathbb{W}},\check{\mathbb{W}}]}{\d \hat{\mathbb{W}}^{\a}} + \d \check{\mathbb{W}}^{\a} \frac{\d \cS[\hat{\mathbb{W}},\check{\mathbb{W}}]}{\d \check{\mathbb{W}}^{\a}} \Bigg \} + \text{c.c.}
\end{align}
The superconformal properties of the superfields \eqref{SGr.8} are: 
\begin{subequations} 
	\label{SGr.10}
	\begin{align}
		K^B \hat{\mathbb{M}}_\a  &= 0 ~, \qquad \mathbb{D} \hat{\mathbb{M}}_\a  = \frac{1}{2} \hat{\mathbb{M}}_\a  ~, 
				\\
		K^B \check{\mathbb{M}}_\a &= 0 ~, \qquad \mathbb{D} \check{\mathbb{M}}_\a = \frac{3}{2} \check{\mathbb{M}}_\a ~. 
			\end{align}
\end{subequations} 
Varying $\cS[\hat{\mathbb{W}},\check{\mathbb{W}}]$ with respect to $\U_i$ yields the equation of motion
\begin{align}
	\label{SGr.11}
	\nabla^\a_j \nabla^{ij} \hat{\mathbb M}_\a = \bar{\nabla}^{\ad i} \bar{\check{\mathbb{M}}}_\ad~.
\end{align}
whose functional form mirrors that of the Bianchi identity \eqref{SGr.7}.

It is clear from the discussion above that the system of equations \eqref{SGr.7} and \eqref{SGr.11} is invariant under the $\sU(1)$ duality rotations
\begin{subequations}
	\label{SGr.12}
	\begin{align}
		\d_\l \hat{\mathbb{W}}_{\a} &= \l \hat{\mathbb{M}}_{\a} ~, \quad  \d_\l \hat{\mathbb{M}}_{\a} = - \l \hat{\mathbb{W}}_{\a} ~, \\
		\d_\l \check{\mathbb{W}}_{\a} &= \l \check{\mathbb{M}}_{\a}  ~, \quad \d_\l \check{\mathbb{M}}_{\a} = - \l \check{\mathbb{W}}_{\a} ~.
	\end{align}
\end{subequations}
One may construct $\sU(1)$ duality-invariant models for $\U_i$.
Their actions satisfy  the self-duality equation
\bea
\label{SGr.13}
\mathrm{Im} \int \rd^4 x \rd^4\q \, \cE \, \Big \{ \hat{\mathbb{W}}^{\a}  \check{\mathbb{W}}_{\a}
+ \hat{\mathbb{M}}^{\a} \check{\mathbb{M}}_{\a} \Big \}  = 0 ~,
\eea
which must hold for unconstrained fields $\hat{\mathbb{W}}_{\a}$ and $\check{\mathbb{W}}_{\a}$.
The simplest solution of this equation is the free action\footnote{The $\cN=1$ superspace reduction of this action is described in appendix \ref{appendixB}.}
\bea
\label{SGr.14}
\cS_{\mathrm{Free}}[\hat{\mathbb{W}}, \check{\mathbb{W}}] = \int \rd^4 x \rd^4\q \, \cE \, \hat{\mathbb{W}}^{\a} \check{\mathbb{W}}_{\a} + \text{c.c.}
\eea


\subsection{Self-duality under Legendre transformations}  \label{section9.3}

We begin by describing a Legendre transformation for a generic theory with action 
$\cS[\hat{\mathbb{W}},\check{\mathbb{W}}]$.
For this we introduce the parent action
\begin{align}
	\label{SGr.15}
	\cS[\hat{\mathbb{W}},\check{\mathbb{W}},\hat{\mathfrak{W}},\check{\mathfrak{W}}] = \cS[\hat{\mathbb{W}},\check{\mathbb{W}}] +
	 \Big \{
	  \ri \int \rd^4 x \rd^4 \q \, \cE \, \big(
	  \hat{\mathbb{W}}^\a \check{\mathfrak{W}}_\a 
	  + \hat{\mathfrak{W}}^{\a} \check{\mathbb{W}}_\a \big)+ \text{c.c.} 
\Big \} ~.
\end{align}
Here $\hat{\mathbb{W}}_\a$ and $\check{\mathbb{W}}_\a$ are chiral, but otherwise unconstrained superfields, while $\hat{\mathfrak{W}}_\a$ and $\check{\mathfrak{W}}_\a$ take the form
\begin{align}
	\label{SGr.16}
	 \check{\mathfrak{W}}_\a := \bar{\nabla}^{4} \nabla_\a^i {\U}_i^{\rm D} ~, \qquad \hat{\mathfrak{W}}_\a := \bar{\nabla}^{4} \nabla^{ij} \nabla_{\a i} \bar{\U}_j^{\rm D} ~,
\end{align}
where $\U_i^{\rm D}$ is a Lagrange multiplier superfield. Indeed, upon varying \eqref{SGr.15} with respect to 
$\U_i^{\rm D}$ one obtains the Bianchi identity \eqref{SGr.7}, and its general solution is given by eq. \eqref{SGr.1}, for some primary iso-spinor superfield $\U_i$ defined modulo the gauge transformations \eqref{SGr.3} and characterised by the superconformal properties \eqref{SGr.4}. 
As a result the second term in \eqref{SGr.15} becomes a total derivative, and we end up with the original model. Alternatively, if we first vary the parent action with respect to $\hat{\mathbb{W}}_\a$ and $\check{\mathbb{W}}_\a$, the equations of motion are
\begin{align}
	\label{SGr.17}
	\hat{\mathbb{M}}_{\a} = - \hat{\mathfrak{W}}_{\a}~, \qquad \check{\mathbb{M}}_{\a} = - \check{\mathfrak{W}}_{\a}~,
\end{align}
which we may solve to express $\hat{\mathbb{W}}_\a$ and $\check{\mathbb{W}}_\a$ in terms of the dual field strengths.
Inserting this solution into \eqref{SGr.13}, we obtain the dual model
\begin{align}
	\label{SGr.18}
	\cS_{\mathrm{Dual}}[\hat{\mathfrak{W}},\check{\mathfrak{W}}]
	:= \cS[\hat{\mathbb{W}},\check{\mathbb{W}},\hat{\mathfrak{W}},\check{\mathfrak{W}}] \Big | ~,
\end{align}
where the vertical bar means setting $\hat{\mathbb{W}} = \hat{\mathbb{W}}(\hat{\mathfrak{W}},\check{\mathfrak{W}})$ and 
$\check{\mathbb{W}} = \check{\mathbb{W}}(\hat{\mathfrak{W}},\check{\mathfrak{W}})$.

Now, given an action $\cS[\hat{\mathbb{W}},\check{\mathbb{W}}]$ satisfying the self-duality equation \eqref{SGr.10}, our aim is to show that it satisfies
\begin{align}
	\label{SGr.19}
	\cS_{\mathrm{Dual}}[\hat{\mathbb{W}},\check{\mathbb{W}}] = \cS[\hat{\mathbb{W}},\check{\mathbb{W}}]~,
\end{align}
which means that the corresponding Lagrangian is invariant under Legendre transformations.
 A routine calculation allows one to show that the following functional
\begin{align}
	\label{SGr.20}
	\cS[\hat{\mathbb{W}},\check{\mathbb{W}}] +
	\Big \{
	\frac{\ri}{2} \int \rd^4 x \rd^4 \q \, \cE \, \big(
	\hat{\mathbb{M}}^\a \check{\mathbb{W}}_\a 
	+ \check{\mathbb{M}}^{\a} \hat{\mathbb{W}}_\a \big)+ \text{c.c.} 
	\Big \} ~.
\end{align}
is invariant under \eqref{SGr.9}. The latter may be exponentiated to obtain the finite $\sU(1)$ duality transformations
\begin{subequations}
	\label{SGr.21}
\begin{align}
	\hat{\mathbb{W}}'_\a &= \text{cos} \l \, \hat{\mathbb{W}}_\a + \text{sin} \l \, \hat{\mathbb{M}}_\a ~, \qquad
	\hat{\mathbb{M}}'_\a = - \text{sin} \l \, \hat{\mathbb{W}}_\a + \text{cos} \l \, \hat{\mathbb{M}}_\a  ~, \\
	\check{\mathbb{W}}'_\a &= \text{cos} \l \, \check{\mathbb{W}}_\a + \text{sin} \l \, \check{\mathbb{M}}_\a ~, \qquad
	\check{\mathbb{M}}'_\a = - \text{sin} \l \, \check{\mathbb{W}}_\a + \text{cos} \l \, \check{\mathbb{M}}_\a  ~.
\end{align}
\end{subequations}
Performing such a transformation with $\l = \frac \pi 2$ on \eqref{SGr.20} yields 
\begin{align}
	\label{SGr.22}
	\cS[\hat{\mathbb{W}},\check{\mathbb{W}}] =
	\cS[\hat{\mathfrak{W}},\check{\mathfrak{W}}] -
	\Big \{
	\ri \int \rd^4 x \rd^4 \q \, \cE \, \big(
	\hat{\mathbb{W}}^\a \check{\mathfrak{W}}_\a 
	+ \hat{\mathfrak{W}}^{ \a} \check{\mathfrak{W}}_\a \big) + \text{c.c.} 
	\Big \} ~.
\end{align}
Upon inserting this expression into \eqref{SGr.15}, we obtain \eqref{SGr.19}.


\section{Discussion and conclusions} \label{section10}

In this paper we have reviewed the self-duality equations for (super)fields of arbitrary (super)spin and supersymmetry type, as the necessary and sufficient conditions for a theory to possess $\sU(1)$ duality invariance. One may think of the self-duality equations as a ``periodic table'' of self-duality. It is pertinent here to list the self-duality equations corresponding to the two types of (super)fields that most often occur in practice: (i) $m=n\equiv s$; and (ii) $m=n+1 \equiv s+1$.
\begin{itemize} 
\item 
In a $\sU(1)$ duality-invariant theory for the real gauge field $h_{\a(s) \ad(s)}$, with $s>0$,  the self--duality equation is 
\bea
\label{10.1}
\text{Im} \int \rd^4x \, e \,\Big \{ 
\mathbb C^{\a(2s)} \mathbb C_{\a(2s)}
+ \mathbb M^{\a(2s)} \mathbb M_{\a(2s)}
\Big \}  = 0 ~.
\eea

\item
In a $\sU(1)$ duality-invariant theory for the complex gauge field $\f_{\a(s+1) \ad(s)}$, with $s>0$, the self-duality equation is  
\bea
\label{10.2}
\text{Re} \int \rd^4x \, e\,\Big \{ \hat{\mathbb{C}}^{ \a(2s+1)}  \check{\mathbb{C}}_{\a(2s+1)}
+ \hat{\mathbb{M}}^{ \a(2s+1)} \check{\mathbb{M}}_{\a(2s+1)} \Big \}  = 0 ~.
\eea
This equation is the special case of \eqref{ComplexSDE} corresponding to $\D=1$.
The simplest example of such a gauge field is the conformal gravitino, $s=1$.
\item
In a $\sU(1)$ duality-invariant theory for the real gauge superfield $H_{\a(s) \ad(s)}$, with $s \geq 0$,  the self-duality equation is 
\bea
\label{10.3}
\text{Im} \int \rd^{4}x \, \rd^{2\cN} \q \, {\mathcal E} \, \Big \{ {\mathbb{W}}^{ \a(2s+\cN)}  {\mathbb{W}}_{\a(2s+\cN)}
+ \mathbb{M}^{\a(2s+\cN)} {\mathbb{M}}_{\a(2s+\cN)} \Big \}  = 0 ~.
\eea
\item
In a $\sU(1)$ duality-invariant theory for the complex gauge superfield $\U_{\a(s+1) \ad(s)}$, with $s \geq 0$,  the self-duality equation is 
\bea
\label{10.4}
\text{Re} \int \rd^{4}x \, \rd^{2\cN} \q \, {\mathcal E} \, \Big \{ \hat{\mathbb{W}}^{\a(2s +1+\cN)}  \check{\mathbb{W}}_{\a(2s+1+\cN)}
+ \hat{\mathbb{M}}^{\a(2s+1+\cN )} \check{\mathbb{M}}_{\a(2s+1+\cN)} \Big \} = 0 ~.
\eea
This equation is the special case of \eqref{SDeqSUSY} corresponding to $\D=1$.
The simplest example of such a gauge superfield field occurs in the case of $\cN=1$ supersymmetry and  is the conformal gravitino multiplet $\f_\a$ \cite{KMT,KP} corresponding to  $s=1$.
\end{itemize}

It follows from the equations \eqref{10.3} and \eqref{10.4} that chiral field strengths carry at least $\cN$ undotted spinor indices. 
In the case of $\cN=1$ supersymmetry, at least free self-dual superconformal models are known in which the superfields strengths 
${\mathbb W}_{\a(2s+1)}$ and  $\hat{\mathbb{W}}_{\a(2s +2)} $ \& $ \check{\mathbb{W}}_{\a(2s+2)}$ are realised.
In the case of $\cN=2$ supersymmetry, making use of the gauge multiplets \eqref{SCGT} does not allow one to realise 
chiral superfield strengths  $\hat{\mathbb{W}}_{\a} $ \& $ \check{\mathbb{W}}_{\a}$. However, we have seen in Section 
\ref{section9} that such field strengths correspond to the $\cN=2$ superconformal gravitino multiplet. 

The case of $\cN=3$ supersymmetry is quite interesting. Here making use of the gauge multiplets \eqref{SCGT} does not allow one to realise chiral superfield strengths: (i) with one spinor index; and (ii) with two spinor indices. It is remarkable that option (i)
corresponds to the linearised $\cN=3$ conformal supergravity \cite{KR23}.\footnote{Option (ii) still remains mysterious so far.}  
The unconstrained prepotential for $\cN=3$ conformal supergravity is a traceless Hermitian superfield $H=( H^{i}{}_j)$, $\tr \,H =0$, which is defined modulo gauge transformations of the form
\begin{align}
	\d_\z H^i{}_j = \nabla_\a^i \z^\a_j + \bar{\nabla}^\ad_j \bar{\z}_\ad^i - \frac 1 3 \d_j^i \Big( \nabla_\a^k \z^\a_k + \bar{\nabla}^\ad_k \bar{\z}_\ad^k \Big)~.
\end{align}
This transformation law is superconformal provided $H$ is characterised by the  properties:
\begin{align}
	K^B H^i{}_j = 0 ~, \qquad \mathbb{D} H^i{}_j = -2 H^{i}{}_j ~.
\end{align}
Since $H$ is Hermitian, $\mathbb{Y} H^{i}{}_j = 0$.

Associated with $H$ is the gauge-invariant superfield strength 
\begin{align}
	\mathbb{W}_{\a} = \ve_{ikl}\bar{\nabla}^6 \nabla^{jk} \nabla_\a^l H^{i}{}_j~,
\end{align}
which is a primary chiral superfield 
\begin{subequations}
	\label{LinSW}
\begin{align}
	\label{LinSW-a}
K^B \mathbb{W}_\a = 0~, \qquad 
\bar{\nabla}^\ad_i \mathbb{W}_\a = 0~, \qquad
\mathbb{D} \mathbb{W}_\a = \hf \mathbb{W}_\a ~, 
\end{align}	
which obeys the Bianchi identity
\begin{align}	
\label{LinSW-b}
 \ri \ve_{jkl} \nabla^{ik} \nabla^{\a l} \mathbb{W}_\a = \ri \ve^{ikl} \bar{\nabla}_{jk}  \bar{\nabla}_{\ad l} \bar{\mathbb{W}}^\ad ~.
\end{align}
\end{subequations}

Using $\mathbb{W}_\a$ and its conjugate $\bar{\mathbb{W}}_\ad$, one can construct the superconformal gauge-invariant functional
\begin{align}
	\label{LinCSG}
	\cS_\text{LCSG}[\mathbb{W},\bar{\mathbb{W}}] = \hf \int \rd^{4}x \, \rd^{6} \q \, \cE \, \mathbb{W}^\a \mathbb{W}_{\a} + \text{c.c.} ~,
\end{align}
which describes propagation of the linearised conformal supergravity multiplet in conformally flat backgrounds. The overall numerical coefficient in \eqref{LinCSG} is chosen in accordance with the identity
\begin{align}
		\ri \int \rd^{4}x \, \rd^{6} \q \, \cE \, \mathbb{W}^\a \mathbb{W}_{\a} + \text{c.c.} = 0~,
\end{align}
which holds up to a total derivative. This action functional \eqref{LinCSG} extends the one proposed by Siegel in 1981 \cite{Siegel} to an arbitrary conformally flat background. 

It may be shown that the model \eqref{LinCSG} possesses $\sU(1)$ duality invariance. This is most easily seen by noting that the equation of motion for $H^i{}_j$ is
\begin{align}
	\label{10.8}
	\ri \ve_{jkl} \nabla^{ik} \nabla^{\a l} \mathbb{M}_\a = \ri \ve^{ikl} \bar{\nabla}_{jk}  \bar{\nabla}_{\ad l} \bar{\mathbb{M}}^\ad ~, 
\end{align}
where $\mathbb{M}_\a = - \ri \mathbb{W}_\a$.
Hence, the off-shell constraint \eqref{LinSW-a} and dynamical equation above are together invariant under the rigid $\sU(1)$ rotations 
\begin{align}
	\label{U(1)rot}
	\d_\l \mathbb{W}_\a = \l \mathbb{M}_\a ~, \qquad \d_\l \mathbb{M}_\a = - \l \mathbb{M}_\a ~, \qquad \l \in \mathbb{R}~.
\end{align}

The above realisation serves as a starting point in constructing nonlinear self-dual extensions of linearised conformal supergravity. 
Such a duality-invariant nonlinear theory is described by an action $\cS_{\text {NL}}[\mathbb{W},\bar{\mathbb{W}}]$, which is assumed to depend only on $\mathbb{W}_\a$ and its conjugate. Considering this as a functional of the chiral, but otherwise unconstrained primary superfield $\mathbb{W}_\a$ (and its conjugate $\bar{\mathbb{W}}_\ad$), we may define the dual chiral field strength
\begin{align}
	\label{10.10}
	\ri \mathbb{M}_{\a} := \frac{\d \cS_{\text {NL}}[\mathbb{W},\bar{\mathbb{W}}]}{\d \mathbb{W}^\a} ~, \qquad \bar{\nabla}^\ad_i \mathbb{M}_{\a} = 0~, \qquad K^B \mathbb{M}_\a = 0 ~, \qquad \mathbb{D} \mathbb{M}_\a = \hf \mathbb{M}_\a~,
\end{align}
where the variational derivative is defined as follows
\begin{align}
	\d \cS_{\text {NL}}[\mathbb{W},\bar{\mathbb{W}}] = \int \rd^{4}x \, \rd^{6} \q \, \cE \, \d \mathbb{W}^\a \frac{\d \cS_{\text {NL}}[\mathbb{W},\bar{\mathbb{W}}]}{\d \mathbb{W}^\a} + \text{c.c.}
\end{align}
Now, varying $\cS_{\text {NL}}[\mathbb{W},\bar{\mathbb{W}}]$ with respect to $H^i{}_j$ yields the dynamical equations \eqref{10.8}, where $\mathbb{M}_\a$ is defined in eq. \eqref{10.10}. This model may then be shown to possess $\sU(1)$ duality invariance \eqref{U(1)rot} provided the following self-duality equation holds
\begin{align}
	\text{Im} \int \rd^{4}x \, \rd^{6} \q \, \cE \, 
	\Big \{ \mathbb{W}^\a \mathbb{W}_{\a} + \mathbb{M}^\a \mathbb{M}_{\a} \Big \} =0~,
\end{align}
where $\mathbb{W}_\a$ is taken to be a general chiral spinor.


\noindent
{\bf Acknowledgements:}\\
I am very grateful to my collaborators on conformal (super)gravity, conformal higher-spin (super)fields, and nonlinear self-duality for their contributions to the results reviewed in this work. 
My special thanks are extended to Emmanouil Raptakis for fruitful collaboration on the projects 
\cite{KR21-2, Kuzenko:2023ebe, KR23} essential for this  review, as well as on an early version of this work.
Comments on the manuscript by Jessica Hutomo and Ian McArthur are gratefully acknowledged. 
I acknowledge the kind hospitality of the INFN, Sezione di Padova, the Department of Physics at the University of Mons, and the Department of Physics at the University of Munich, during my visits in June--July 2025. 
This work was supported in part by the Australian Research Council, project DP230101629.


\appendix 

\section{Conformal algebra}\label{appendixA}

For completeness, in this appendix we re-derive the conformal algebra \eqref{ConAlg.1} from the algebra of conformal Killing vector fields on Minkowski space ${\mathbb M}^D$ parametrised by Cartesian coordinates $x^a=(x^0, x^1, \dots , x^{D-1})$. Let $\x = \x^b (x) \pa_b $ be a vector field 
on ${\mathbb M}^D$.  It is called conformal Killing if the following equation holds
\bea
\big[ \x, \pa_a \big] = -K_a{}^b[\x] \pa_b -\s[\x] \pa_a~, \qquad K_{ab}[\x]= -K_{ba}[\x]~,
\eea
for some descendants $K_{ab}[\x]$ and $\s[\x]$. The equation implies that 
\bea
K_{ab} [\x]= \pa_{[a}\x_{b]} := \hf (\pa_a \x_b -\pa_b \x_a)~, \qquad \s[\x] = \frac{1}{D} \pa_b \x^b~,
\label{A.2}
\eea 
and the vector field $\x^b$ obeys the standard conformal Killing equation
\bea
\pa_{(a} \x_{b)} := \hf (\pa_a \x_b +\pa_b \x_a) = \eta_{ab} \s[\x]~.
\label{A.3}
\eea
The general solution of \eqref{A.3} is
\bea
\x^a (x) = a^a +\s x^a -K^a{}_b x^b +2x^a b\cdot x - b^a x^2~, 
\label{A.4}
\eea
with $b\cdot x = \eta_{ab} b^a x^b$. The constant parameters in \eqref{A.4} correspond to a spacetime translation ($a^a$), Lorentz transformation ($K_{ab}=-K_{ba}$),  scale transformation ($\s$), and  special conformal transformation ($b^a$).
The explicit expressions for the parameters \eqref{A.2} are:
\bea
K_{ab}[\x] = K_{ab} + 4 b_{[a}x_{b]} ~, \qquad 
\s[\x] = \s + 2 b\cdot x~.
\eea

Given two conformal Killing vector fields $\x_1$ and $\x_2$, their commutator $\x_1, \x_2$ is also conformal Killing. The Lie algebra of conformal Killing vectors on ${\mathbb M}^D$ is isomorphic to $\mathfrak{so}(d,2)$. Let 
$X_{\tilde a} =\big(P_a, M_{ab}, \mathbb{D}, K^a)$ be the generators of  $\mathfrak{so}(d,2)$. The conformal Killing vector fields corresponding to $X_{\tilde a}$ will be denoted $\x_{\tilde a}(X)$. They are given by 
\bea
\x_a(P) = \pa_a~, \quad \x_{ab}(M) = 2x_{[a}\pa_{b]}~, \quad 
\x({\mathbb D} )= x^b \pa_b~, \quad \x_a (K) = x_a x^b \pa_b -x^2 \pa_a~.
\label{A.6}
\eea
The first-order operator corresponding to \eqref{A.4} is 
\bea
\x= \l^{\tilde a} \x_{\tilde a} = a^a \x_a (P) + \hf K^{a b} \x_{ab} (M) +\s \x ( {\mathbb D}) 
+ b^a \x_a(K)~,
\eea
and it is associated with the element $\l^{\tilde a} X_{\tilde a} \in \mathfrak{so}(d,2)$. 
We can work out the commutation relations between the operators $X_{\tilde a}$.
For this we consider a primary dimensionless scalar field with the conformal transformation law 
\bea
\d_\x \F = \l^{\tilde a} X_{\tilde a} \F := \x \F~.
\eea
Applying two subsequent  variations gives
\bea
\d_{\x_2} \d_{\x_1}\F = \l^{\tilde a}_2 \l^{\tilde b}_1 X_{\tilde a} X_{\tilde b} \F 
= \d_{\x_2} \x_1 \F = \x_1 \x_2 \F~,
\eea
and therefore
\bea
\big[ \d_{\x_2} , \d_{\x_1} \big] \F =  \l^{\tilde a}_2 \l^{\tilde b}_1 \big[ X_{\tilde a} X_{\tilde b} \big] \F = - \big[\x_2, \x_1\big]~.
\eea
Using this definition and making use of the first-order operators \eqref{A.6}, one can read off the commutation relations  \eqref{ConAlg.1}.


\section{Conformal differential operators and composites}\label{appendixB}

The formalism of conformal gravity, which has been reviewed in Section \ref{Section2}, is quite powerful for constructing conformal differential operators and primary composite fields. 

Let $\mathfrak L$ be a space of conformal primary fields of fixed tensor type and dimension. A differential operator 
$\mathfrak O$ on this space is defined to be conformal if
\bea
K^a {\mathfrak O} \J =0~, \qquad  \forall \J \in {\mathfrak L}~.
\eea 
Here we consider simple examples of conformal operators.

Let $\f$ be a primary scalar field
\begin{align}
	K^a \f= 0 ~, \qquad \mathbb{D} \f = \D \f~.
\end{align} 
With $\Box_c := \nabla^a \nabla_a$ we obtain 
\bea
K_b \Box_c \f = 4(\D +1 - D/2 ) \nabla_b \F~.
\eea
Operator $\Box_c$ is conformal for $\D = D/2-1$. 
The degauged form of this operator is
\bea
 \Box_c \f = (\cD^a \cD_a - 2\D {\mathfrak f}^a{}_a)\f ~, \qquad {\mathfrak f}^a{}_a = -\hf P^a{}_a
= - \frac{1}{4(D-1)} R~.
\eea

What about the higher-derivative operator $\Box_c \Box_c $? A short calculation gives 
\bea
K_b \Box_c \Box_c \f = 8(\D +2 - D/2 ) \nabla_b \Box_c \F~.
\eea
We see that the operator $\Box_c \Box_c $ is conformal for $\D= D/2-2$.
This is the famous Fradkin-Tseytlin-Paneitz operator \cite{FT1982, Paneitz}.\footnote{
This operator was discovered by Fradkin and Tseytlin in 1981 \cite{FT1982}
and re-discovered by Paneitz in 1983 \cite{Paneitz}.
It is known as the Paneitz operator in the mathematics literature.
 The same operator  was used by Riegert in 1984 \cite{Riegert} for the
  purpose of integrating the Weyl anomaly. } 
In the $D=4$ case, the degauging of  $\Box_c \Box_c $ gives
\begin{align} 
\label{D_0}
 \Box_c \Box_c  = \Box \Box -  \mathcal{D}^a \big(
	2 {R}_{ab} \,\mathcal{D}^b 
	- \tfrac{2}{3}{R} \,\mathcal{D}_a
	\big)~, \qquad \Box = \cD^a \cD_a~.
\end{align}
It follows that 
\bea
\int \rd^4 x \, e \, \psi \Box_c \Box_c \phi 
= \int \rd^4 x \, e \,\Big\{ \Box\psi \Box \phi +2 (R^{ab} -\frac 13 \eta^{ab}R ) \cD_a \psi \cD_b \phi\Big\}~,
\label{Fourth-Order1}
\eea 
for any primary scalar fields $\psi$ and $\phi$ of dimension $0$.

Now let $\Psi $ be a nowhere vanishing primary scalar field of dimension $\Delta \neq 0$, eq. \eqref{compensator}, in four dimensions,  $D=4$. 
One may check that $ \Box_c \Box_c \ln \J$ is a primary field  \cite{Butter:2013lta}, 
\bea
K_b   \Box_c \Box_c \ln \J = 8 \Box_c \nabla_b (\mathbb D +2 -2) \ln \Psi =0~.
\eea
The degauging of  $\Box_c \Box_c \ln \Psi$ leads to 
\begin{align} \label{D_00}
 \Box_c \Box_c  \ln \Psi =& \Box \Box \ln \Psi -  \mathcal{D}^a \big(
	2 {R}_{ab} \,\mathcal{D}^b \ln \Psi
	- \tfrac{2}{3}{R} \,\mathcal{D}_a \ln \Psi
	\big)\non  \\
	&+ \D \Big( \frac 16 \Box R - \frac 12 R^{ab}R_{ab} + \frac 16 R^2\Big) ~.
\end{align}
It follows that the primary field $ C^{abcd} C_{abcd} + 4 \D^{-1}  \Box_c \Box_c  \ln \Psi $, introduced in   \cite{Butter:2013lta},
differs from the Gauss-Bonnet invariant 
\bea
C^{abcd} C_{abcd} - 2 R^{ab} R_{ab} +\frac 23 R^2
\eea
by a total derivative. It should be pointed out that the expression in the first line of \eqref{D_00} vanishes in a dilatation gauge $ \Psi = \text{const}$.

Let $\tau $ be a primary complex dimensionless scalar, $\mathbb{D} \tau =0$. Associated with $\tau $ and its complex conjugate $\bar \tau$ is the following composite primary real field of dimension $+4$:
\begin{align} 
\Upsilon = (6-D) \Box_c \tau \Box_c \bar \tau 
+2(D-2) \Big( \nabla^a\tau \nabla_a\Box_c \bar \tau + \nabla^a \bar \tau \nabla_a\Box_c  \tau \Big)
+(D-2)^2 \nabla^a \nabla^b \tau \nabla_a \nabla_b \bar \tau~,
\label{CompositePrimary}
\end{align}
$K^a \Upsilon =0$.
The degauged form of this field is 
\begin{align} 
\Upsilon = (6-D) \Box \tau \Box \bar \tau 
&+2(D-2) \Big( \cD^a\tau \cD_a\Box \bar \tau + \cD^a \bar \tau \cD_a\Box  \tau +4(D-2) {\mathfrak f}^{ab}\cD_a \tau \cD_b\bar \tau
\Big) \non \\
&+(D-2)^2 \cD^a \cD^b \tau \cD_a \cD_b \bar \tau~.
\end{align}
In the $D=4$ case, we obtain 
\bea
- \hf \int \rd^4 x \, e \, \U = \int \rd^4 x \, e \,\Big\{ \Box\tau \Box \bar \tau +2 (R^{ab} -\frac 13 \eta^{ab}R ) \cD_a \tau \cD_b\bar \tau\Big\}~.
\label{Fourth-Order2}
\eea 
Comparing the relations \eqref{Fourth-Order1} and \eqref{Fourth-Order2}, we see that the conformal operator $\Box_c \Box_c$ and the composite primary field \eqref{CompositePrimary} allow us to generate the same fourth-order conformal functional in four dimensions.
Some  $\sigma$-model applications of the composite primary field \eqref{Fourth-Order2} are described in  section 
 \ref{section4.4.2}. 

In conclusion, we follow \cite{BKNT} and consider a conformal higher-derivative operator is six dimensions, $D=6$, 
 \bea
 \Box_c^3 - \frac{8}{3} (\nabla^{b} \nabla^{d} C_{abcd}) \nabla^{a} \nabla^{c}~,
\eea
compare with the result in \cite{Wunsch} for primary covariants in six dimensions. This operator is conformal on the space of primary dimensionless scalar fields. 
Given a nowhere vanishing primary scalar field  $\Psi $  of dimension $\Delta \neq 0$, its descendant \cite{BKNT}
\bea
 \Big( \Box_c^3 - \frac{8}{3} (\nabla^{b} \nabla^{d} C_{abcd}) \nabla^{a} \nabla^{c} \Big) \ln \J~
\label{LOGconstruction}
\eea
proves to be primary. It may be shown \cite{BKNT} that the $D=6$ Euler invariant can be presented as linear combination 
of \eqref{LOGconstruction} and the composite 
($C^3$ and $C\Box C$) 
primary fields \eqref{CCC} and \eqref{CBoxC} constructed from the Weyl tensor.


\section{The $\cN$-extended superconformal algebra} 
\label{appendixC}

In this appendix, we spell out our conventions for the $\cN$-extended superconformal algebra of Minkowski superspace,
$\mathfrak{su}(2,2|\cN)$.\footnote{Our normalisation of the generators of $\mathfrak{su}(2,2|\cN)$ is similar to \cite{FT}.}
The superalgebra $\mathfrak{su}(2,2|\cN)$ is spanned by the super-translation $P_A=(P_a, Q_\a^i ,\bar Q^\ad_i)$, Lorentz $M_{ab}$,  $R$-symmetry
$\mathbb{Y}$ and $\mathbb{J}^{i}{}_j$, dilatation $\mathbb{D}$,  and the special superconformal $K^A=(K^a, S^\a_i ,\bar S_\ad^i)$ generators.
The commutation relations for the conformal subalgebra  $\mathfrak{su}(2,2)$ of $\mathfrak{su}(2,2|\cN)$ are given in eq. \eqref{ConAlg.1}.

The $R$-symmetry group $\sU(\cN)_R$ is generated by the $\sU(1)_R$ $(\mathbb{Y})$ and $\sSU(\cN)_R$ $(\mathbb{J}^i{}_j)$ generators, which commute with all elements of the conformal algebra. Amongst themselves, they obey the commutation relations
\begin{align}
	[\mathbb{J}^{i}{}_j,\mathbb{J}^{k}{}_l] = \d^i_l \mathbb{J}^k{}_j - \d^k_j \mathbb{J}^i{}_l ~.
\end{align}

The superconformal algebra is then obtained by extending the translation generator to $P_A=(P_a,Q_\a^i,\bar{Q}^\ad_i)$ and the special conformal generator to $K^A=(K^a,S^\a_i,\bar{S}_\ad^i)$. The commutation relations involving the $Q$-supersymmetry generators with the bosonic ones are:
\begin{subequations} 
	\bea
	\big[M_{ab}, Q_\g^i \big] &=& (\s_{ab})_\g{}^\d Q_\d^i ~,\quad 
	\big[M_{ab}, \bar Q^\gd_i \big] = (\tilde{\s}_{ab})^\gd{}_\dd \bar Q^\dd_i~,\\
	\big[\mathbb{D}, Q_\a^i \big] &=& \hf Q_\a^i ~, \quad
	\big[\mathbb{D}, \bar Q^\ad_i \big] = \hf \bar Q^\ad_i ~, \\
	\big[\mathbb{Y}, Q_\a^i \big] &=&  \frac{4-\cN}{\cN} Q_\a^i ~, \quad
	\big[\mathbb{Y}, \bar Q^\ad_i \big] = \frac{\cN-4}{\cN} \bar Q^\ad_i ~, \label{2.19c} \\
	\big[\mathbb{J}^i{}_j, Q_\a^k \big] &=&  - \d^k_j Q_\a^i + \frac{1}{\mathcal{N}} \d^i_j Q_\a^k ~, \quad
	\big[\mathbb{J}^i{}_j, \bar Q^\ad_k \big] = \d^i_k \bar Q^\ad_j - \frac{1}{\mathcal N} \d^i_j \bar Q^\ad_k ~,  \\
	\big[K^a, Q_\b^i \big] &=& -\ri (\s^a)_\b{}^\bd \bar{S}_\bd^i ~, \quad 
	\big[K^a, \bar{Q}^\bd_i \big] = 
	-\ri ({\s}^a)^\bd{}_\b S^\b_i ~.
	\eea
\end{subequations}
At the same time, the commutation relations involving the $S$-supersymmetry generators 
with the bosonic operators are: 
\begin{subequations}
	\bea
	\big [M_{ab} , S^\g_i \big] &=& - (\s_{ab})_\b{}^\g S^\b_i ~, \quad
	\big[M_{ab} , \bar S_\gd^i \big] = - (\ts_{ab})^\bd{}_\gd \bar S_\bd^i~, \\
	\big[\mathbb{D}, S^\a_i \big] &=& -\hf S^\a_i ~, \quad
	\big[\mathbb{D}, \bar S_\ad^i \big] = -\hf \bar S_\ad^i ~, \\
	\big[\mathbb{Y}, S^\a_i \big] &=&  \frac{\cN-4}{\cN} S^\a_i ~, \quad
	\big[\mathbb{Y}, \bar S_\ad^i \big] =  \frac{4-\cN}{\cN} \bar S_\ad^i ~,  \label{2.20c}\\
	\big[\mathbb{J}^i{}_j, S^\a_k \big] &=&  \d^i_k S^\a_j - \frac{1}{\mathcal{N}} \d^i_j S^\a_k ~, \quad
	\big[\mathbb{J}^i{}_j, \bar S_\ad^k \big] = - \d_j^k \bar S_\ad^i + \frac{1}{\mathcal N} \d^i_j \bar S_\ad^k ~,  \\
	\big[ S^\a_i , P_b \big] &=& \ri (\s_b)^\a{}_\bd \bar{Q}^\bd_i ~, \quad 
	\big[\bar{S}_\ad^i , P_b \big] = 
	\ri ({\s}_b)_\ad{}^\b Q_\b^i ~.
	\eea
\end{subequations}
Finally, the anti-commutation relations of the fermionic generators are: 
\begin{subequations}
	\bea
	\{Q_\a^i , \bar{Q}^\ad_j \} &=& - 2 \ri \d^i_j (\s^b)_\a{}^\ad P_b=- 2 \ri \d^i_j  P_\a{}^\ad~, \\
	\{ S^\a_i , \bar{S}_\ad^j \} &=& 2 \ri  \d_i^j (\s^b)^\a{}_\ad K_b=2 \ri \d_i^j  K^\a{}_\ad
	~, \\
	\{ S^\a_i , Q_\b^j \} &=& \d_i^j \d^\a_\b \Big(2 \mathbb{D} - \mathbb{Y} \Big) - 4 \d_i^j  M^\a{}_\b 
	+ 4 \d^\a_\b  \mathbb{J}^j{}_i ~, \\
	\{ \bar{S}_\ad^i , \bar{Q}^\bd_j \} &=& \d_j^i \d^\bd_\ad \Big(2 \mathbb{D} + \mathbb{Y} \Big) + 4 \d_j^i  \bar{M}_\ad{}^\bd 
	- 4 \d_\ad^\bd  \mathbb{J}^i{}_j  ~. \label{C.4d}
	\eea
\end{subequations}


\section{Degauged conformal superspace with flat connection}
\label{appendixD}

According to eq. \eqref{CFSS.3}, under an infinitesimal special superconformal gauge transformation $\mathscr{K} = \Lambda_{B} K^{B}$, the dilatation connection transforms as follows
\bea
\d_{\mathscr{K}} B_{A} = - 2 \Lambda_{A} ~.
\eea
As a result, it is possible to impose the gauge
$B_{A} = 0$, completely fixing
the special superconformal gauge freedom.\footnote{Actually, there is a class of residual gauge transformations which preserve this gauge. They lead to the super-Weyl transformations of the degauged geometry.} Hence, the corresponding connection is no longer required for the covariance of $\nabla_A$ under the residual gauge freedom and
may be
extracted from $\nabla_{A}$,
\begin{subequations}\label{super-degauging}
\bea
\nabla_{A} &=& \cD_{A} - \mathfrak{F}_{AB} K^{B} ~. \label{ND}
\eea
Here the operator $\cD_{A} $ involves only the Lorentz and $R$-symmetry connections
\bea
\cD_A = E_A{}^M \partial_M - \frac{1}{2} \O_A{}^{bc} M_{bc} - \Phi_A{}^j{}_k \mathbb{J}^{k}{}_j - \ri \F_A \mathbb{Y}~.
\eea
\end{subequations}

The next step is to relate the special superconformal connection
$\mathfrak{F}_{AB}$  to the torsion tensor associated with $\cD_A$. To do this, one can  make use of the relation
\bea
\label{4.3}
[ \cD_{A} , \cD_{B} \} &=&  [ \nabla_{A} , \nabla_{B} \} + \big(\cD_{A} \mathfrak{F}_{BC} - (-1)^{\e_A \e_B} \cD_{B} \mathfrak{F}_{AC} \big) K^C + \mathfrak{F}_{AC} [ K^{C} , \nabla_B \} \non \\
&& - (-1)^{\e_A \e_B} \mathfrak{F}_{BC} [ K^{C} , \nabla_A \} - (-1)^{\e_B \e_C} \mathfrak{F}_{AC} \mathfrak{F}_{BD} [K^D , K^C \} ~.
\eea
In conjunction with the algebra \eqref{CSSCFlat}, this leads to a set of consistency conditions that are equivalent to the Bianchi identities of $\sU(\cN)$ superspace \cite{Howe} with vanishing super-Weyl tensor. Their solution expresses the components of $\mathfrak{F}_{AB}$ in terms of the torsion tensor of $\sU(\cN)$ superspace and determines the algebra $[ \cD_{A} , \cD_{B} \}$. We omit such an analysis here and instead simply present the geometry of $\cD_A$ below. The interested reader is referred to \cite{KKR} for the complete analysis.

\subsection{$\cN=1$ case}

In the $\cN=1$ case, the algebra of covariant derivatives \eqref{4.3} may be brought to the form\footnote{We emphasise that this algebra will not coincide with \eqref{4.3}. This is because we have simplified the geometry by performing the shift $\cD_{\a \ad} \rightarrow \cD_{\a \ad} + \frac{\ri}{2} G^{\b}{}_{\ad} M_{\a \b} - \frac{\ri}{2} G_{\a}{}^{\bd} \bar{M}_{\ad \bd} - \frac{\ri}{4} G_{\a \ad} \mathbb{Y}$.}
\begin{subequations} \label{U(1)algebra}
	\bea
	\{ \cD_{\a}, \cD_{\b} \} &=& -4{\bar R} M_{\a \b}~, \qquad
	\{\cDB_{\ad}, \cDB_{\bd} \} =  4R {\bar M}_{\ad \bd}~, \label{U(1)algebra.a}\\
	&& {} \qquad \{ \cD_{\a} , \cDB_{\ad} \} = -2{\rm i} \cD_{\a \ad} ~, 
	\label{U(1)algebra.b}	\\
	\big[ \cD_{\a} , \cD_{ \b \bd } \big]
	& = &
	{\rm i}
	{\ve}_{\a \b}
	\Big({\bar R}\,\cDB_\bd + G^\g{}_\bd \cD_\g
	- \cD^\g G^\d{}_\bd  M_{\g \d}
	\Big)
	+ {\rm i} \cDB_{\bd} {\bar R}  M_{\a \b}
	\non \\
	&&
	-\frac{\ri}{3} \ve_{\a\b} \bar X^\gd \bar M_{\gd \bd} - \frac{\ri}{6} \ve_{\a\b} \bar X_\bd \mathbb{Y}
	~, \label{U(1)algebra.c}\\
	\big[ {\bar \cD}_{\ad} , \cD_{\b\bd} \big]
	& = &
	- {\rm i}
	\ve_{\ad\bd}
	\Big({R}\,\cD_{\b} + G_\b{}^\gd \cDB_\gd
	- \cDB^{\gd} G_{\b}{}^{\dd}  \bar M_{\gd \dd}
	\Big) 
	- {\rm i} \cD_\b R  {\bar M}_{\ad \bd}
	\non \\
	&&
	+\frac{\ri}{3} \ve_{\ad \bd} X^{\g} M_{\g \b} - \frac{\ri}{6} \ve_{\ad\bd} X_\b \mathbb{Y}
	~. \label{U(1)algebra.d}
	\eea
\end{subequations}
Here $R$ is a chiral scalar superfield
\begin{subequations}
	\label{N=1BIs}
	\bea
	\bar{\cD}_{\ad} R = 0 ~, \qquad \mathbb{Y} R = -6 R~, 
	\eea
	while $X_\a$ is the chiral field strength of a $\sU(1)$ vector multiplet
	\bea
	\bar{\cD}_{\ad} X_{\a} = 0 ~, \qquad \cD^\a X_\a = \bar{\cD}_\ad \bar{X}^\ad~, \qquad \mathbb{Y} X_\a = - 3X_\a~,
	\eea
	and $G_{\a \ad}$ is a real vector superfield. These are related via
	\bea
	X_{\a} &=& \cD_{\a}R - \bar{\cD}^{\ad}G_{\a \ad} ~, \label{Bianchi1} \\
	{\rm i} \cD_{(\a}{}^{\gd} G_{\b ) \gd} &=&  \frac{1}{3} \cD_{(\a} X_{\b)} ~.
	\eea
\end{subequations}
This supergeometry is a $\sU(1)$ superspace \cite{Howe,GGRS} with vanishing super-Weyl tensor.


\subsection{$\cN>1$ case}

As was shown in \cite{KKR}, it follows from eq. \eqref{4.3} that for $\cN>1$ the algebra of degauged spinor covariant derivatives takes the form:\footnote{In the $\cN=2$ case, the torsion tensor $Y_{\a \b}^{ij}$ is reducible and should be replaced with $\hf \ve^{ij} Y_{\a \b}$.}
\begin{subequations}
	\label{U(N)algebra}
	\bea
	\{ \cD_\a^i , \cD_\b^j \}
	&=&
	4 S^{ij}  M_{\a\b} 
	+4\ve_{\a\b} Y^{ij}_{\g\d}  M^{\g\d}  
	-4\ve_{\a \b} S^{k[i} \mathbb{J}^{j]}{}_k
	+ 8{Y}_{\a\b}^{k(i}  \mathbb{J}^{j)}{}_k
	~,
	\\
	\label{U(N)algebra-b}
	\{ \cD_\a^i , \bar{\cD}^\bd_j \}
	&=&
	- 2 \ri \d_j^i\cD_\a{}^\bd
	+4\Big(
	\d^i_jG^{\g\bd}
	+\ri G^{\g\bd}{}^i{}_j
	\Big) 
	M_{\a\g} 
	+4\Big(
	\d^i_jG_{\a\gd}
	+\ri G_{\a\gd}{}^i{}_j
	\Big)  
	\bar{M}^{\bd\gd}
	\non\\
	&&
	+8 G_\a{}^\bd \mathbb{J}^i{}_j
	+4\ri\d^i_j G_\a{}^\bd{}^{k}{}_j \mathbb{J}^i{}_{k}
	-2\Big(
	\d^i_jG_\a{}^\bd
	+\ri G_\a{}^\bd{}^i{}_j
	\Big)
	\mathbb{Y} 
	~.
	\eea
	\esubeq
	The dimension-1 superfields introduced above have the following symmetry properties:  
	\bea
	S^{ij}=S^{ji}~, \qquad Y_{\a\b}^{ij}=Y_{\b\a}^{ij}=-Y_{\a \b}^{ji}~, \qquad {G_{\a \ad}{}^{i}{}_i} = 0~,
	\eea
	and satisfy the reality conditions
	\bea
	\overline{S^{ij}} =  \bar{S}_{ij}~,\quad
	\overline{Y_{\a\b}^{ij}} = \bar{Y}_{\ad\bd ij}~,\quad
	\overline{G_{\b\ad}} = G_{\a\bd}~,\quad
	\overline{G_{\b\ad}{}^{i}{}_j} = - G_{\a\bd}{}^j{}_{i}
	~.~~~~~~
	\eea
	The ${\sU}(1)_R$ charges of the complex superfields are:\footnote{We note that these torsions are uncharged for $\cN=4$. This follows from $\mathbb{Y}$ acting as a central charge in this case.}
	\bea
	{\mathbb Y} S^{ij}=\frac{2(4-\cN)}{\cN}S^{ij}~,\qquad
	{\mathbb Y}  Y^{ij}_{\a\b}=\frac{2(4-\cN)}{\cN}Y^{ij}_{\a\b}~.
	\eea
	Further, they satisfy the Bianchi identities:
	\begin{subequations}\label{BI-U2}
		\bea
		\cD_{\a}^{(i}S^{jk)}&=&0~, \quad
		\bar{\cD}_{\ad i}S^{jk} - \ri\cD^{\b (j}G_{\b\ad}{}^{k)}{}_i = \frac{1}{\cN+1} \d_i^{(j} \Big( 2 \bar{\cD}_{\ad l} S^{k)l} - \ri {\cD}^{\b |l|} G_{\b \ad}{}^{k)}{}_l\Big) ~, ~~~
		\\
		\cD_{(\a}^{(i}Y_{\b\g)}^{j)k}&=&0~, \quad
		\cD^{\b k} Y_{\a \b}^{ij} = - \cD_\a^{[i} S^{j]k}~, \quad
		\bar{\cD}_j^\bd Y_{\a \b}^{ij} = 2 \cD_{(\a}^i G_{\b)}{}^{\bd} - \ri \frac{\cN-2}{\cN+1} \cD_{(\a}^j G_{\b)}{}^{\bd i}{}_j~,~~~ \\
		\cD_{(\a}^{(i}G_{\b)\bd}{}^{j)}{}_k&=&\frac{1}{\cN+1} \cD_{(\a}^l G_{\b) \bd}{}^{(i}{}_l \d^{j)}_k~, 
		\qquad
		\cD_{(\a}^{[i}G_{\b)\bd}{}^{j]}{}_k=-\frac{1}{\cN-1} \cD_{(\a}^l G_{\b) \bd}{}^{[i}{}_l \d^{j]}_k~, 
		\\
		\cD_\a^iG^{\a \bd}&=&
		\frac{\ri}{2(\cN+1)} \Big( \frac{\cN+2}{\cN-1} \cD_\a^j G^{\a \bd i}{}_j + \ri \bar{\cD}^{\bd}_j S^{ij} \Big) ~.
		\eea
	\end{subequations}
	This defines a $\sU(\cN)$ superspace with vanishing super-Weyl tensor. For $\cN \leq 4$, it is the conformally flat limit of the supergeometry due to \cite{Howe}.

\begin{footnotesize}

\end{footnotesize}


\begin{thebibliography}{66}




\bibitem{FvNF}
D.~Z.~Freedman, P.~van Nieuwenhuizen and S.~Ferrara,
``Progress toward a theory of supergravity,''
Phys. Rev. D \textbf{13}, 3214 (1976).

\bibitem{DZ}
S.~Deser and B.~Zumino,
``Consistent supergravity,''
Phys. Lett. B \textbf{62}, 335 (1976).


\bibitem{SalamSezgin} 
A.~Salam and E.~Sezgin, {\it Supergravities in Diverse Dimensions}, Vol. 1 \& 2, 
North-Holland/World Scientific, 1989.



\bibitem{FvN}
S.~Ferrara and P.~van Nieuwenhuizen,
``Consistent supergravity with complex spin 3/2 gauge fields,''
Phys.\ Rev.\ Lett.\  {\bf 37}, 1669 (1976).

\bibitem{Faddeev:1967fc}
L.~D.~Faddeev and V.~N.~Popov,
``Feynman diagrams for the Yang-Mills field,''
Phys. Lett. B \textbf{25}, 29 (1967).

\bibitem{BRS}
C.~Becchi, A.~Rouet and R.~Stora,
``Renormalization of the Abelian Higgs-Kibble model,''
Commun. Math. Phys. \textbf{42}, 127 (1975);
``Renormalization of gauge theories,''
Annals Phys. \textbf{98}, 287 (1976).

\bibitem{Tyutin:1975qk}
I.~V.~Tyutin,
``Gauge invariance in field theory and statistical physics in operator formalism,''
Lebedev Institute Preprint {\bf 39} (1975),
[arXiv:0812.0580 [hep-th]].

\bibitem{Batalin:1981jr}
I.~A.~Batalin and G.~A.~Vilkovisky,
``Gauge algebra and quantization,''
Phys. Lett. B \textbf{102}, 27 (1981).

\bibitem{BV}
I.~A.~Batalin and G.~A.~Vilkovisky,
``Quantization of gauge theories with linearly dependent generators,''
Phys. Rev. D \textbf{28}, 2567 (1983)
[erratum: Phys. Rev. D \textbf{30}, 508 (1984)].

\bibitem{HT} 
M. Henneaux and C. Teitelboim, {\it Quantization of Gauge Systems},  Princeton University
Press, 1994.

\bibitem{Duff:1986hr}
M.~J.~Duff, B.~E.~W.~Nilsson and C.~N.~Pope,
``Kaluza-Klein supergravity,''
Phys. Rept. \textbf{130}, 1 (1986).

\bibitem{Appelquist:1987nr}
T.~Appelquist, A.~Chodos and P.~G.~O.~Freund,
{\it Modern Kaluza-Klein Theories}, Addison-Wesley, 1987. 

\bibitem{Duff:2025tot}
M.~J.~Duff, B.~E.~W.~Nilsson and C.~N.~Pope,
``Kaluza-Klein supergravity 2025,''
[arXiv:2502.07710 [hep-th]].

\bibitem{FSZ}
S.~Ferrara, J.~Scherk and B.~Zumino,
``Algebraic properties of extended supergravity theories,''
Nucl.\ Phys.\  B {\bf 121}, 393 (1977).


\bibitem{Cremmer:1977tt}
E.~Cremmer, J.~Scherk and S.~Ferrara,
``SU(4) invariant supergravity theory,''
Phys. Lett. B \textbf{74}, 61 (1978).

\bibitem{CJ}
E.~Cremmer and B.~Julia,
``The N=8 supergravity theory. 1. The Lagrangian,''
Phys. Lett. B \textbf{80}, 48 (1978);
``The SO(8) supergravity,''
Nucl.\ Phys.\  B {\bf 159},  141 (1979).

\bibitem{GZ1}
M. K.~Gaillard and B.~Zumino,
``Duality rotations for interacting fields,''
Nucl.\ Phys.\  B {\bf 193},  221 (1981). 

\bibitem{Zumino}
B.~Zumino, ``Duality rotations,''
in {\it Quantum Structure of Space and Time}, M.~J. Duff and
C.~J. Isham (Eds.), Cambridge University Press, 1982, pp. 363--373.


\bibitem{GR1}
G.~W.~Gibbons and D.~A.~Rasheed,
``Electric-magnetic duality rotations in nonlinear electrodynamics,''
Nucl.\ Phys.\  B {\bf 454}, 185 (1995) 
[arXiv:hep-th/9506035].

\bibitem{GR2}
G. W.~Gibbons and D. A.~Rasheed,
``SL(2,R) invariance of non-linear electrodynamics
coupled to an axion and a dilaton,''
Phys.\ Lett.\  B {\bf 365}, 46 (1996) 
[hep-th/9509141].

\bibitem{GZ2}
M.~K.~Gaillard and B.~Zumino,
``Self-duality in nonlinear electromagnetism,''
in {\it Supersymmetry and Quantum Field Theory},
J.~Wess and V.~P.~Akulov (Eds.), Springer Verlag, 1998, pp. 121--129 [arXiv:hep-th/9705226].

\bibitem{GZ3}
M.~K.~Gaillard and B.~Zumino,
``Nonlinear electromagnetic self-duality
and Legendre transformations,'' in {\it Duality and
	Supersymmetric Theories}, D.~I.~Olive and
P.~C.~West (Eds.), Cambridge University Press,
1999, pp. 33--48 [hep-th/9712103].

\bibitem{Tanii}
Y.~Tanii,
``Introduction to supergravities in diverse dimensions,''
hep-th/9802138.

\bibitem{ArakiT}
M.~Araki and Y.~Tanii,
``Duality symmetries in non-linear gauge theories,''
Int.\ J.\ Mod.\ Phys.\  A {\bf 14}, 1139 (1999) 
[hep-th/9808029].

\bibitem{BMZ}
D.~Brace, B.~Morariu and B.~Zumino,
``Duality invariant Born-Infeld theory,''
in {\it The Many Faces of the Superworld: Yury Golfand
Memorial Volume}, M. Shifman (Ed.),
World Scientific, 2000, p. 103--110 [hep-th/9905218].

\bibitem{ABMZ} 
P.~Aschieri, D.~Brace, B.~Morariu and B.~Zumino,
``Nonlinear self-duality in even dimensions,''
Nucl.\ Phys.\  B {\bf 574}, 551 (2000) 
[hep-th/9909021].


\bibitem{KT1}
S.~M.~Kuzenko and S.~Theisen,
``Supersymmetric duality rotations,''
JHEP {\bf 0003}, 034 (2000)
[arXiv:hep-th/0001068].



\bibitem{KT2}
S.~M.~Kuzenko and S.~Theisen,
``Nonlinear self-duality and supersymmetry,''
Fortsch.\ Phys.\  {\bf 49}, 273 (2001) [arXiv:hep-th/0007231].

%
\bibitem{Fronsdal}
C.~Fronsdal,
``Massless fields with integer spin,''
Phys.\ Rev.\  D {\bf18},   3624 (1978).

\bibitem{FF}
J.~Fang and C.~Fronsdal,
``Massless fields with half-integral spin,''
Phys.\ Rev.\  D {\bf 18}, 3630 (1978).

\bibitem{Fronsdal2}
C.~Fronsdal,
``Singletons and massless, integral-spin fields on de Sitter space,''
Phys.\ Rev.\  D {\bf 20},  848 (1979).

\bibitem{FF2}
J.~Fang and C.~Fronsdal,
``Massless, half-integer-spin fields in de Sitter space,''
Phys.\ Rev.\  D {\bf 22},  1361 (1980).

\bibitem{Fradkin:1986qy}
E.~S.~Fradkin and M.~A.~Vasiliev,
``Cubic interaction in extended theories of massless higher spin fields,''
Nucl. Phys. B \textbf{291}, 141 (1987).

\bibitem{Fradkin:1987ks}
E.~S.~Fradkin and M.~A.~Vasiliev,
``On the Gravitational Interaction of Massless Higher Spin Fields,''
Phys. Lett. B \textbf{189}, 89 (1987).

\bibitem{Vasiliev:1990en}
M.~A.~Vasiliev,
``Consistent equation for interacting gauge fields of all spins in (3+1)-dimensions,''
Phys. Lett. B \textbf{243}, 378 (1990).


\bibitem{FT} 
E.~S.~Fradkin and A.~A.~Tseytlin,  ``Conformal supergravity,''
Phys.\ Rept.\  {\bf 119}, 233 (1985).

\bibitem{Tseytlin} 
A.~A.~Tseytlin,
``On limits of superstring in AdS(5) x S**5,''
Theor.\ Math.\ Phys.\  {\bf 133}, 1376 (2002)
[Teor.\ Mat.\ Fiz.\  {\bf 133}, 69 (2002)]
[hep-th/0201112].


\bibitem{Segal}
A.~Y.~Segal,
``Conformal higher spin theory,''
Nucl. Phys. B \textbf{664}, 59 (2003)
[arXiv:hep-th/0207212 [hep-th]].


\bibitem{KR21-2}
S.~M.~Kuzenko and E.~S.~N.~Raptakis,
``Duality-invariant superconformal higher-spin models,''
Phys. Rev. D \textbf{104}, no.12, 125003 (2021)
[arXiv:2107.02001 [hep-th]].

\bibitem{Kuzenko:2023ebe}
S.~M.~Kuzenko and E.~S.~N.~Raptakis,
``Self-duality for N-extended superconformal gauge multiplets,''
Nucl. Phys. B \textbf{997}, 116378 (2023)
[arXiv:2308.10660 [hep-th]].

\bibitem{Minkowski} H. Minkowski, ``Die Grundgleichungen f\"ur die elektromagnetischen Vorg\"ange in bewegten K\"orpern,'' Nachrichten der K. Gesellschaft der Wissenschaften zu G\"ottingen. Mathematisch-physikalische Klasse, 53–111 (1908);
English Translation: ``The Fundamental equations for electromagnetic processes in moving bodies,'' 
in {\it Spacetime: Minkowski's Papers on Spacetime Physics},  V. Petkov (Ed.), Minkowski Institute Press, 2020, pp. 93--167.

\bibitem{Schrodinger:1935oqa}
E.~Schr{\"o}dinger,
``Contributions to Born's new theory of the electromagnetic field,''
Proc. Roy. Soc. Lond. A \textbf{150},  465 (1935).

\bibitem{BI} 
  M.~Born and L.~Infeld,
  ``Foundations of the new field theory,''
  Proc.\ Roy.\ Soc.\ Lond.\ A {\bf 144}, 425 (1934).

\bibitem{B-B} I. Bialynicki-Birula, ``Nonlinear electrodynamics: Variations on a theme by Born and Infeld,'' in {\it Quantum Theory of Particles and Fields}, 
B. Jancewicz and J. Lukierski (Eds.), 
World Scientific, 1983, pp. 31--48. 


\bibitem{Fradkin:1985qd}
E.~S.~Fradkin and A.~A.~Tseytlin,
``Nonlinear electrodynamics from quantized strings,''
Phys. Lett. B \textbf{163}, 123 (1985).

\bibitem{Leigh:1989jq}
R.~G.~Leigh,
``Dirac-Born-Infeld Action from Dirichlet sigma model,''
Mod. Phys. Lett. A \textbf{4}, 2767 (1989).


\bibitem{Tanii2} Y. Tanii, {\it Introduction to Supergravity}, Springer, 2014.


\bibitem{AFZ}
P.~Aschieri, S.~Ferrara and B.~Zumino,
``Duality rotations in nonlinear electrodynamics and in extended supergravity,''
Riv.\ Nuovo Cim.\  {\bf 31}, 625 (2008)
[arXiv:0807.4039 [hep-th]].



\bibitem{KMcC}
S.~M.~Kuzenko and S.~A.~McCarthy,
``Nonlinear self-duality and supergravity,''
JHEP {\bf 0302}, 038 (2003)
[hep-th/0212039].  

\bibitem{KMcC2}
S.~M.~Kuzenko and S.~A.~McCarthy,
``On the component structure of N=1 supersymmetric nonlinear electrodynamics,''
JHEP \textbf{05}, 012 (2005)
[arXiv:hep-th/0501172 [hep-th]].

\bibitem{K12} 
S.~M.~Kuzenko,
``Nonlinear self-duality in N=2 supergravity,''
JHEP {\bf 1206}, 012 (2012)
[arXiv:1202.0126 [hep-th]].


\bibitem{BG}
J.~Bagger and A.~Galperin,
``A new Goldstone multiplet for partially 
broken supersymmetry,''
Phys.\ Rev.\ D {\bf 55}, 1091 (1997) 
[arXiv:hep-th/9608177].

\bibitem{RT}
M.~Ro\v{c}ek and A.~A.~Tseytlin,
``Partial breaking of global D = 4 supersymmetry, 
constrained  superfields, and 3-brane actions,''
Phys.\ Rev.\ D {\bf 59},  106001 (1999) 
[arXiv:hep-th/9811232].

\bibitem{CF}
S.~Cecotti and S.~Ferrara,
``Supersymmetric Born-Infeld Lagrangians,''
Phys.\ Lett.\ B {\bf 187}, 335 (1987).


\bibitem{BG2}
J.~Bagger and A.~Galperin,
``The tensor Goldstone multiplet for partially broken supersymmetry,''
Phys.\ Lett.\  {\bf B412}, 296 (1997) 
[hep-th/9707061].


\bibitem{KT-M16} 
  S.~M.~Kuzenko and G.~Tartaglino-Mazzucchelli,
  ``Nilpotent chiral superfield in N=2 supergravity and 
  partial rigid supersymmetry breaking,''
  JHEP {\bf 1603}, 092 (2016)
  [arXiv:1512.01964 [hep-th]].


\bibitem{FS}
 G.~Festuccia and N.~Seiberg,
``Rigid supersymmetric theories in curved superspace,''
JHEP {\bf 1106}, 114 (2011) [arXiv:1105.0689 [hep-th]].

 \bibitem{NappiW} 
  C.~R.~Nappi and E.~Witten,
  ``A WZW model based on a nonsemisimple group,''
  Phys.\ Rev.\ Lett.\  {\bf 71}, 3751 (1993)
  [hep-th/9310112].


\bibitem{Ketov}
S.~V.~Ketov,
``A manifestly N=2 supersymmetric Born-Infeld action,''
Mod. Phys. Lett. A \textbf{14}, 501 (1999)
[arXiv:hep-th/9809121 [hep-th]];
``Born-Infeld-Goldstone superfield actions for gauge-fixed D5- and D3-branes in 6d,''
Nucl.\ Phys.\  {\bf B553}, 250 (1999) 
[hep-th/9812051].

\bibitem{BIK1}
S.~Bellucci, E.~Ivanov and S.~Krivonos,
``N=2 and N=4 supersymmetric Born-Infeld theories from nonlinear realizations,''
Phys. Lett. B \textbf{502}, 279 (2001)
[arXiv:hep-th/0012236 [hep-th]].


\bibitem{BIK2}
S.~Bellucci, E.~Ivanov and S.~Krivonos,
``Towards the complete N = 2 superfield Born-Infeld 
action with partially broken N = 4 supersymmetry,''
Phys.\ Rev.\ D {\bf 64},    025014 (2001) 
[arXiv:hep-th/0101195].

\bibitem{BCFKR} 
  J.~Broedel, J.~J.~M.~Carrasco, S.~Ferrara, R.~Kallosh and R.~Roiban,
  ``N=2 supersymmetry and U(1)-duality,''
  Phys.\ Rev.\ D {\bf 85}, 125036 (2012)
  [arXiv:1202.0014 [hep-th]].


\bibitem{Ivanov:2013maa}
E.~A.~Ivanov and B.~M.~Zupnik,
``Self-dual $\mathcal N=2$ Born-Infeld theory through auxiliary superfields,''
JHEP \textbf{05}, 061 (2014)
[arXiv:1312.5687 [hep-th]].


\bibitem{VA}
D.~V.~Volkov and V.~P.~Akulov,
``Possible universal neutrino interaction,''
  {JETP Lett.\  {\bf 16}, 438 (1972)}   
  [Pisma Zh.\ Eksp.\ Teor.\ Fiz.\   {\bf 16},  621 (1972)]; 
  ``Is the neutrino a Goldstone particle?,''
  Phys.\ Lett.\  B {\bf 46}, 109 (1973).

\bibitem{AV}
V.~P. Akulov and D.~V. Volkov, ``Goldstone fields with spin 1/2,''
   Theor. Math. Phys. {\bf 18}, 28 (1974)  28 [Teor. Mat. Fiz. {\bf 18}, 39 (1974)].
 
\bibitem{IZ_N3} 
E.~A.~Ivanov and B.~M.~Zupnik,
``N=3 supersymmetric Born-Infeld theory,''
Nucl.\ Phys.\ B {\bf 618}, 3 (2001)
[hep-th/0110074].

\bibitem{IZ1} 
E.~A.~Ivanov and B.~M.~Zupnik,
``New representation for Lagrangians of self-dual nonlinear electrodynamics,''
in {\it Supersymmetries and Quantum Symmetries. Proceedings of the 16th Max Born Symposium, SQS'01: Karpacz, Poland, September 21--25, 2001}, E. Ivanov (Ed.), Dubna, 2002, pp. 235--250 
[hep-th/0202203].

\bibitem{IZ2} 
E.~A.~Ivanov and B.~M.~Zupnik,
``New approach to nonlinear electrodynamics: Dualities as symmetries of interaction,''
Phys.\ Atom.\ Nucl.\  {\bf 67}, 2188 (2004)
[Yad.\ Fiz.\  {\bf 67}, 2212 (2004)]
[hep-th/0303192].


\bibitem{BLST}
I.~Bandos, K.~Lechner, D.~Sorokin and P.~K.~Townsend,
``A non-linear duality-invariant conformal extension of Maxwell's equations,''
Phys. Rev. D \textbf{102}, 121703 (2020)
[arXiv:2007.09092 [hep-th]].




\bibitem{K13} 
S.~M.~Kuzenko,
``Duality rotations in supersymmetric nonlinear electrodynamics revisited,''
JHEP {\bf 1303}, 153 (2013)
[arXiv:1301.5194 [hep-th]].



\bibitem{ILZ}
E.~Ivanov, O.~Lechtenfeld and B.~Zupnik,
``Auxiliary superfields in N=1 supersymmetric self-dual electrodynamics,''
JHEP \textbf{05}, 133 (2013)
[arXiv:1303.5962 [hep-th]].


\bibitem{BN} 
  G.~Bossard and H.~Nicolai,
 ``Counterterms vs. dualities,''
  JHEP {\bf 1108}, 074 (2011)
  [arXiv:1105.1273 [hep-th]].


\bibitem{CKR} 
 J.~J.~M.~Carrasco, R.~Kallosh and R.~Roiban,
  ``Covariant procedures for perturbative non-linear deformation of duality-invariant theories,''
Phys.\ Rev.\ D {\bf 85}, 025007 (2012) [arXiv:1108.4390 [hep-th]].

\bibitem{Chemissany:2011yv} 
W.~Chemissany, R.~Kallosh and T.~Ortin,
``Born-Infeld with higher derivatives,''
Phys.\ Rev.\ D {\bf 85}, 046002 (2012)
[arXiv:1112.0332 [hep-th]].

\bibitem{IZ3}
E.~A.~Ivanov and B.~M.~Zupnik,
``Bispinor auxiliary fields in duality-invariant electrodynamics revisited,''
Phys. Rev. D \textbf{87}, no.6, 065023 (2013)
[arXiv:1212.6637 [hep-th]].



\bibitem{BLST2}
I.~Bandos,  K.~Lechner, D.~Sorokin and P.~K.~Townsend, 
``ModMax meets Susy,'' 
JHEP \textbf{10}, 031 (2021)
[arXiv:2106.07547 [hep-th]].

\bibitem{K21}
S.~M.~Kuzenko,
``Superconformal duality-invariant models and $\mathcal{N} = 4$ SYM effective action,''
JHEP \textbf{09}, 180 (2021)
[arXiv:2106.07173 [hep-th]].

\bibitem{Kuzenko:2023ysh}
S.~M.~Kuzenko and I.~N.~McArthur,
``A supersymmetric nonlinear sigma model analogue of the ModMax theory,''
JHEP \textbf{05}, 127 (2023)
[arXiv:2303.15139 [hep-th]].

\bibitem{KTvN}
M.~Kaku, P.~K.~Townsend and P.~van Nieuwenhuizen,
``Gauge theory of the conformal and superconformal group,''
Phys. Lett. B \textbf{69}, 304 (1977).

\bibitem{KTvN2} 
  M.~Kaku, P.~K.~Townsend and P.~van Nieuwenhuizen,
  ``Properties of conformal supergravity,''
  Phys.\ Rev.\ D {\bf 17}, 3179 (1978).


\bibitem{MacDowell:1977jt}
S.~W.~MacDowell and F.~Mansouri,
``Unified geometric theory of gravity and supergravity,''
Phys. Rev. Lett. \textbf{38}, 739 (1977)
[erratum: Phys. Rev. Lett. \textbf{38}, 1376 (1977)].


\bibitem{ButterN=1}
D.~Butter,
``N=1 conformal superspace in four dimensions,''
Annals Phys. \textbf{325}, 1026-1080 (2010)
[arXiv:0906.4399 [hep-th]].

\bibitem{ButterN=2} 
D.~Butter,
``N=2 conformal superspace in four dimensions,''
JHEP {\bf 1110}, 030 (2011)
[arXiv:1103.5914 [hep-th]].



\bibitem{KR23}
S.~M.~Kuzenko and E.~S.~N.~Raptakis,
``$ \mathcal{N} $ = 3 conformal superspace in four dimensions,''
JHEP \textbf{03}, 026 (2024)
[arXiv:2312.07242 [hep-th]].

\bibitem{ButterN=4}
D.~Butter, F.~Ciceri and B.~Sahoo,
``$N=4$ conformal supergravity: the complete actions,''
JHEP \textbf{01}, 029 (2020)
[arXiv:1910.11874 [hep-th]].


\bibitem{BKNT-M1} 
D.~Butter, S.~M.~Kuzenko, J.~Novak and G.~Tartaglino-Mazzucchelli,
``Conformal supergravity in three dimensions: New off-shell formulation,''
JHEP {\bf 1309}, 072 (2013)
[arXiv:1305.3132 [hep-th]].


\bibitem{BKNT-M2}
D.~Butter, S.~M.~Kuzenko, J.~Novak and G.~Tartaglino-Mazzucchelli,
``Conformal supergravity in three dimensions: Off-shell actions,''
JHEP \textbf{10}, 073 (2013)
[arXiv:1306.1205 [hep-th]].


\bibitem{KNT-M}
S.~M.~Kuzenko, J.~Novak and G.~Tartaglino-Mazzucchelli,
``N=6 superconformal gravity in three dimensions from superspace,''
JHEP \textbf{01}, 121 (2014)
[arXiv:1308.5552 [hep-th]].

\bibitem{BKNT-M15}
  D.~Butter, S.~M.~Kuzenko, J.~Novak and G.~Tartaglino-Mazzucchelli,
``Conformal supergravity in five dimensions: New approach and applications,''
JHEP {\bf 1502}, 111 (2015).
[arXiv:1410.8682 [hep-th]].


\bibitem{BKNT}
  D.~Butter, S.~M.~Kuzenko, J.~Novak and S.~Theisen,
  ``Invariants for minimal conformal supergravity in six dimensions,''
  JHEP {\bf 1612}, 072 (2016)
  [arXiv:1606.02921 [hep-th]].

\bibitem{Kennedy:2025nzm}
C.~Kennedy and G.~Tartaglino-Mazzucchelli,
``Six-dimensional $ \mathcal{N} $ = (2, 0) conformal superspace,''
JHEP \textbf{08}, 215 (2025)
[arXiv:2506.01630 [hep-th]].

\bibitem{Boulanger:2004eh}
N.~Boulanger,
``A Weyl-covariant tensor calculus,''
J. Math. Phys. \textbf{46}, 053508 (2005)
[arXiv:hep-th/0412314 [hep-th]].


\bibitem{BEG}
T. N.  Bailey, M. G.  Eastwood, A. R.  Gover, 
``Thomas's structure bundle for conformal, projective and related structures,''
Rocky Mt. J. Math. {\bf 24}, 1191 (1994).

\bibitem{Gover} A. R. Gover, ``Invariant theory and calculus for conformal geometries,''
Adv. Math. {\bf 163}, 206 (2001). 

\bibitem{Thomas} T. Y. Thomas,
{\it The Differential Invariants of Generalized Spaces},  Cambridge University Press, 
1934.

  
\bibitem{BPB} 
  L.~Bonora, P.~Pasti and M.~Bregola,
  ``Weyl cocycles,''
  Class.\ Quant.\ Grav.\  {\bf 3}, 635 (1986).
  

\bibitem{DS}
S.~Deser and A.~Schwimmer,
  ``Geometric classification of conformal anomalies in arbitrary dimensions,''
  Phys.\ Lett.\ B {\bf 309}, 279 (1993)
  [hep-th/9302047].


\bibitem{KMM} 
  D.~R.~Karakhanian, R.~P.~Manvelyan and R.~L.~Mkrtchian,
  ``Trace anomalies and cocycles of Weyl and diffeomorphism groups,''
  Mod.\ Phys.\ Lett.\ A {\bf 11}, 409 (1996)
  [hep-th/9411068]. 
  


\bibitem{Deser70} 
S.~Deser, 
``Scale invariance and gravitational coupling,''
Annals Phys.\  {\bf 59}, 248 (1970).

\bibitem{Zumino70} B. Zumino, 
``Effective Lagrangians and broken symmetries," 
in {\it Lectures on Elementary Particles and Quantum Field Theory,
Vol. 2}, S. Deser, M. Grisaru and H. Pendleton (Eds.),
Cambridge, Mass. 1970, pp. 437-500.



\bibitem{BK} I.~L.~Buchbinder and S.~M.~Kuzenko,
{\it Ideas and Methods of Supersymmetry and
	Supergravity or a Walk Through Superspace}, IOP, Bristol, 1998.

\bibitem{WB} J.~Wess and J.~Bagger,
{\it Supersymmetry and Supergravity},
Princeton University Press, 1992.


\bibitem{KP}
S.~M.~Kuzenko and M.~Ponds,
``Conformal geometry and (super)conformal higher-spin gauge theories,''
JHEP \textbf{05}, 113 (2019)
[arXiv:1902.08010 [hep-th]].


\bibitem{Vasiliev}
M.~A.~Vasiliev,
``Bosonic conformal higher-spin fields of any symmetry,''
Nucl. Phys. B \textbf{829}, 176 (2010)
[arXiv:0909.5226 [hep-th]].


\bibitem{AF} 
P.~Aschieri and S.~Ferrara,
``Constitutive relations and Schroedinger's formulation of nonlinear electromagnetic theories,''
JHEP {\bf 1305}, 087 (2013)
[arXiv:1302.4737 [hep-th]].

\bibitem{AFT} 
P.~Aschieri, S.~Ferrara and S.~Theisen,
``Constitutive relations, off shell duality rotations and the hypergeometric form of Born-Infeld theory,''
Springer Proc.\ Phys.\  {\bf 153}, 23 (2014)
[arXiv:1310.2803 [hep-th]].



\bibitem{Hatsuda:1999ys}
M.~Hatsuda, K.~Kamimura and S.~Sekiya,
``Electric magnetic duality invariant Lagrangians,''
Nucl. Phys. B \textbf{561}, 341 (1999)
[arXiv:hep-th/9906103 [hep-th]].

\bibitem{Kuzenko:2024zra}
S.~M.~Kuzenko and E.~S.~N.~Raptakis,
``Higher-derivative deformations of the ModMax theory,''
JHEP \textbf{06}, 162 (2024)
[arXiv:2404.09108 [hep-th]].

\bibitem{Osborn} 
  H.~Osborn,
  ``Local couplings and Sl(2,R) invariance for gauge theories at one loop,''
  Phys.\ Lett.\ B {\bf 561}, 174 (2003)
  [hep-th/0302119].

\bibitem{BPT} 
I.~L.~Buchbinder, N.~G.~Pletnev and A.~A.~Tseytlin, ``Induced N=4 conformal supergravity,'' Phys.\ Lett.\ B {\bf 717}, 274 (2012)
[arXiv:1209.0416 [hep-th]].


\bibitem{Grasso:2023hmv}
D.~T.~Grasso, S.~M.~Kuzenko and J.~R.~Pinelli,
``Weyl invariance, non-compact duality and conformal higher-derivative sigma models,''
Eur. Phys. J. C \textbf{83}, no.3, 206 (2023)
[arXiv:2301.00577 [hep-th]].

\bibitem{Hjelmeland:1997eg}
S.~E.~Hjelmeland and U.~Lindstr\"om,
``Duality for the no-nspecialist,''
[arXiv:hep-th/9705122 [hep-th]].



\bibitem{Kuzenko:2019nlm}
S.~M.~Kuzenko,
``Manifestly duality-invariant interactions in diverse dimensions,''
Phys. Lett. B \textbf{798}, 134995 (2019)
[arXiv:1908.04120 [hep-th]].


\bibitem{Ferko:2023wyi}
C.~Ferko, S.~M.~Kuzenko, L.~Smith and G.~Tartaglino-Mazzucchelli,
``Duality-invariant nonlinear electrodynamics and stress tensor flows,''
Phys. Rev. D \textbf{108}, no.10, 106021 (2023)
[arXiv:2309.04253 [hep-th]].

\bibitem{Conti:2018jho}
R.~Conti, L.~Iannella, S.~Negro and R.~Tateo,
``Generalised Born-Infeld models, Lax operators and the $ \mathrm{T}\overline{\mathrm{T}} $ perturbation,''
JHEP \textbf{11}, 007 (2018)
[arXiv:1806.11515 [hep-th]].

\bibitem{Babaei-Aghbolagh:2022uij}
H.~Babaei-Aghbolagh, K.~B.~Velni, D.~M.~Yekta and H.~Mohammadzadeh,
``Emergence of non-linear electrodynamic theories from $T\bar T$-like deformations,''
Phys. Lett. B \textbf{829}, 137079 (2022)
[arXiv:2202.11156 [hep-th]].

\bibitem{Ferko:2022iru}
C.~Ferko, L.~Smith and G.~Tartaglino-Mazzucchelli,
``On current-squared flows and ModMax theories,''
SciPost Phys. \textbf{13}, no.2, 012 (2022)
[arXiv:2203.01085 [hep-th]].

\bibitem{Ferko:2024zth}
C.~Ferko, S.~M.~Kuzenko, K.~Lechner, D.~P.~Sorokin and G.~Tartaglino-Mazzucchelli,
``Interacting chiral form field theories and $ T\overline{T} $-like flows in six and higher dimensions,''
JHEP \textbf{05}, 320 (2024)
[arXiv:2402.06947 [hep-th]].

\bibitem{FL} 
E.~S.~Fradkin and V.~Y.~Linetsky,
``Cubic interaction in conformal theory of integer higher-spin fields in 
four dimensional space-time,''  Phys.\ Lett.\ B {\bf 231}, 97 (1989).

\bibitem{FL2}
E.~S.~Fradkin and V.~Y.~Linetsky,
``Superconformal higher spin theory in the cubic approximation,''
Nucl. Phys. B \textbf{350}, 274 (1991).

\bibitem{Murcia:2025psi}
{\'A}.~J.~Murcia,
``Novel duality-invariant theories of electrodynamics,''
Phys. Rev. D \textbf{112}, no.10, L101902 (2025)
[arXiv:2507.16502 [hep-th]].

\bibitem{Kuzenko:2026kvr}
S.~M.~Kuzenko,
``On nonlinear self-duality in $4p$ dimensions,''
[arXiv:2601.13022 [hep-th]].

\bibitem{Bandos:2020hgy}
I.~Bandos, K.~Lechner, D.~Sorokin and P.~K.~Townsend,
``On p-form gauge theories and their conformal limits,''
JHEP \textbf{03}, 022 (2021)
[arXiv:2012.09286 [hep-th]].


\bibitem{GGRS}
S.~J.~Gates Jr., M.~T.~Grisaru, M.~Ro\v{c}ek and W.~Siegel,
\textit{Superspace Or One Thousand and One Lessons in Supersymmetry},
Benjamin/Cummings (Reading, MA), 1983,
[arXiv:hep-th/0108200 [hep-th]].


\bibitem{Hutchings:2024qqf}
D.~Hutchings and M.~Ponds,
``Spin-(s, j) projectors and gauge-invariant spin-s actions in maximally symmetric backgrounds,''
JHEP \textbf{07}, 292 (2024)
[arXiv:2401.04523 [hep-th]].

\bibitem{PenroseR} R. Penrose and W. Rindler, 
{\it Spinors and Space-Time: Volume 2, Spinor and Twistor Methods in Space-Time Geometry}, Cambridge University Press,  
1986.


\bibitem{Cederwall:2025ywy}
M.~Cederwall, J.~Hutomo, S.~M.~Kuzenko, K.~Lechner and D.~P.~Sorokin,
``Some remarks on invariants,''
[arXiv:2509.14350 [hep-th]].

\bibitem{KRTM1}
S.~M.~Kuzenko, E.~S.~N.~Raptakis and G.~Tartaglino-Mazzucchelli,
``Superspace approaches to $\mathcal{N}=1$ supergravity,''
in: {\it Handbook of Quantum Gravity}, C. Bambi,  L. Modesto, I. Shapiro, I. (Eds.) Springer, Singapore (2024), 
\href{https://link.springer.com/referenceworkentry/10.1007/978-981-19-3079-9_40-1}
[arXiv:2210.17088 [hep-th]].

\bibitem{KRTM2}
S.~M.~Kuzenko, E.~S.~N.~Raptakis and G.~Tartaglino-Mazzucchelli,
``Covariant superspace approaches to ${\cal N}=2$ supergravity,''
in: {\it Handbook of Quantum Gravity}, C. Bambi,  L. Modesto, I. Shapiro, I. (Eds.) Springer, Singapore (2024), 
\href{https://link.springer.com/referenceworkentry/10.1007/978-981-19-3079-9_44-1}
[arXiv:2211.11162 [hep-th]].


\bibitem{KR21}
S.~M.~Kuzenko and E.~S.~N.~Raptakis,
``Extended superconformal higher-spin gauge theories in four dimensions,''
JHEP \textbf{12}, 210 (2021)
[arXiv:2104.10416 [hep-th]].

\bibitem{ERThesis}
E.~S.~N.~Raptakis, 
``Aspects of superconformal symmetry,'' PhD thesis, UWA, 2023 
[arXiv:2403.02700 [hep-th]].

\bibitem{KKR}
N.~E.~Koning, S.~M.~Kuzenko and E.~S.~N.~Raptakis,
``Embedding formalism for ${\mathcal N}$-extended AdS superspace in four dimensions,''
JHEP \textbf{11}, 063 (2023)
[arXiv:2308.04135 [hep-th]].

\bibitem{KKR2}
N.~E.~Koning, S.~M.~Kuzenko and E.~S.~N.~Raptakis,
``The anti-de Sitter supergeometry revisited,''
JHEP \textbf{02}, 175 (2025)
[arXiv:2412.03172 [hep-th]].

\bibitem{KT-M2009}
S.~M.~Kuzenko and G.~Tartaglino-Mazzucchelli,
``Different representations for the action principle in 4D N = 2 supergravity,''
JHEP \textbf{04} (2009), 007
[arXiv:0812.3464 [hep-th]].


\bibitem{HST}
P.~S.~Howe, K.~S.~Stelle and P.~K.~Townsend,
``Supercurrents,''
Nucl. Phys. B \textbf{192}, 332 (1981).

\bibitem{FZ2}
S.~Ferrara and B.~Zumino,
``Structure of conformal supergravity,''  Nucl.\ Phys.\  B {\bf 134}, 301 (1978).

\bibitem{KMT}
S.~M.~Kuzenko, R.~Manvelyan and S.~Theisen,
``Off-shell superconformal higher spin multiplets in four dimensions,''
JHEP \textbf{07}, 034 (2017)
[arXiv:1701.00682 [hep-th]].


\bibitem{KPR}
S.~M.~Kuzenko, M.~Ponds and E.~S.~N.~Raptakis,
``New locally (super)conformal gauge models in Bach-flat backgrounds,''
JHEP \textbf{08}, 068 (2020)
[arXiv:2005.08657 [hep-th]].


\bibitem{KT}
S.~M.~Kuzenko and S.~Theisen,
``Correlation functions of conserved currents in N=2 superconformal theory,''
Class. Quant. Grav. \textbf{17}, 665 (2000)
[arXiv:hep-th/9907107 [hep-th]].

%

\bibitem{SG81}
W.~Siegel and S.~J.~Gates Jr.,
``Superprojectors,''
Nucl. Phys. B \textbf{189}, 295 (1981).


\bibitem{GSW}
R.~Grimm, M.~Sohnius and J.~Wess,
``Extended supersymmetry and gauge theories,''
Nucl.\ Phys.\  B {\bf 133}, 275 (1978).


\bibitem{Mezincescu}
L.~Mezincescu,
``On the superfield formulation of O(2) supersymmetry,''
Dubna preprint JINR-P2-12572 (June, 1979).


\bibitem{ButterK}
D.~Butter and S.~M.~Kuzenko,
``New higher-derivative couplings in 4D N = 2 supergravity,''
JHEP {\bf 1103}, 047 (2011) [arXiv:1012.5153 [hep-th]].

\bibitem{ButterK2}
D.~Butter and S.~M.~Kuzenko,
``N=2 AdS supergravity and supercurrents,''
JHEP \textbf{07}, 081 (2011)
[arXiv:1104.2153 [hep-th]].


\bibitem{HKR}
D.~Hutchings, S.~M.~Kuzenko and E.~S.~N.~Raptakis,
``The $\mathcal{N}=2$ superconformal gravitino multiplet,''
Phys. Lett. B \textbf{845}, 138132 (2023)
[arXiv:2305.16029 [hep-th]].

\bibitem{Ivanov:2024bsb}
E.~Ivanov and N.~Zaigraev,
``N=2 superconformal gravitino in harmonic superspace,''
Phys. Lett. B \textbf{862}, 139333 (2025)
[arXiv:2412.14822 [hep-th]].

\bibitem{Siegel}
W.~Siegel,
``On-shell O($N$) supergravity in superspace,''
Nucl. Phys. B \textbf{177}, 325 (1981).

\bibitem{Howe}
P.~S.~Howe,
``A superspace approach to extended conformal supergravity,''
Phys.\ Lett.\ B {\bf 100}, 389 (1981);
``Supergravity in superspace,''  Nucl.\ Phys.\  B {\bf 199}, 309 (1982).

\bibitem{FT1982} 
  E.~S.~Fradkin and A.~A.~Tseytlin,
  ``Asymptotic freedom in extended conformal supergravities,''
  Phys.\ Lett.\ B {\bf 110}, 117 (1982);
  ``One-loop beta function in conformal supergravities,''
  Nucl.\ Phys.\ B {\bf 203}, 157 (1982).

\bibitem{Paneitz}
  S.~M.~Paneitz,
 ``A quartic conformally covariant differential operator for 
 arbitrary pseudo-Riemannian manifolds,'' MIT preprint, March 1983; 
 published posthumously in:  SIGMA {\bf 4} (2008), 036,
  [arXiv:0803.4331 [math.DG]].
  

\bibitem{Riegert}
  R.~J.~Riegert,
  ``A non-local action for the trace anomaly,''
  Phys.\ Lett.\ B {\bf 134} (1984) 56.

\bibitem{Butter:2013lta}
D.~Butter, B.~de Wit, S.~M.~Kuzenko and I.~Lodato,
``New higher-derivative invariants in N=2 supergravity and the Gauss-Bonnet term,''
JHEP \textbf{12}, 062 (2013)
[arXiv:1307.6546 [hep-th]].

\bibitem{Wunsch} 
  V.~W\"unsch,
  ``Some new conformal covariants,''  Journal of Analysis and Its Applications, {\bf 19},    339 (2000).


\end{thebibliography}
\end{document}